\documentclass[twocolumn]{aastex63}
\usepackage{graphicx,times}    
\usepackage{amssymb,amsmath}
\usepackage{amsfonts}
\usepackage{subfigure}
\usepackage{natbib}
\shorttitle{Sample article}
\shortauthors{Zhang et al.}
\begin{document}
\title{Ba-enhanced dwarf and subgiant stars in the LAMOST Galactic surveys}
\correspondingauthor{Hua-Wei Zhang, Xiao-Wei Liu}
\email{zhanghw@pku.edu.cn, x.liu@ynu.edu.cn}
\author{Meng Zhang}
\affiliation{Department of Astronomy, School of Physics, Peking University, Beijing 100871, P.R. China\\}
\affiliation{Kavli institute of Astronomy and Astrophysics, Peking University, Beijing 100871, P.R. China\\}
\affiliation{South-Western Institute For Astronomy Research, Yunnan University, Kunming 650500, P.R. China\\}
\affiliation{National Astronomical Observatories, Chinese Academy of Sciences, Beijing 100012, P.R. China\\}
\author{Maosheng Xiang}
\affiliation{National Astronomical Observatories, Chinese Academy of Sciences, Beijing 100012, P.R. China\\}
\affiliation{Institute for Frontiers in Astronomy and Astrophysics, Beijing Normal University, Beijing 102206, P.R. China\\}
\affiliation{Max-Planck Institute for Astronomy, K$\ddot{o}$nigstuhl 17, D-69117 Heidelberg, Germany \\}
\author{Hua-Wei Zhang}
\affiliation{Department of Astronomy, School of Physics, Peking University, Beijing 100871, P.R. China\\}
\affiliation{Kavli institute of Astronomy and Astrophysics, Peking University, Beijing 100871, P.R. China\\}
\author{Yuan-Sen Ting}
\affiliation{Research School of Astronomy \& Astrophysics, Australian National University, Cotter Rd., Weston, ACT 2611, Australia\\}
\affiliation{School of Computing, Australian National University, Acton, ACT 2601, Australia\\}
\author{Ya-qian Wu}
\affiliation{National Astronomical Observatories, Chinese Academy of Sciences, Beijing 100012, P.R. China\\}
\author{Xiao-Wei Liu}
\affiliation{South-Western Institute For Astronomy Research, Yunnan University, Kunming 650500, P.R. China\\}
\begin{abstract}
Ba-enhanced stars are interesting probes of stellar astrophysics and Galactic formation history.
In this work, we investigate the chemistry and kinematics for a large sample of Ba-enhanced ([Ba/Fe]$>$1.0) dwarf and subgiant stars with $5000 < T_{\rm eff }< 6700$\,K from LAMOST. We find that both stellar internal evolution process and external mass exchange due to binary evolution are responsible for the origins of the Ba-enhancement of our sample stars.
About one third of them exhibit C and N enhancement and ultraviolet brightness excess, indicating they are products of binary evolution.
The remaining Ba-enhanced stars with normal C and N abundances are mostly warm stars with $T_{\rm eff} > 6000$\,K. They are likely consequences of stellar internal elemental transport processes, but they show very different elemental patterns to the hotter Am/Fm stars. Our results reveal a substantially lack of high-[$\alpha$/Fe] Ba-enhanced stars in the [Fe/H]--[$\alpha$/Fe] plane, which we dub as a {\em high-$\alpha$ desert}. We suggest it is due to a lower efficiency for producing Ba-enhanced stars by low-mass AGB progenitors in binary systems. Our results call for detailed modellings for these Ba-enhanced stellar peculiars, in the context of both stellar internal elemental transport and external mass accretion. 
\end{abstract}

\section{Introduction}
\label{sec:intro}
 Barium (Ba) is an s-process element, which is mainly synthesized in the interiors of asymptotic giant branch (AGB) stars (\citealt{Bus1999, Her2005,Kap2011}). 
 Ba-rich materials in the interior of AGB stars are transported to their surface via the third dredge-up process (TDP), leading to the presence of Ba-enhanced AGB stars, which have been discovered over half of a century ago (\citealt{Bid1951, Esc2019, Jor2019}).

Stars at evolutionary phases other than AGB, such as main-sequence (MS) and red giant branch (RGB), have also been observed exhibiting Ba enhancement \citep[e.g.,][]{Nor1991,Nor1991A, Nor1994, Jor1992, Nor2000}.
For these stars, different mechanisms have been proposed to explain their Ba enhancement. 
The most frequently mentioned mechanism is the binary evolution, in which the stars obtain Ba-rich materials from their AGB companions via mass transfer (\citealt{Web1986, Han1995}) or stellar wind accretion (\citealt{Bof1988, Jor1992, Han1995}). 
Indeed, many Ba-rich stars are found to be in binary systems, confirmed via either radial velocity variations (\citealt{Mac1972,Gri1980,Gri1981,McC1983,Jor1988,McC1990,Jor1998, Ud1998a, Ud1998b}) or  ultraviolet (UV) brightness excess (\citealt{Boh1984, Boh2000,Gra2011}).
The AGB stars could synthesize not only Ba but also C.
Ba-rich stars originated from binary evolution might also exhibit C enhancement, but other species such as the $\alpha$-elements and iron-peak elements are not expected to be altered  (\citealt{All2006,Yang2016, Dec2016}).
Currently, literature works of Ba-enhanced stars have mostly focused on giant stars, whereas the dwarfs are relatively less well explored.

Another mechanism producing Ba-enhanced stars is via element transport by stellar internal evolution. For stars with an intermediate initial mass ($\gtrsim1.4\,M_\odot$), the competition between gravity and radiative acceleration may lead to an increase of surface Ba abundance by up to 1000 times (\citealt{Gha2018, Xiang2020}). Ba-enhanced stars due to this mechanism do not have to be in a binary system. But their abundances of other element species, such as $\alpha$ elements and iron peak elements, may exhibit deviations from their initial values. For instance, [Mg/Fe] are suggested to exhibit a decrease \citep{Vic2010,Xiang2020,Cha2021,Guo2021}. Ba-enhanced stars due to this mechanism have been shown to be common for main-sequence and subgiant stars of intermediate mass, but it remains to be further explored for the case of low-mass stars. While stellar evolution models have predicted that the element transport effect for low-mass stars ($<1.3\,M_\odot$) are insignificant (\citealt{Tal2006, Vic2010, Mic2011}), detailed modelings are mainly focused on stars of around solar metallicity, and on elements with atomic mass number smaller than Zn, but the elemental transport effect for Ba awaits more detailed modeling.    

In this work, we present a study of the chemistry and kinematics for a large sample of Ba-enhanced ([Ba/Fe] $ > 1.0$\,dex) dwarf and subgiant stars from the LAMOST spectroscopic surveys (\citealt{Zhao2012, Deng2012, Liu2014}). The LAMOST survey has taken low-resolution ($R\sim1800$) spectra for about 6 million stars in its fifth data release (DR5)\footnote{http://dr5.lamost.org}. Stellar abundances for 16 individual elements have been derived from these spectra (\citealt{Xiang2019}). The large spectral sample size is crucial for a systematic characterization of Ba-enhanced stars.
From the LAMOST spectra for 0.4 million giant stars, \citet{Nor2019} identified 895 s-process-rich candidates.
Based on the elemental abundance determinations for the LAMOST DR5 sample, \citet{Xiang2020} have studied the chemistry and kinematics of metal-rich ([Fe/H] $>-0.2$\,dex), hot ($\gtrsim6700$\,K, A/F type) stars of the Ba-enhanced dwarfs and subgiants. Our goal in this work is to carry out a detailed analysis of chemistry and kinematics for all Ba-enhanced dwarf and subgiant stars with $T_{\rm eff} < 6700$\,K, for which stellar abundances are determined from the LAMOST spectra with higher precision and accuracy than the hotter stars.

The paper is organized as follows.
In Section\,2, we describe the selection of the Ba-enhanced star samples of LAMOST.
Results of chemistry and kinematics are presented in Section\,3.
We then discuss their origins in Section\,4.
Section\,5 presents the conclusion.

\section{The Ba-enhanced star sample} 
\label{section}
The value-added catalog of \citet{Xiang2019} provides the abundances of 16 elements, including C, N, O, Na, Mg, Al, Si, Ca, Ti, Cr, Mn, Fe, Co, Ni, Cu, and Ba, for 6 million stars deduced from the LAMOST DR5 low-resolution spectra. The abundances are derived with {\it The DD-Payne}, which is a hybrid method that combines the data-driven approach with priors of spectral flux gradients from stellar atmospheric models \citep{Ting2017a, Xiang2019}.

To ensure the precision, we restrict our analysis only for stars with the spectral signal-to-noise ratio (S/N) higher than 50, for which the statistical uncertainties in abundance estimates are about 0.05\,dex for [Fe/H], [Mg/Fe], [Ca/Fe], [Ti/Fe], [Cr/Fe], [Ni/Fe]; 0.1\,dex for [C/Fe], [N/Fe], and 0.2-0.3\,dex for [Ba/Fe].

\begin{figure*}[htb!]
\centering   
\subfigure
{
	\begin{minipage}{0.5\linewidth}
	\centering          
	\includegraphics[width=0.95\columnwidth]{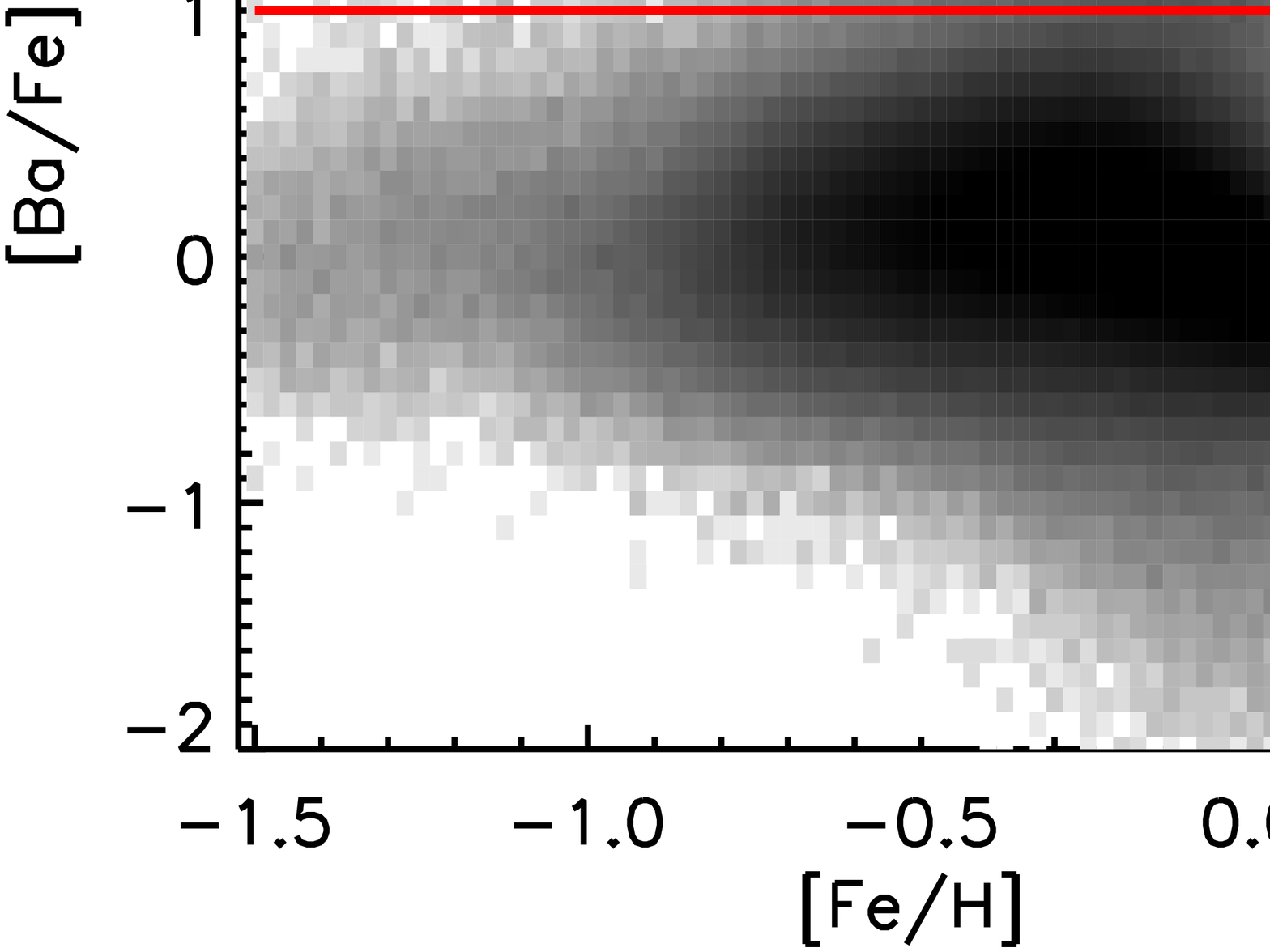}
	\end{minipage}
}	
\subfigure
{
	\begin{minipage}{0.42\linewidth}
	\centering     
	\includegraphics[width=0.95\columnwidth]{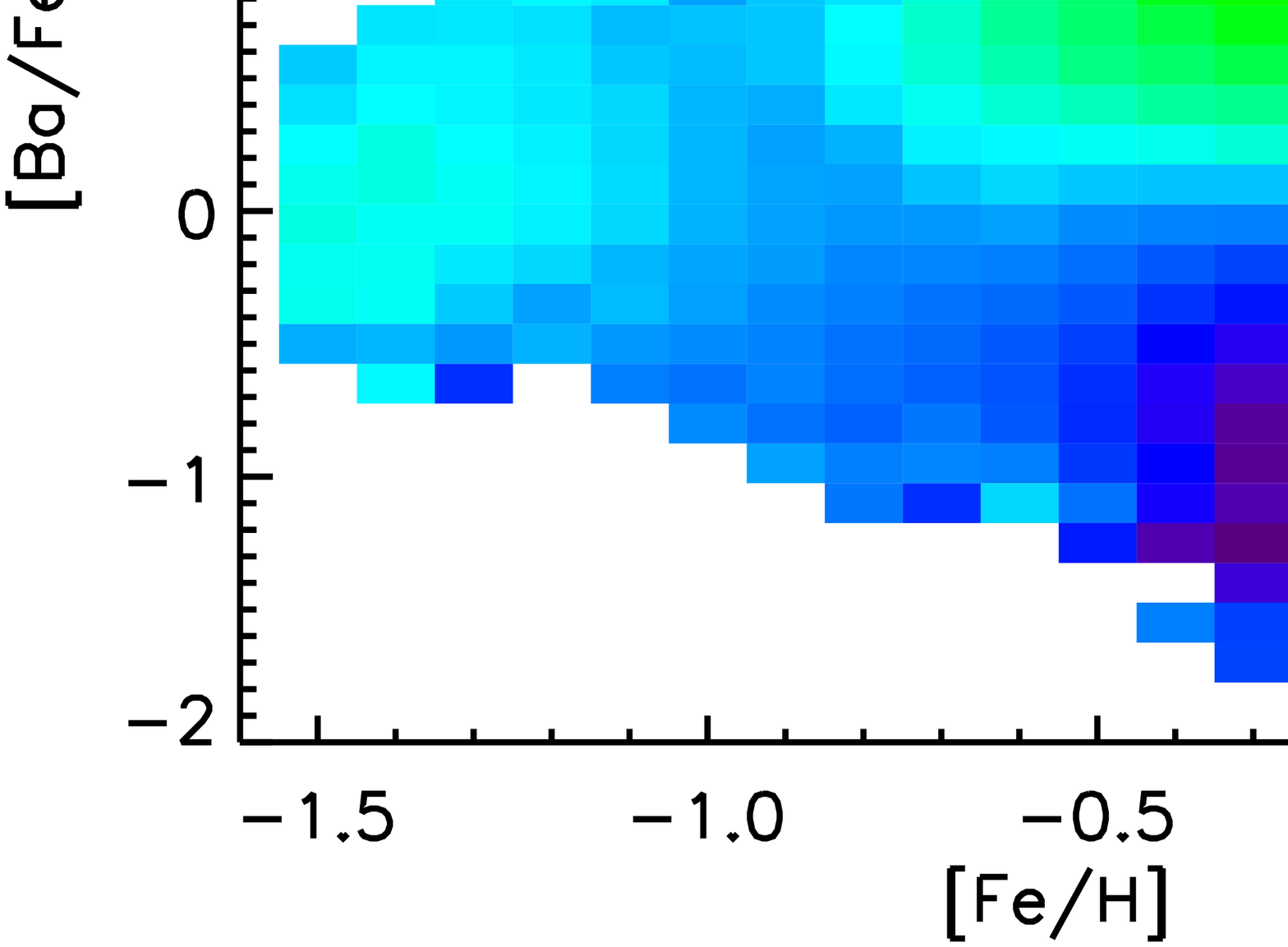}
	\end{minipage}
}
\subfigure
{
	\begin{minipage}{0.45\linewidth}
	\centering  
	\includegraphics[width=0.95\columnwidth]{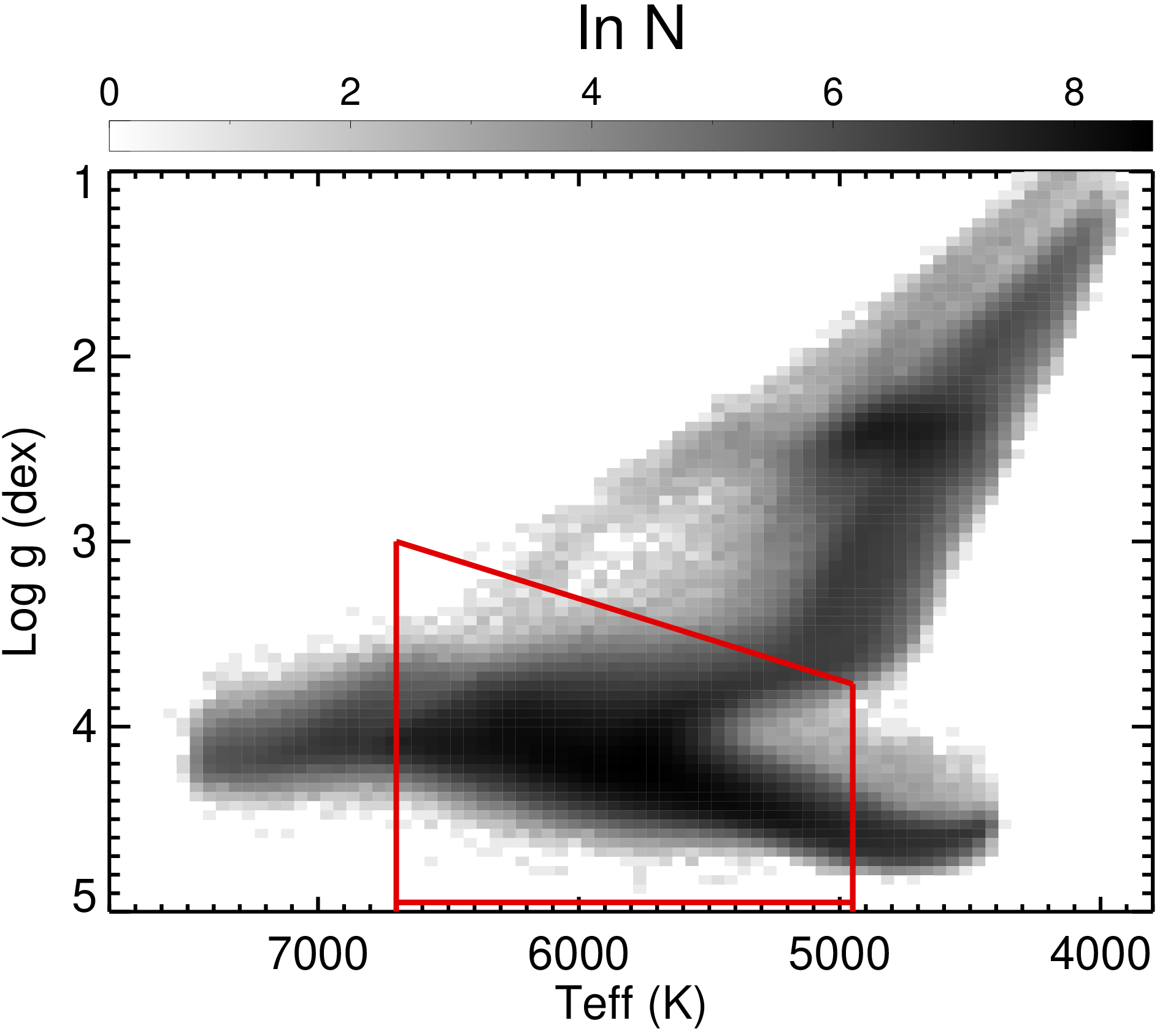}  
	\end{minipage}
}	
\subfigure
{
	\begin{minipage}{0.45\linewidth}
	\centering  
	\includegraphics[width=0.95\columnwidth]{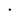}  
	\end{minipage}
}
\vspace{2.em}
\caption{Distribution of the LAMOST DD-Payne stellar parameters and abundances in the [Fe/H]--[Ba/Fe] and $T_{\rm eff}$--$\log\,g$ (Kiel diagram) planes. {$\it Top-left$}: Number density distributions in the [Fe/H]--[Ba/Fe] plane for dwarfs and subgiants with $5000 < T_{\rm eff} < 7500$\,K. The horizontal red line delineates the criterion we adopt to define a Ba-enhanced star (${\rm [Ba/Fe]} > 1.0$).  {$\it Top-right$}: Similar to the $Top-left$ panel, but color-coded by the mean effective temperatures for stars in each [Fe/H] and [Ba/Fe] bins. 
$Bottom$: Stellar number density distribution in the Kiel diagram. The box enclosed by the red solid lines is used to select our dwarf and subgiant sample stars in this work.
\label{fig1}}
\end{figure*}

The top left panel of Fig.\,\ref{fig1} shows the stellar density distribution in the [Fe/H]--[Ba/Fe] plane for dwarfs and subgiants with ${\rm [Fe/H]} > -1.5$\,dex. These stars are selected with $5000 < T_{\rm eff} < 7500$\,K and $\log\,g >-0.00045\times T_{eff}+3.05 $.

The vast majority of the stars have [Ba/Fe]$\sim0$, and the Ba abundance values may spread more than 4 orders of magnitudes, with [Ba/Fe] values varying from $-2.0$ to $2.0$\,dex. The [Ba/Fe] values exhibit a clear trend with temperature. More Ba-enhanced stars tend to have higher $T_{\rm eff}$ and the Ba-poor ([Ba/Fe]$\lesssim -1.0$) stars mostly have lower $T_{\rm eff}$ (top-right panel of Fig.\,\ref{fig1}). The most metal-rich ([Fe/H]$ > -0.2$) [Ba/Fe]-enhanced stars that have a typical temperature higher than 6700\,K have been studied by \citet{Xiang2020}, who suggested these hot stars are chemically peculiar A/F stars with intermediate mass ($> 1.4\,M_\odot$), and the enhanced Ba abundances are results of stellar internal elemental transports. Except for the A/F Ba-enhanced stars, there is also a significant proportion of Ba-enhanced stars with lower $T_{\rm eff}$.

In this study, we will focus on stars for relatively cool dwarf and subgiant stars, defined as
\begin{equation}
  \begin{cases}
    5000 < T_{\rm eff} < 6700\,K, \\
    \log\,g >-0.00045\times T_{eff} (K)+3.05,\\
    [\rm{Fe}/\rm{H}]>-1.5, \\
    \end{cases}
 \label{eq1}
\end{equation}
which are delineated in the Kiel diagram by the red box in the bottom panel of Fig.\,1. These criteria leads to 1,010,984 dwarfs and subgiants for all Ba abundances. Among them, 8468 are Ba-enhanced stars with ${\rm [Ba/Fe]} > 1.0$. Compared to the warm (A/F) Ba-enhanced stars, these relatively cool sample stars have more robust elemental abundance measurements from the LAMOST spectra.

\begin{figure*}[htb!]
\centering
\vspace{1.em}

\includegraphics[width=\textwidth]{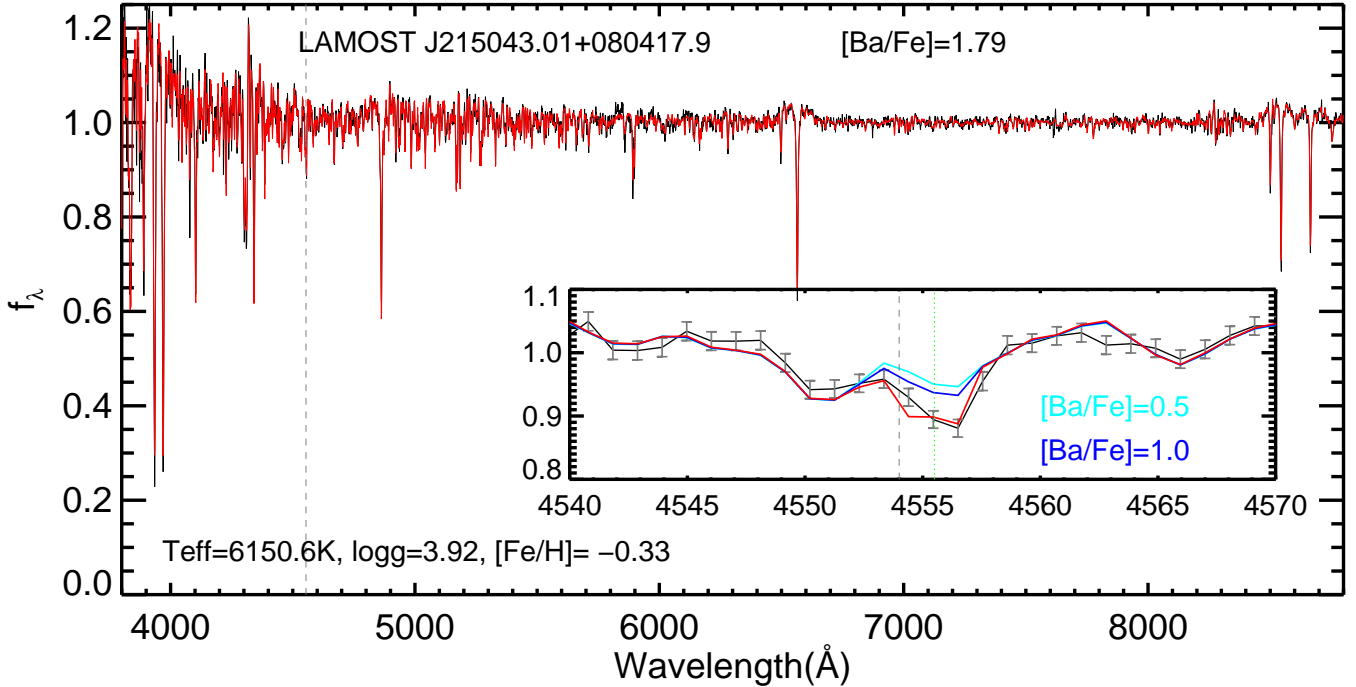}
\vspace{2.em}
\caption{LAMOST spectrum of a Ba-enhanced star.
The black one is the observed LAMOST spectrum, and the red one is the DD-Payne model for the best-fitting parameters. The spectrum is normalized to a smoothed version of its own, in the same way as \citet{Xiang2019}. The zoom-in plot highlights the Ba\,4554\,${\rm{\AA}}$ line. The cyan and blue lines show the DD-Payne model spectra with [Ba/Fe]$=$0.5 and 1.0\,dex, respectively. The observational data errors are shown by the grey error bars.
\label{fig2}}
\end{figure*}

As an illustration, Fig.\,\ref{fig2} shows the LAMOST spectrum of a Ba-enhanced star. The spectrum shows a strong Ba\,{\textsc{ii}}~4554~${\rm{\AA}}$ line, suggesting that the Ba enhancement is a genuinely recognizable feature in the LAMOST low-resolution spectra.

\section{Results}
\label{section3}
In this section we present the chemistry (Section 3.1), kinematics (Section 3.2), UV photometry (Section 3.3), and occurrence rate across the HR diagram (Section 3.4) of the Ba-enhanced dwarfs and subgiants.
\subsection{Chemistry}
Elemental abundance patterns offer us clues to understand the origin of the Ba-enhanced stars. 
For example, as mentioned above, the Ba-enhanced stars produced by binary evolution may usually have overabundance of carbon, while the Ba-enhanced stars produced by stellar internal atomic transports may have peculiar Mg and iron-peak elemental abundances.

\subsubsection{Carbon and Nitrogen}
As an s-process element, Ba is mainly synthesized in the interior of AGB stars.
The third dredge-up (TDU) process would bring primary nucleosynthesis products including C, and the s-process elements from combined H-shell and He-shell burning to their surface. If a main-sequence star has accreted Ba-rich material from an AGB companion, we expect it might gain C-rich materials from its companion. Moreover, the quantitative pattern of the C and N abundances depends on the mass of its AGB donator. For a low-mass AGB star ($\lesssim 3.0\,M_\odot$), the C enhancement is most prominent.
As a consequence of the hot bottom burning process, the high-mass AGB companion would have enhanced N abundance \citep{Boo1993}.

\begin{figure*}[htb]
\centering   
\vspace{1.5em}
\subfigure
{
	\begin{minipage}{0.45\linewidth}
	\centering  
	\includegraphics[width=0.9\columnwidth]{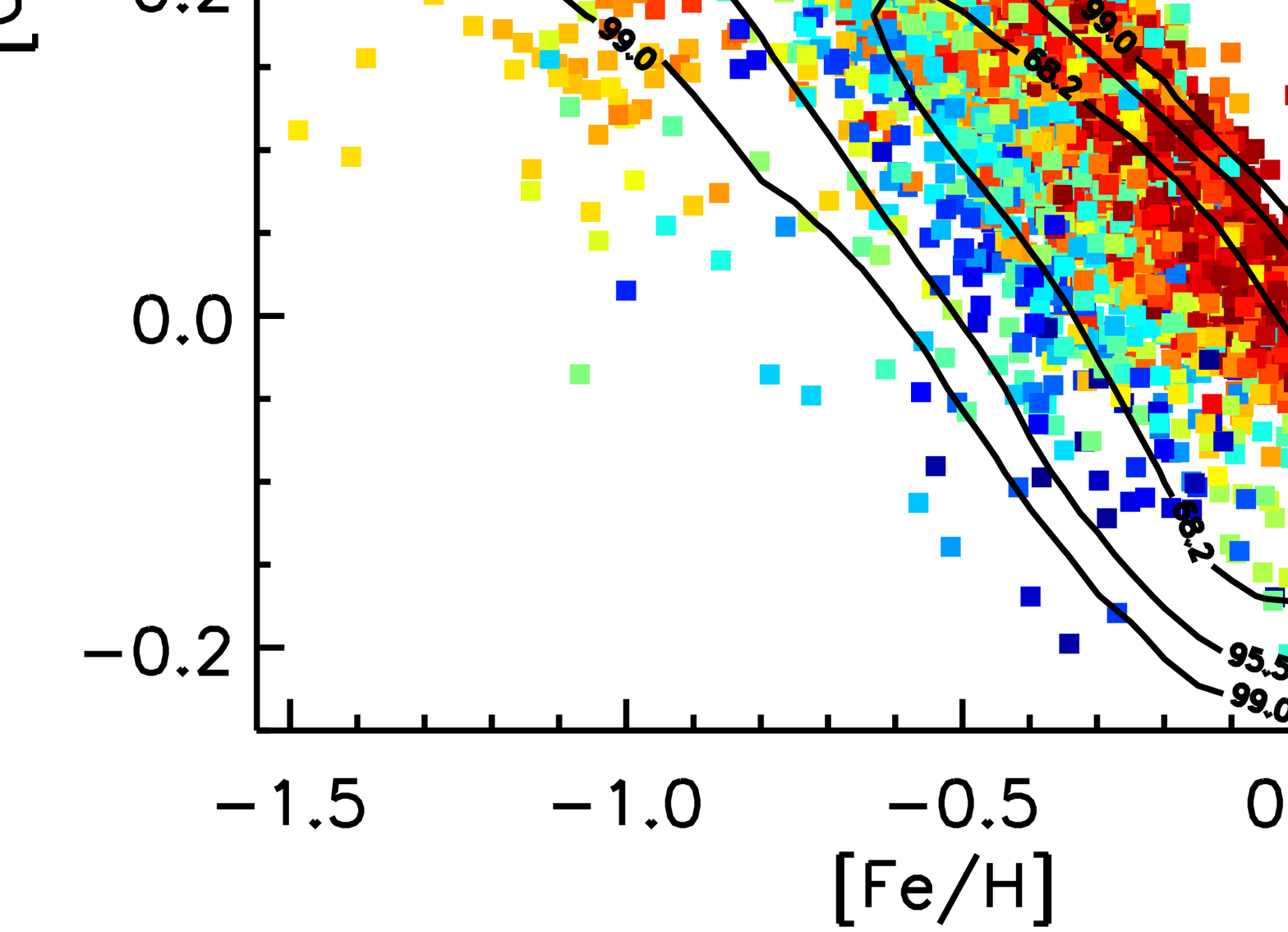} 
	\end{minipage}
}      
\subfigure
{
	\begin{minipage}{0.45\linewidth}
	\centering     
	\includegraphics[width=0.9\columnwidth]{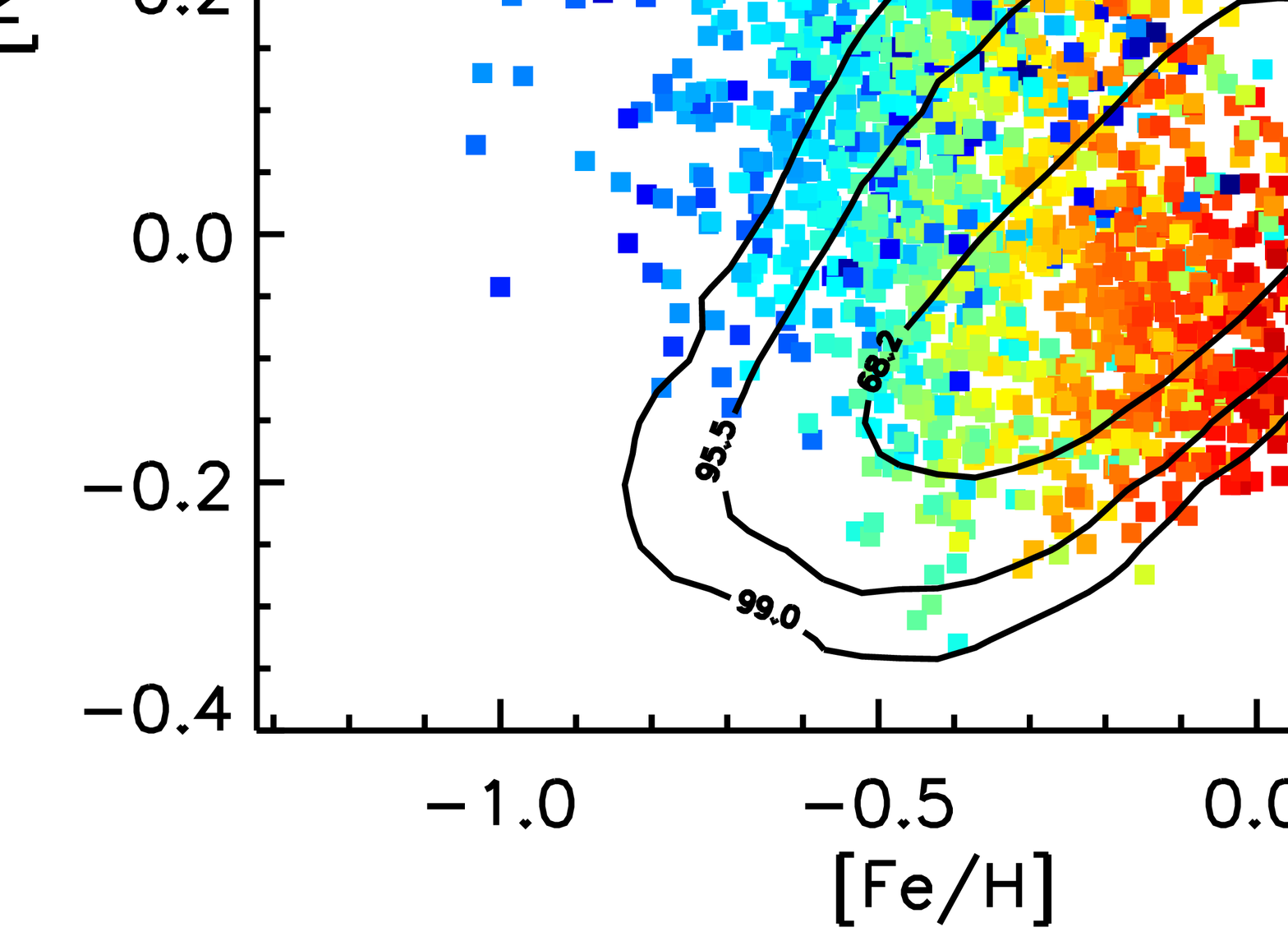}  
	\end{minipage}
}
\subfigure
{
	\begin{minipage}{0.45\linewidth}
	\centering     
	\includegraphics[width=0.9\columnwidth]{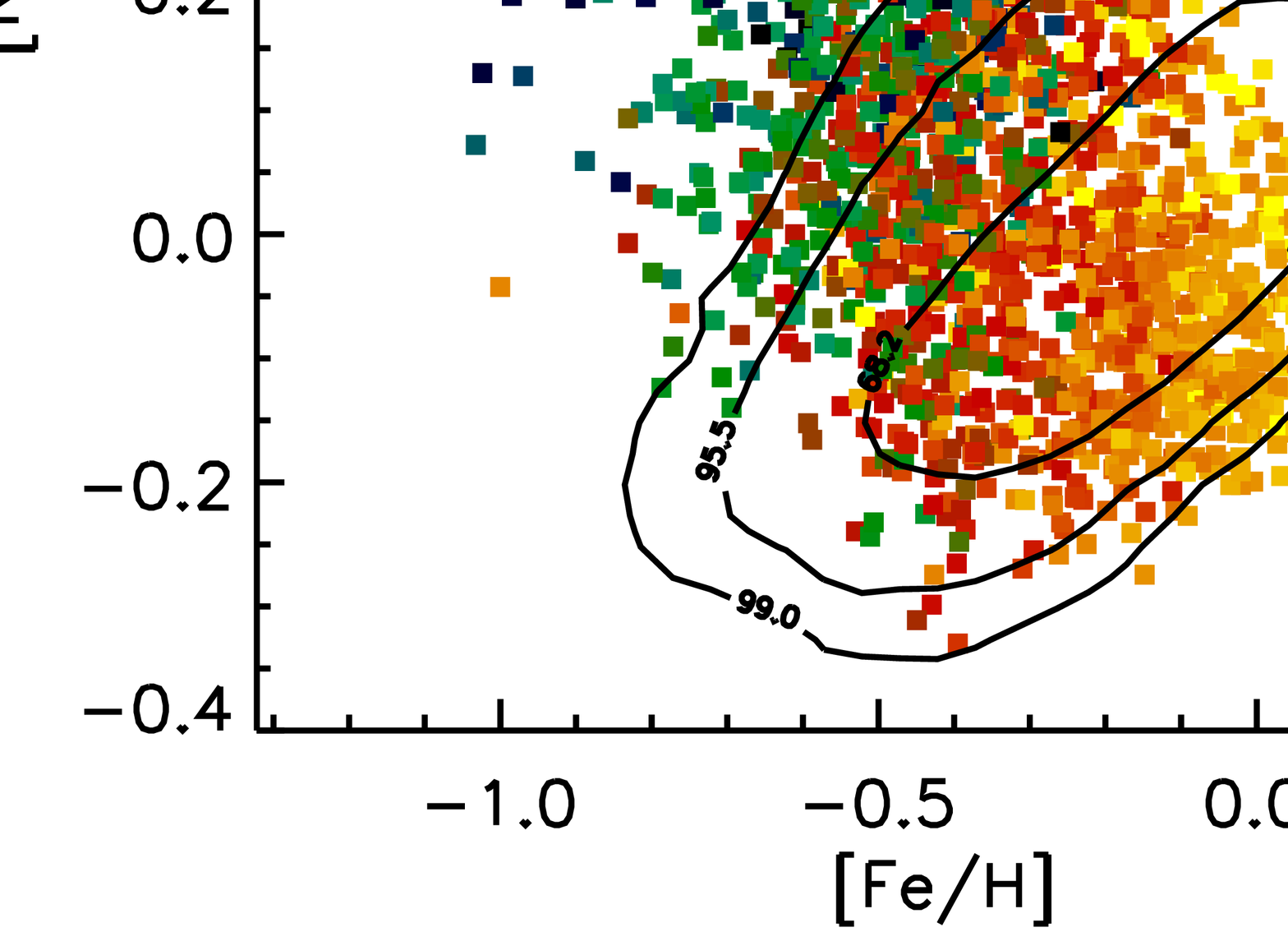}  
	\end{minipage}
}
\subfigure
{
	\begin{minipage}{0.45\linewidth}
	\centering     
	\includegraphics[width=0.9\columnwidth]{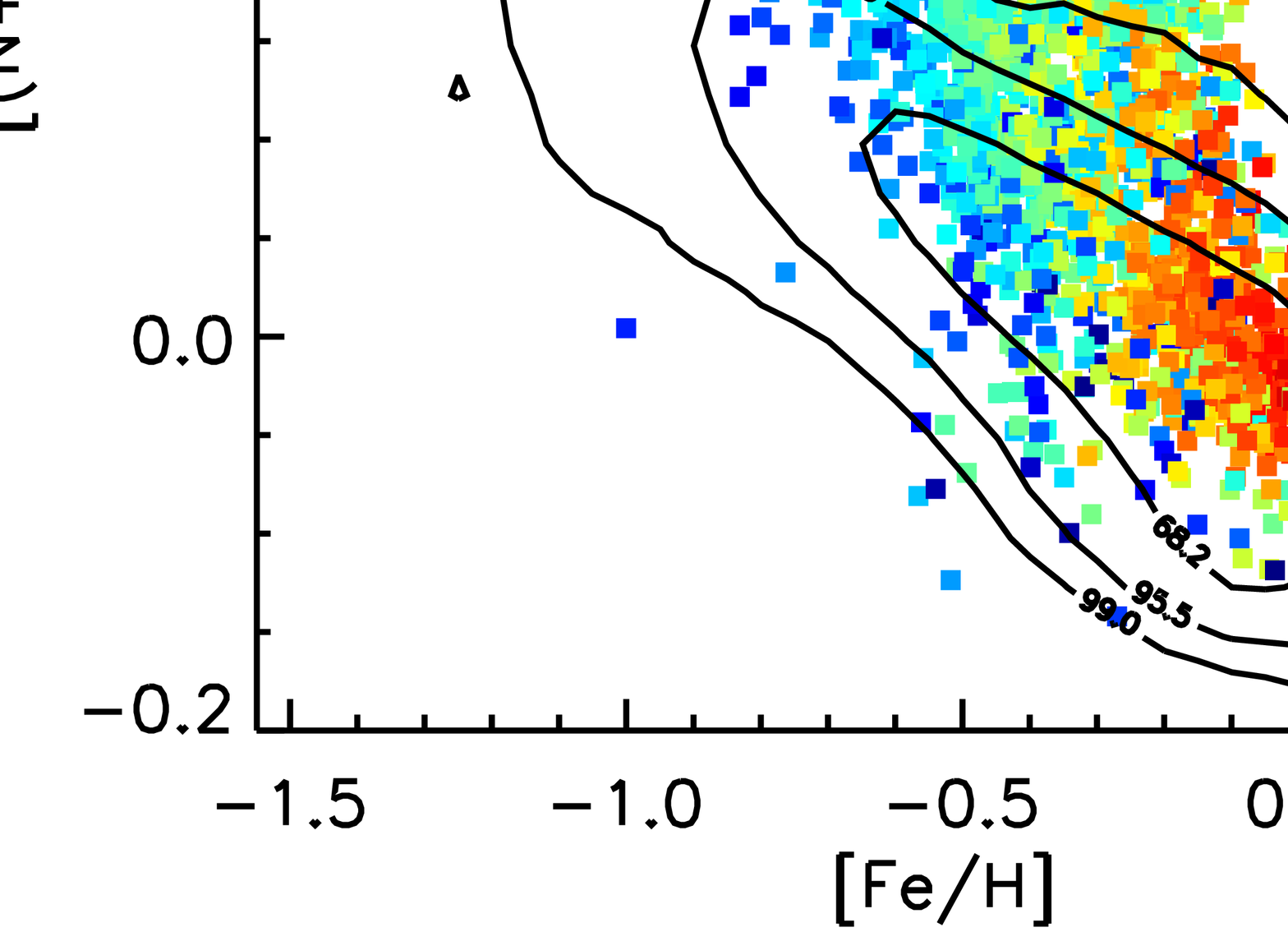}  
	\end{minipage}
}
\vspace{2.em}
\caption{Distributions of Ba-enhanced stars in [C/Fe]--[Fe/H] ({\em top left}), [N/Fe]--[Fe/H] ({\em top right} and {\em bottom left}), and [Fe/H]--[(N+C)/Fe] plane ({\em bottom right}). Ba-enhanced stars are color-coded by [C/Fe] in the {\em bottom-left} panel, and in the other panels, they are color-coded by effective temperature. The black curves show the equal-density contours enclosing 68.3, 95.5 and 99.7\% of the Ba-normal stars (${\rm [Ba/Fe]} < 0.5$). The horizontal dashed lines in the {\em top-left} panel and the {\em bottom-right} panel mark a constant [C/Fe] abundance ratio of 0.3 and [(N+C)/Fe] abundance ratio of 0.2, respectively, which we adopted to split the Ba-enhanced sample stars into sub populations (see text).
\label{fig3}}
\end{figure*}

The top-left panel of Fig.\,\ref{fig3} shows that many Ba-enhanced stars exhibit much higher [C/Fe] values than those of the Ba-normal stars. About 25.7\% of them have ${\rm [C/Fe]} > 0.3$. Many of the Ba-enhanced stars also have higher [N/Fe] values (top-right panel of Fig.\,\ref{fig3}). These Ba-enhanced stars with enhanced C and/or N abundances are relatively cool stars ($T_{\rm eff}\lesssim6000$\,K), while the warm stars mostly have normal C and N abundances.
This implies that the Ba-enhanced stars may have different origins, as the former (with C and N enhancement) are likely results of binary evolution. 
It should be noticed that for many of the metal-poor stars ([Fe/H]$\lesssim-0.5$), the [N/Fe] value is not available as their N features are too weak \citep{Xiang2019}, so that there are fewer stars presented in the top-right panel (and also in the bottom-left panel) than those in the top-left panel.

As shown in the left-bottom panel of Fig.\,\ref{fig3}, a proportion of Ba-enhanced stars with N enhancement have [C/Fe]$\lesssim{0.2}$. About one-third Ba-enhanced stars have enhanced total C and N abundances ([(N+C)/Fe] \footnote{Defined as $\rm{[(N+C)/Fe]}\equiv \log_{10}\left(\frac{(N_{N}+N_{C})/N_{Fe}}{(N_{N}+N_{C})_{\odot}/N_{Fe,\odot}}\right)$} larger than 0.2\,dex), which is shown in the bottom-right panel. Others have [(N+C)/Fe] values similar to those of the Ba-normal stars.

To distinguish Ba-enhanced stars which could be originated from different mechanisms, we parse the Ba-enhanced sample into two sets.
One set is a population of Ba-enhanced stars with [(N+C)/Fe] abundance ratios larger than 0.2, or the Ba-enhanced stars with [C/Fe] values larger than 0.3 for which have unavailable [N/Fe] measurements from LAMOST spectra.
With enhanced C and N abundances, these Ba-enhanced stars are presumed to be formed via external mechanism, i.e., binary interaction.
The other set of Ba-enhanced stars are with C and N abundances similar to those of the Ba-normal stars.
We select the Ba-enhanced stars with [(N+C)/Fe]$ < 0.1$ or [C/Fe]$ < 0.2$\,dex for those with unavailable [N/Fe] measurements. Note here we have ignored the Ba-enhanced stars with $0.1< {\rm [(N+C)/Fe]} < 0.2$, to avoid ambiguity.


\subsubsection{Other elements}
\begin{figure*}[hb!]
 \centering
 \vspace{2.em}    
 \includegraphics[width=0.9\textwidth]{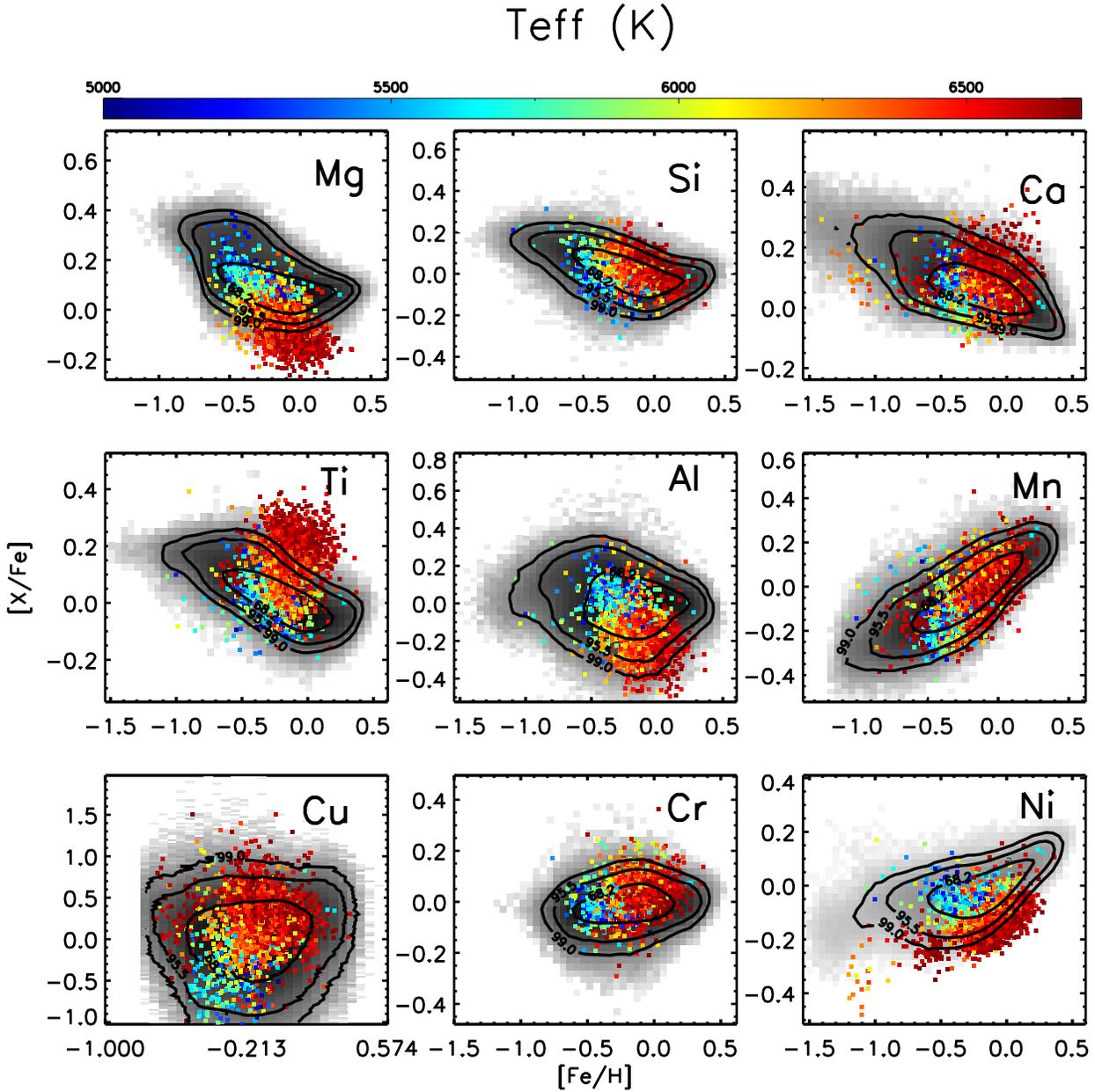} 
 \vspace{1.5em} 
 \caption{Stellar distributions in [X/Fe]--[Fe/H] planes. The black-grey background overlapped with contours shows the distribution of Ba-normal stars, and the color-coded dots are for Ba-enhanced stars with normal C and N abundances ([C/Fe]$ < 0.2$\,dex or [(N+C)/Fe]$ < 0.1$\,dex). Colors represent effective temperature. The majority of these Ba-enhanced stars are warm stars with $T_{\rm eff} > 6000K$, and they show depleted [Mg/Fe] and [Ni/Fe] but enhanced [Ti/Fe] ratios. The stellar distributions of some elements, e.g., the Cu and Cr, are incomplete in [Fe/H] coverage, because there are no available measurements of those elements for the metal-poor stars (see text).
 \label{fig4a}}
 \end{figure*}

\begin{figure*}[hb!]
 \centering
  \vspace{2.em} 
 \centering

 \includegraphics[width=0.95\textwidth]{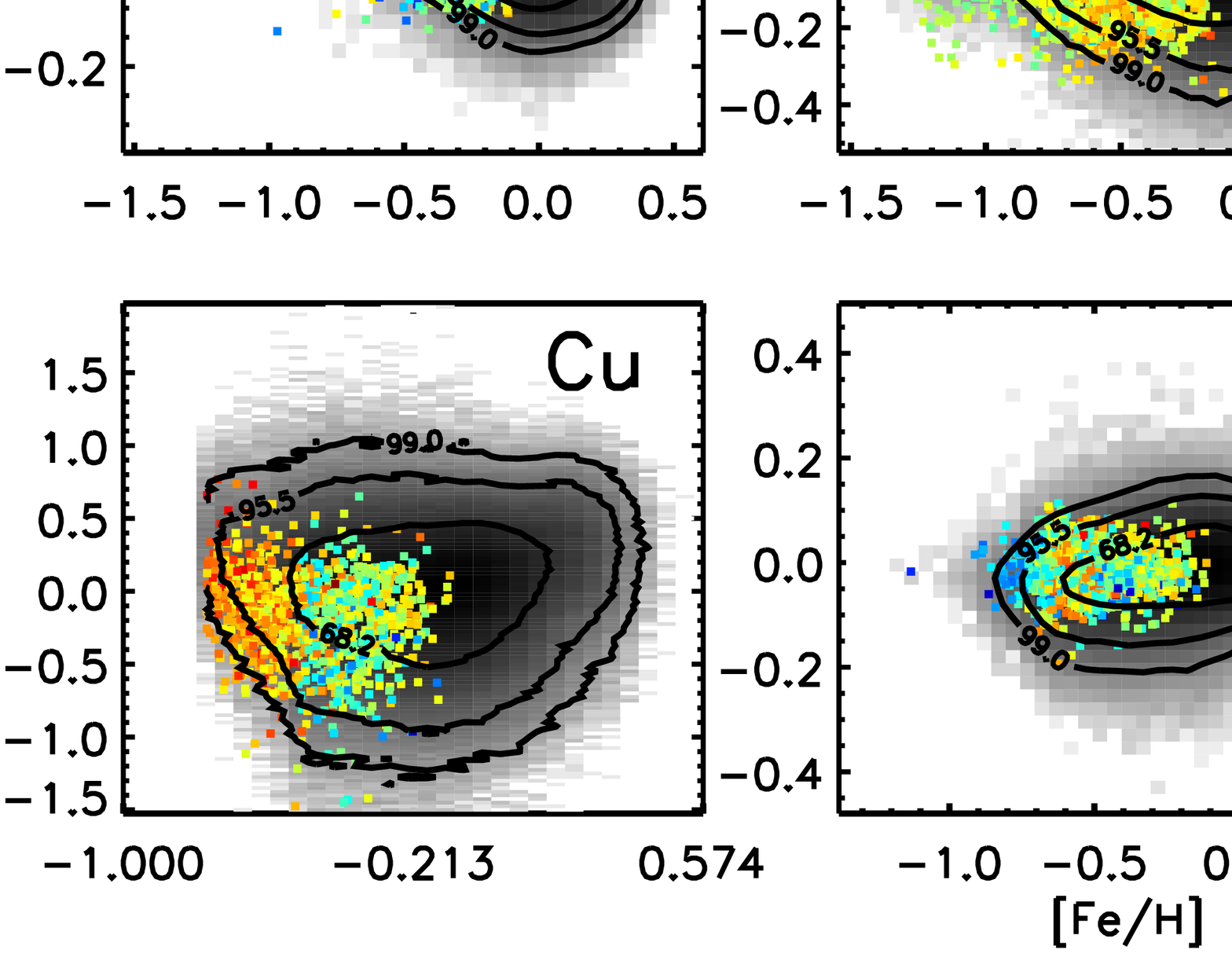}
 \vspace{2.em}
 \caption{Same as Fig.\,\ref{fig4a}, but for Ba-enhanced stars with high C and N abundance ratios (${\rm [(N+C)/Fe]} > 0.2$\,dex). Most of these Ba-enhanced stars have effective temperatures lower than 6000\,K. They exhibit  Mg, Si, Ca, Ti, Al, Mn, Cu, Cr, and Ni abundances in line with those of the Ba-normal stars. 
 \label{fig4b}}
 \end{figure*}

 Figs.\,\ref{fig4a} and \ref{fig4b} show the distributions of those two populations of Ba-enhanced stars, i.e., the N+C (or C) enhanced stars and the C and N normal stars, respectively, in the [X/Fe]--[Fe/H] plane, where X refers to Mg, Si, Ca, Ti, Al, Mn, Cu, Cr, and Ni.
 
 Fig.\,\ref{fig4a} illustrates that, for the Ba-enhanced stars with normal C and N abundances which is at the warmer end ($T_{\rm eff}\gtrsim6000$\,K), their [Mg/Fe] ratios exhibit a clear depletion of $\sim0.2$\,dex compared to those of the Ba-normal stars with similar metallicity. This could be a signature of stellar internal atomic diffusion in Am stars \citep[e.g.,][]{Mic1970, Mic2011, Mic2015, Xiang2020}.
 
 It also shows that the warm Ba-enhanced stars tend to have higher [Ca/Fe] and [Ti/Fe] values than those of the Ba-normal stars, and their [Ni/Fe] values are clearly lower than those of the Ba-normal stars. For the other elements, the differences between the Ba-enhanced and Ba-normal stars are not obvious (see the differential elemental patterns in Sect.4). 
 
 These elemental patterns of [Ca/Fe], [Ti/Fe] and [Ni/Fe] are different from models for Am stars in previous work \citep{Tal2006, Vic2010, Mic2011}. The atomic diffusion in their models would cause lower [Mg/Fe] and [Ca/Fe] values, and a higher [Ni/Fe] value, but the [Ti/Fe] value would not varied obviously. We suspect these differences might be related to different masses and ages of our sample stars with respect to the Am stars as modelled in literature.
 Since our sample stars have lower effective temperatures, it means they should have lower masses or older ages. This needs to be modelled quantitatively.

Fig.\,\ref{fig4b} shows that the Ba-enhanced stars with enhanced N+C (or C) abundance have ${\rm [Fe/H]}\lesssim-0.2$. 
Most of these stars have indiscernible [X/Fe] values compared to the Ba-normal stars except for the relatively warm stars ($T_{\rm eff}\gtrsim6000$\,K). This is in line with the scenario of binary evolution, as any AGB companion as the Ba-rich material donor would also have donated C and N-rich materials while keep other elemental abundances unchanged.

For the relatively warm stars of $T_{\rm eff}\gtrsim6000$\,K, they exhibit a depletion of [Mg/Fe] and a slight depletion of [Ni/Fe]. Similar to those warm Ba-enhanced stars in Fig.\,\ref{fig4a}, i.e., the N+C normal Ba-enhanced stars. 
Given these element patterns, it is possible that both the binary evolution 
mechanism and the stellar internal elemental transport may be response for the presence of these stars. We note again that for some of the elements, such as Cu and Cr, not all stars have available abundance estimates from the LAMOST DD-Payne catalog, so that stars presented in different panels of Figs.\,4 and 5 are not always identical.
 
Interestingly, Fig.\,\ref{fig4b} also shows that there seems to be a lack of Ba-enhanced stars with high [Mg/Fe] and [Ca/Fe] values, which will be discussed in  Sect.\,3.1.3.

\subsubsection{The ``high-[$\alpha$/Fe] desert"}
\begin{figure}[htb!]
\centering
\vspace{1.em} 
\centering
\includegraphics[width=0.98\columnwidth]{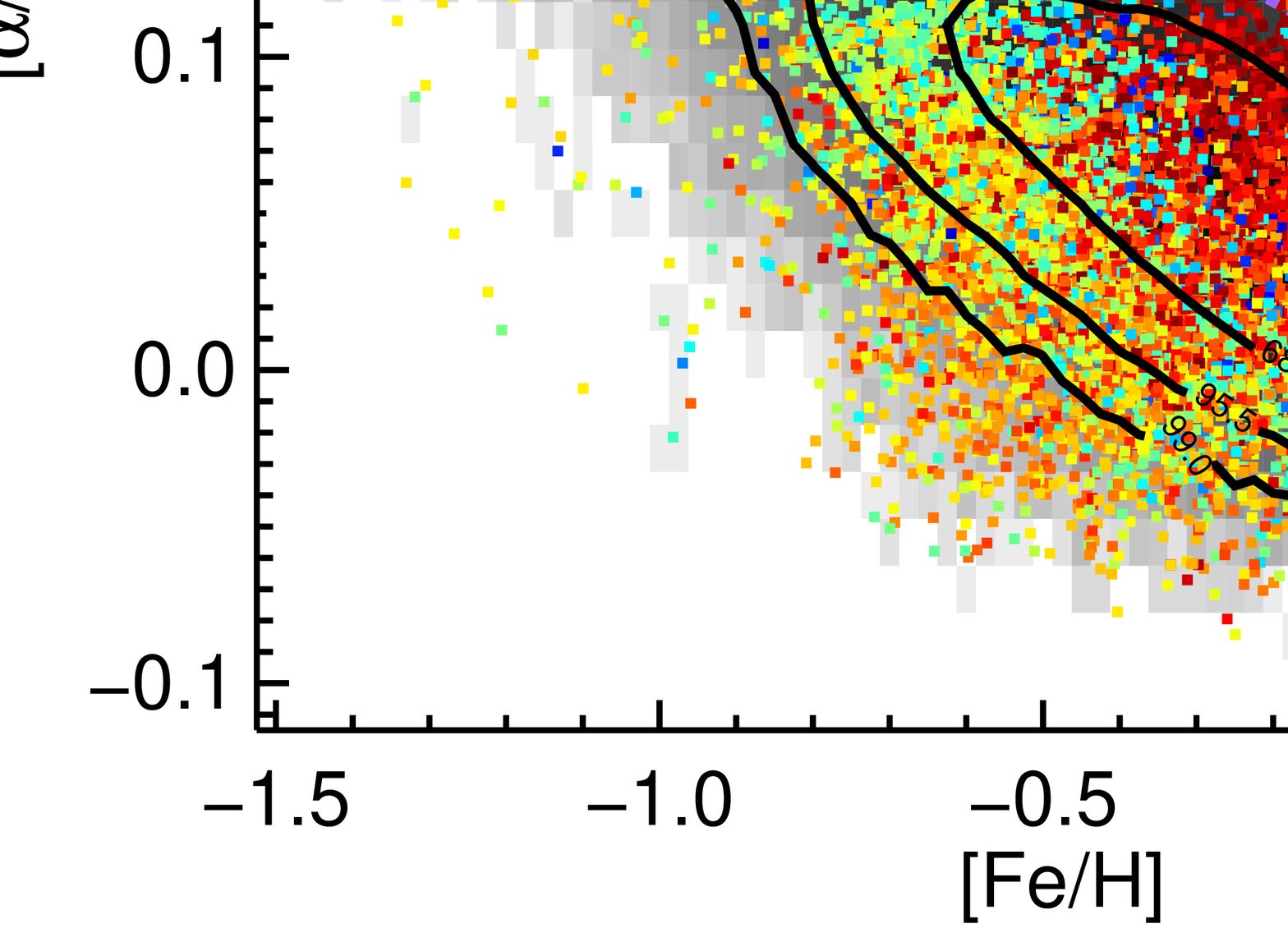}
\vspace{2.em}
\caption{Distribution of the Ba-enhanced stars in the [Fe/H]--[$\alpha$/Fe] plane, color-coded by effective temperature. The black-grey background with contours show the density distribution of the Ba-normal stars. The dashed lines in purple mark an empirical distinction of the low-$\alpha$, chemical thin disk from the high-$\alpha$, chemical thick disk (see text). The solid box in purple delineates the location of a ``desert" of high-$\alpha$ Ba-enhanced stars.
\label{fig5}}
\end{figure}

Fig.\,\ref{fig5} shows the stellar distribution in the 
 [$\alpha$/Fe]--[Fe/H] plane, where the [$\alpha$/Fe] is an average estimate of [Mg/Fe], [Ca/Fe], [Si/Fe], [Ti/Fe] values \citep{Xiang2019}. As a comparison, the grey background with contours shows the distribution of Ba-normal stars. The distribution is a superposition of two [$\alpha$/Fe]--[Fe/H] sequences -- a sequence of higher [$\alpha$/Fe] values, which is the chemical thick disk, and a sequence of lower [$\alpha$/Fe] values, which is the chemical thin disk \citep[e.g.,][]{Ben2003, Lee2011, Hay2013, Hay2015b}. This two-sequence features can be well recovered with the LAMOST data set \citep[e.g.,][]{Xiang2017, Wu2019, Huang2020, Zhang2021}, and the detailed morphology of the [$\alpha$/Fe]--[Fe/H] distribution varies with the spatial location, as high-[$\alpha$/Fe] stars dominate at the inner disk and at larger vertical height to the Galactic disk mid-plane, and the low-[$\alpha$/Fe] stars dominate at the outer disk or at the disk close to the mid-plane  \citep[e.g.,][]{Hay2015b, Wang2019}. Our sample is dominated by stars in the outer disk due to the survey's footprint, so that the low-[$\alpha$/Fe] sequence is prominent while the high-[$\alpha$/Fe] sequence is visible as an excess of high-[$\alpha$/Fe] stars in the range of $-1\lesssim{\rm [Fe/H]}\lesssim0$. Similar to \citet{Zhang2021}, we adopt a broken line to separate the high-[$\alpha$/Fe] and low-[$\alpha$/Fe] sequences (see Fig.\,\ref{fig5}).

\begin{figure}[htb!]
\centering      
	\includegraphics[width=0.95\columnwidth]{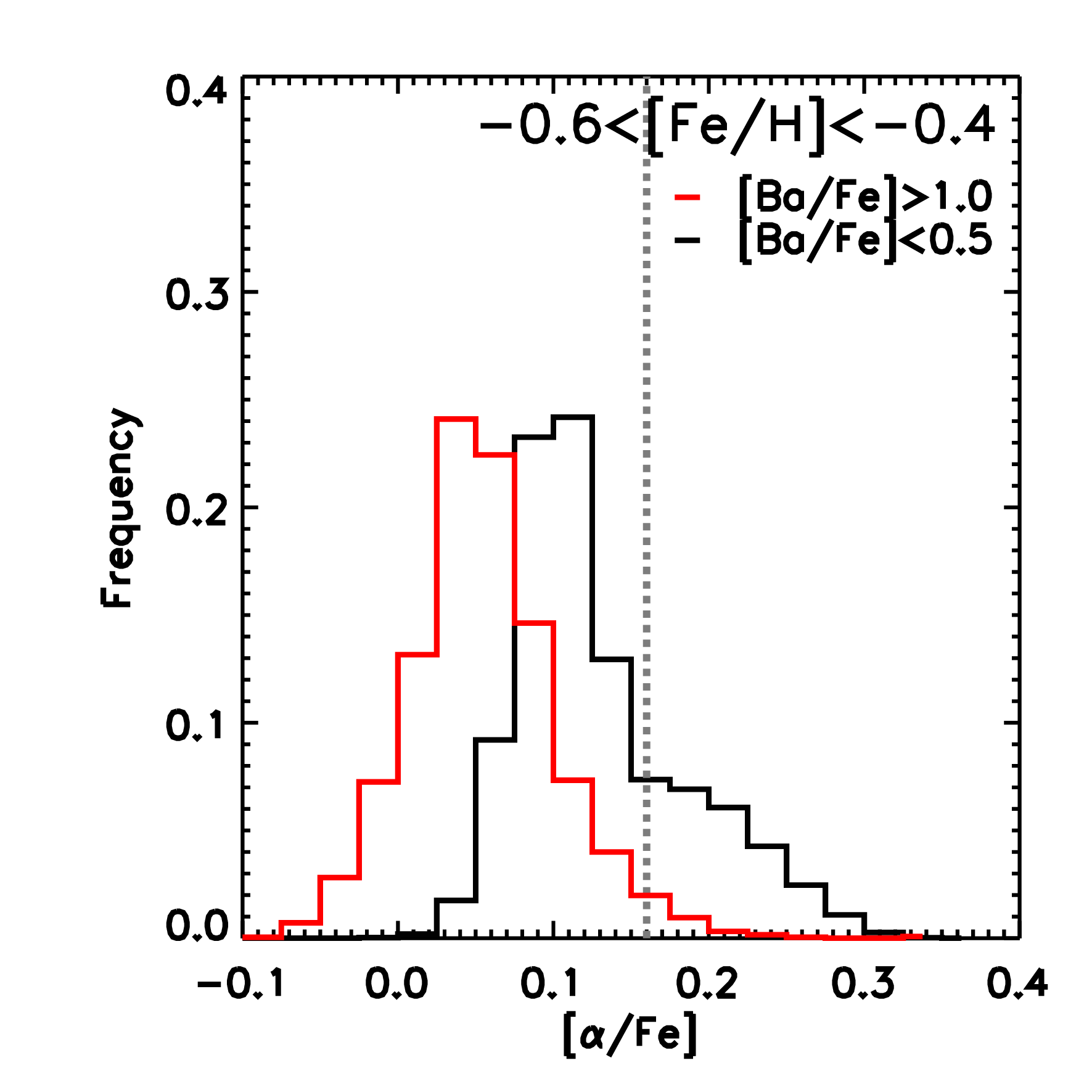}
\vspace{2.em}
\caption{[$\alpha$/Fe] distribution of the Ba-enhanced (red) and Ba-normal (black) stars, both with $-0.6 < {\rm [Fe/H]} < -0.4$. The Ba-enhanced stars have systematically lower [$\alpha$/Fe] values, and there are few Ba-enhanced stars with ${\rm [\alpha/Fe]} > 0.16$ (delineated by the vertical dotted line).
\label{fig6}}
\end{figure}

Fig.\,\ref{fig5} shows that the Ba-enhanced stars tend to distribute along the low-[$\alpha$/Fe], chemical thin disk sequence, whereas few of them are in the high-[$\alpha$/Fe], chemical thick disk sequence, particularly for stars with ${\rm [Fe/H]}\gtrsim-0.6$.
The lack of Ba-enhanced stars of high-[$\alpha$/Fe] is further demonstrated in Fig.\,\ref{fig6}, which shows the [$\alpha$/Fe] distribution for stars with $-0.6 < {\rm [Fe/H]} < -0.4$. The Ba-normal stars spread a wide range of [$\alpha$/Fe] values, and about 25\% of them have an [$\alpha$/Fe]  higher than 0.16. However, most of the Ba-enhanced stars have [$\alpha$/Fe] values of $\sim0.05$, and the proportion of high-[$\alpha$/Fe] stars ($>0.16$) is negligible. We deem this phenomenon as a high-[$\alpha$/Fe] {\em desert} of Ba-enhanced stars for convenience.

Such a high-[$\alpha$/Fe] desert phenomenon was not known. We conjecture that it is either due to the property of the AGB donators, or any intrinsic high-[$\alpha$/Fe] Ba-enhanced stars might have gone through some poorly known processes that decreased their surface [$\alpha$/Fe]. We will discuss these possibilities in Section\,4.3.

\subsection{Kinematics}
We cross-matched our sample with the third Gaia early data release (EDR\,3, \citealt{Gaia2021}) using a 3" angular distance criterion, and found that around 96\% of our sample stars have a Gaia eDR3 counterpart, which gives astrometric information, including celestial position, parallax and proper motions. Using the Gaia eDR3 positions, proper motions, the LAMOST line-of-sight velocity, and the Gaia geometric distance estimates of \citep{Ba2021}, we derive the orbital parameters of these stars with the $Galpy$ module \citealt{Bo2015}. We adopt the Galactic potential model of MWPotential2014. The Sun is assumed to be located at $X=-8$\,kpc, $Y=0$, and $Z=0$. The solar motion with respect to the LSR is adopted to be $(U,V,W)_\odot$ = (7.01, 10.13, 4.95) km/s \citep{Huang2015b}.     

\begin{figure}[htb]
\centering        
\includegraphics[width=0.95\columnwidth]{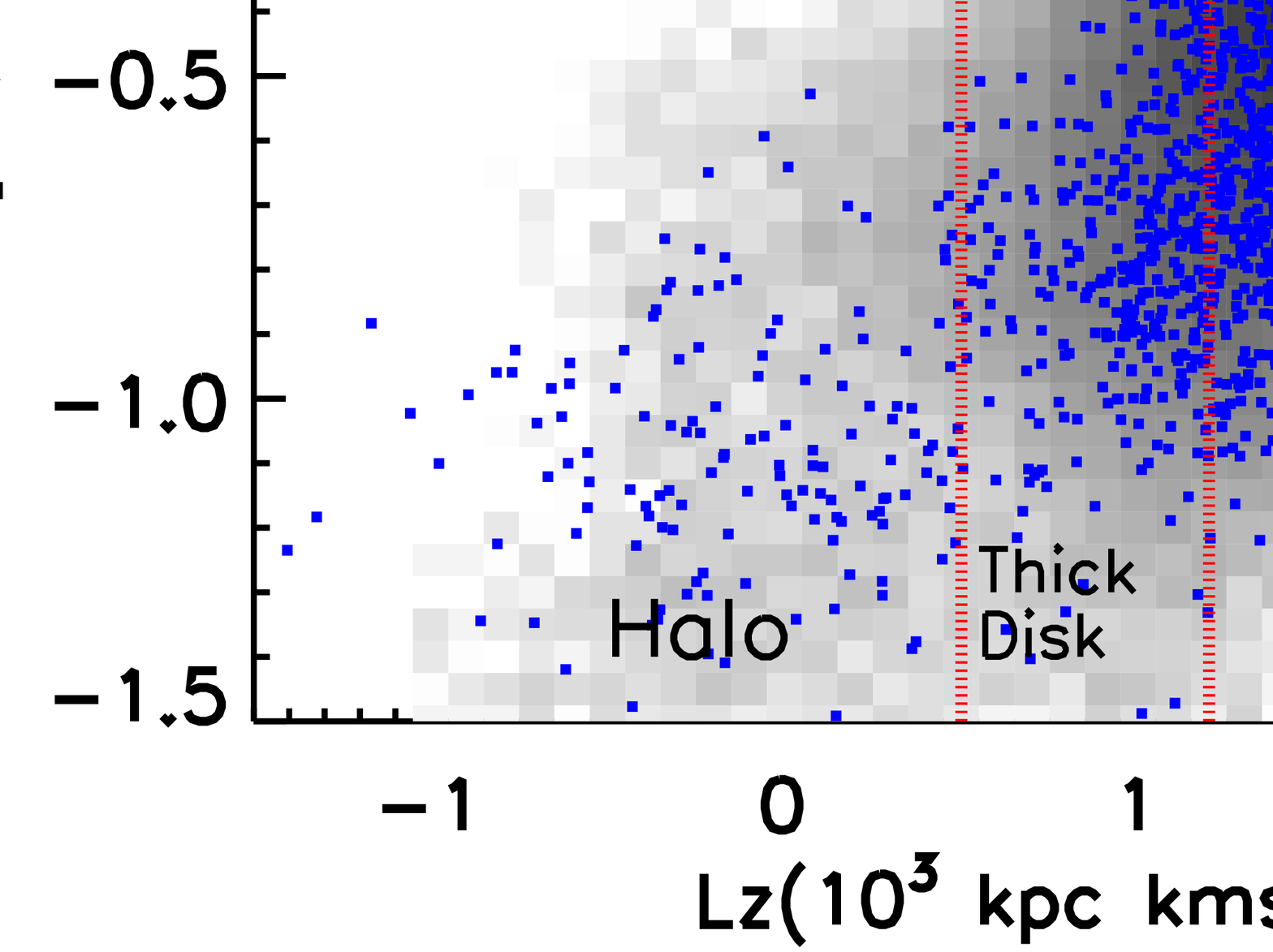}
\vspace{2.em}
\caption{Stellar distributions in the angular momentum ($L_Z$)--[Fe/H] plane. The blue dots show the Ba-enhanced stars, and the grey-black background show the density of Ba-normal stars. Most of the Ba-enhanced stars are the disk stars, and the regimes of the thin and thick disks are marked. Some Ba-enhanced stars with $L_Z\lesssim0$ belong to the halo. 
\label{fig11}}
\end{figure}

Fig.\,\ref{fig11} shows the stellar distributions in the $L_Z$--[Fe/H] plane, where $L_Z$ is the orbital angular momentum. The figure illustrates that the angular momentum of the Ba-enhanced is distributed in a broad range, similar to that of the Ba-normal stars. Although most of the Ba-enhanced stars are expected to belong to the Galactic disk components, which have an angular momentum value of $L_{Z} > 500$\,kpc$\cdot$km/s, some of them can be halo stars of small angular momentum (e.g. $L_{Z} < 500$\,kpc$\cdot$km/s).

Such a uniform distribution suggests that the Ba-enhanced stars in our sample cannot be simply explained to be consequences of accreting dwarf galaxies which experienced a different chemical evolution history to our Milky Way. Instead, they are most likely formed through stellar evolution processes, such as elemental transport due to stellar atomic diffusion and binary evolution, as we introduced above. 

\subsection{UV photometric properties}
Barium is mainly synthesized in AGB stars. Fig.\,\ref{fig3} shows that about one third of the Ba-enhanced stars have enhanced C and N abundances. These Ba-enhanced dwarf and subgiant stars might have accreted Ba-rich materials from their AGB companions which have evolved to WDs now. However, the WDs are too faint to be observed in the optical band. Here we investigate their colors of ultraviolet and optical band to search for signature of the WD companions, as the latter are usually luminous in ultraviolet band. Using Near-ultraviolet ($NUV$) photometry from the Galaxy Evolution Explorer (Galex) DR5 \citep{Bia2011} and the $G$ magnitude from the Gaia EDR3 \citep{Gaia2021}, we can obtain $NUV-G$ color with photometric error smaller than 0.1\,mag. To obtain the intrinsic color $(NUV-G)_0$, we need to correct for the interstellar extinction. 
We adopt the extinction derived using method described in our previous work \citep[e.g.][]{Xiang2019, Xiang2022}. In brief summary, we derive the reddening $E(B-V)$ through the star-pair method: given the spectroscopic parameters (effective temperature, surface gravity and metallicity), we estimate their intrinsic colors in optical and near infrared bands with an empirical pair method, similar to \citet{Yuan2013, Yuan2015}, and convert to $E(B-V)$ using the total-to-selective extinction coefficients for individual pass-bands, which are derived from the Fitzpatrick extinction curve \citep{Fit1999}. Typical uncertainty of the $E(B-V)$ estimates is 0.01-0.02\,mag.

\begin{figure*}[htb!]
\centering
\vspace{0.5cm}
\subfigtopskip=10pt 
\subfigbottomskip=10pt
\subfigure
{\begin{minipage}{0.45\linewidth}
	\includegraphics[width=0.9\columnwidth]{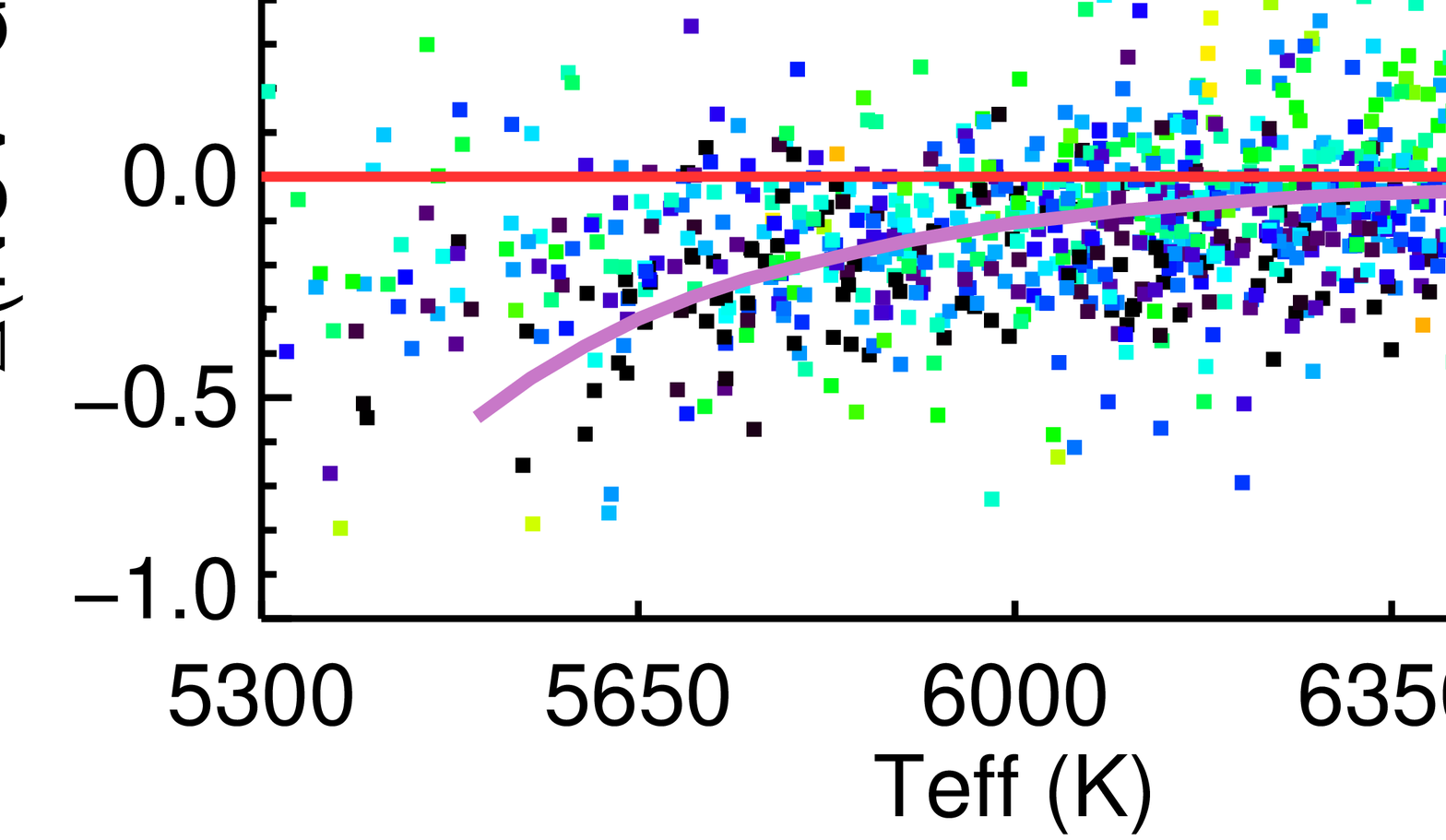}
	\end{minipage}
	}	
\quad
\subfigure
{\begin{minipage}{0.45\linewidth}
	\includegraphics[width=0.9\columnwidth]{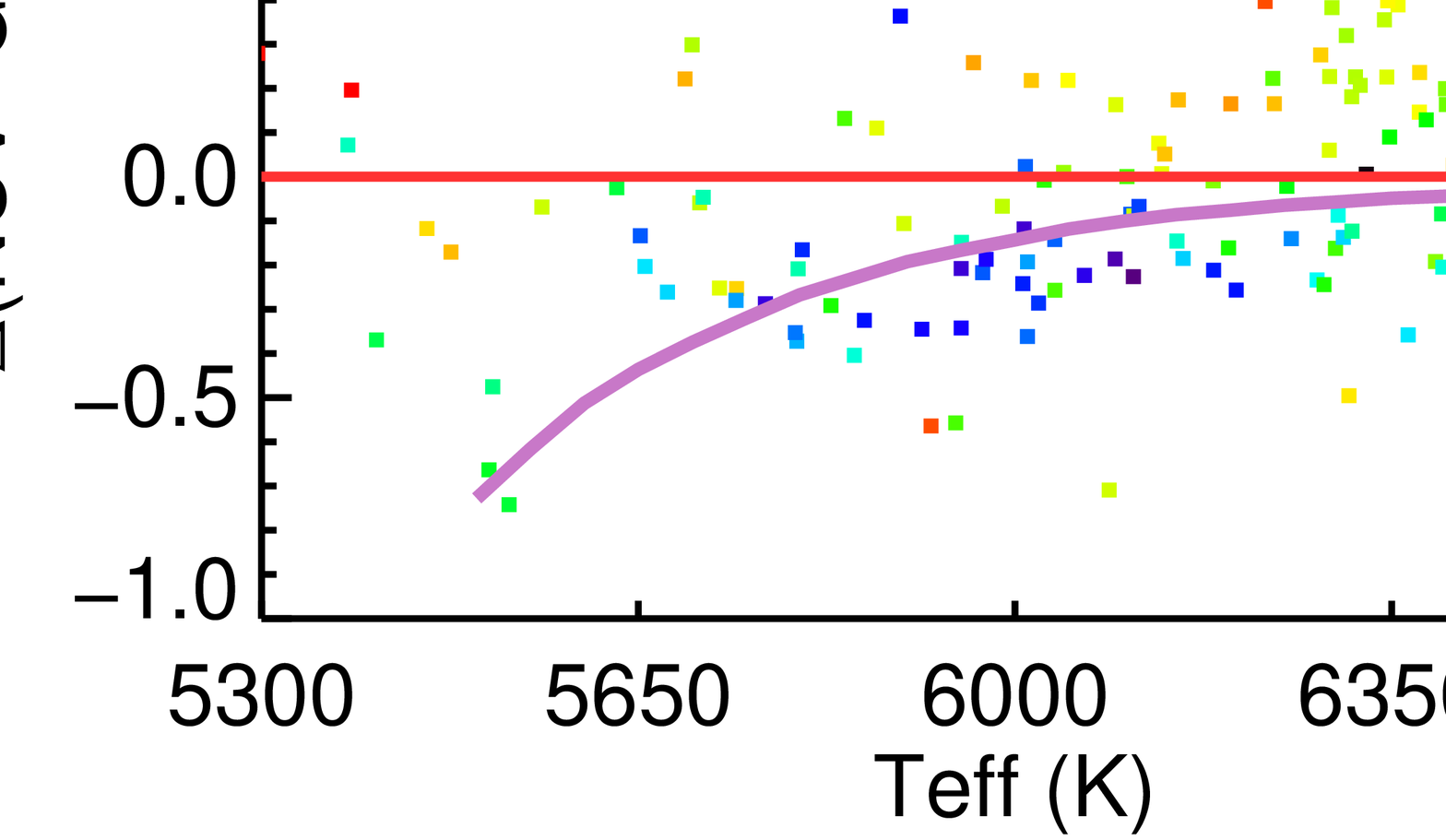}
\end{minipage}}	
\caption{Stellar distributions in $T_{\rm eff}$ versus $(NUV-G)_0$ plane for stars with $-0.6<{\rm [Fe/H]}<-0.4$ ({\em left}) and ${\rm [Fe/H]}>-0.2$ ({\em right}). The grey-black background shows the distributions for Ba-normal stars, and the colored symbols show the Ba-enhanced stars color-coded by their [(N+C)/Fe] ratios. The red line shows the medium color of the Ba-normal stars as a function temperature. The purple line shows the predicted colors from simple models of binary systems that hold a white dwarf companion star (see text for the model details). 
The bottom panels show the $(NUV-G)_0$ difference  between the Ba-enhanced stars and the mean color of the Ba-normal stars. The median error in $(NUV-G)_0$ for our sample stars is shown at the top-left corner of the {\em bottom-left} panel. The trend of $(NUV-G)_0$ excess for the Ba-enhanced stars that exhibit C and N enhancement is consistent with the model prediction from the contribution of white dwarf companions.
 \label{fig111}}
 \end{figure*}

As shown in the top panels of Fig.\,\ref{fig111}, the median of the Ba-normal stars in the magenta line shows the $(NUV-G)_0$ color various with the effective temperature.
The colored symbols show the Ba-enhanced stars with different [(N+C)/Fe] values.
The bottom panels show the differences of $(NUV-G)_0$ color of Ba-enhanced stars with enhanced [(N+C)/Fe] abundance ratios tend to have bluer $(NUV-G)_0$ colors than those of the Ba-normal stars.
It suggests those Ba-enhanced stars might have WD companions.
We use the Gaia EDR3 WD catalog by \citet{Gen2021} and cross it with Galex.
After correcting the interstellar extinction using the similar method as we shown above, we found that the absolute magnitude in NUV band of a WD star is about 10.0-11.0 mag and in G band is about 10.5-11.5 mag.
The purple line shows the predicted colors for modelled binary systems that hold a WD companion with 10.5 mag in NUV band and 10.75 mag in G band. The deviation of $(NUV-G)_0$ color from the mean color of the Ba-normal (presumbly single) stars becomes smaller for warmer sample stars. This is because the UV light from the WD companion becomes less prominent compared to the luminous primary in the warmer case.
At the metal-rich side, the majority of Ba-enhanced stars, mostly with normal C and N abundances, exhibit $(NUV-G)_0$ values similar to those of the Ba-normal stars. For these metal-rich population, the binary evolution is unlikely a decisive reason to account for the chemical peculiarity. 
The right panel of Fig.\,\ref{fig111} shows that, at the warm temperature end ($T_{\rm eff}>6400$~K), most of the Ba-enhanced stars are redder in $(NUV-G)_0$ than the mean value of the Ba-normal stars. We found this is because they have higher surface metallicity.

\subsection{Incidence across the HR diagram}
\begin{figure}[htb!]
\centering   
\subfigure
{
	\begin{minipage}{0.45\linewidth}
	\centering          
	\includegraphics[width=0.96\columnwidth]{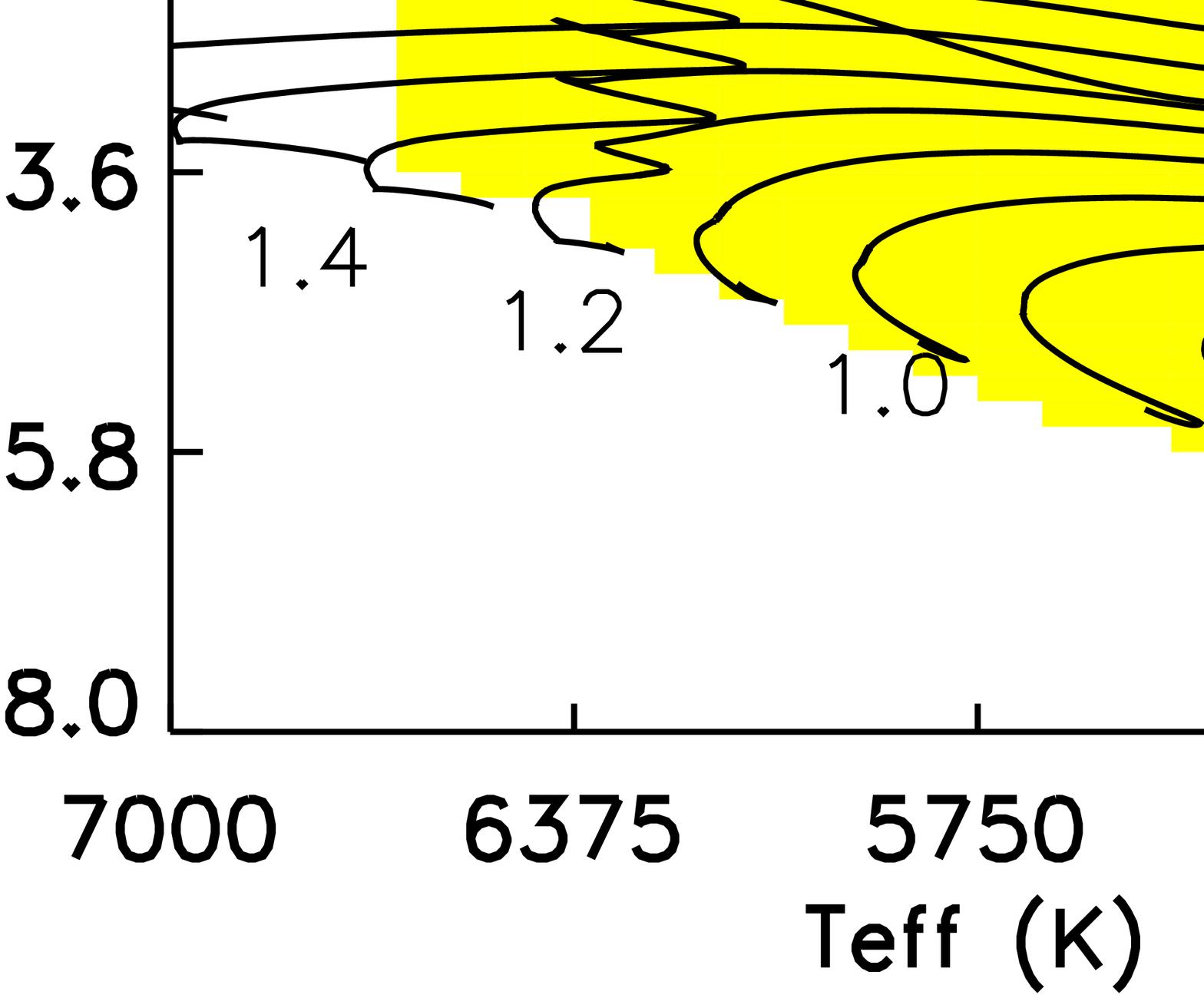}  
	\end{minipage}
}	
\subfigure
{
	\begin{minipage}{0.45\linewidth}
	\centering     
	\includegraphics[width=0.96\columnwidth]{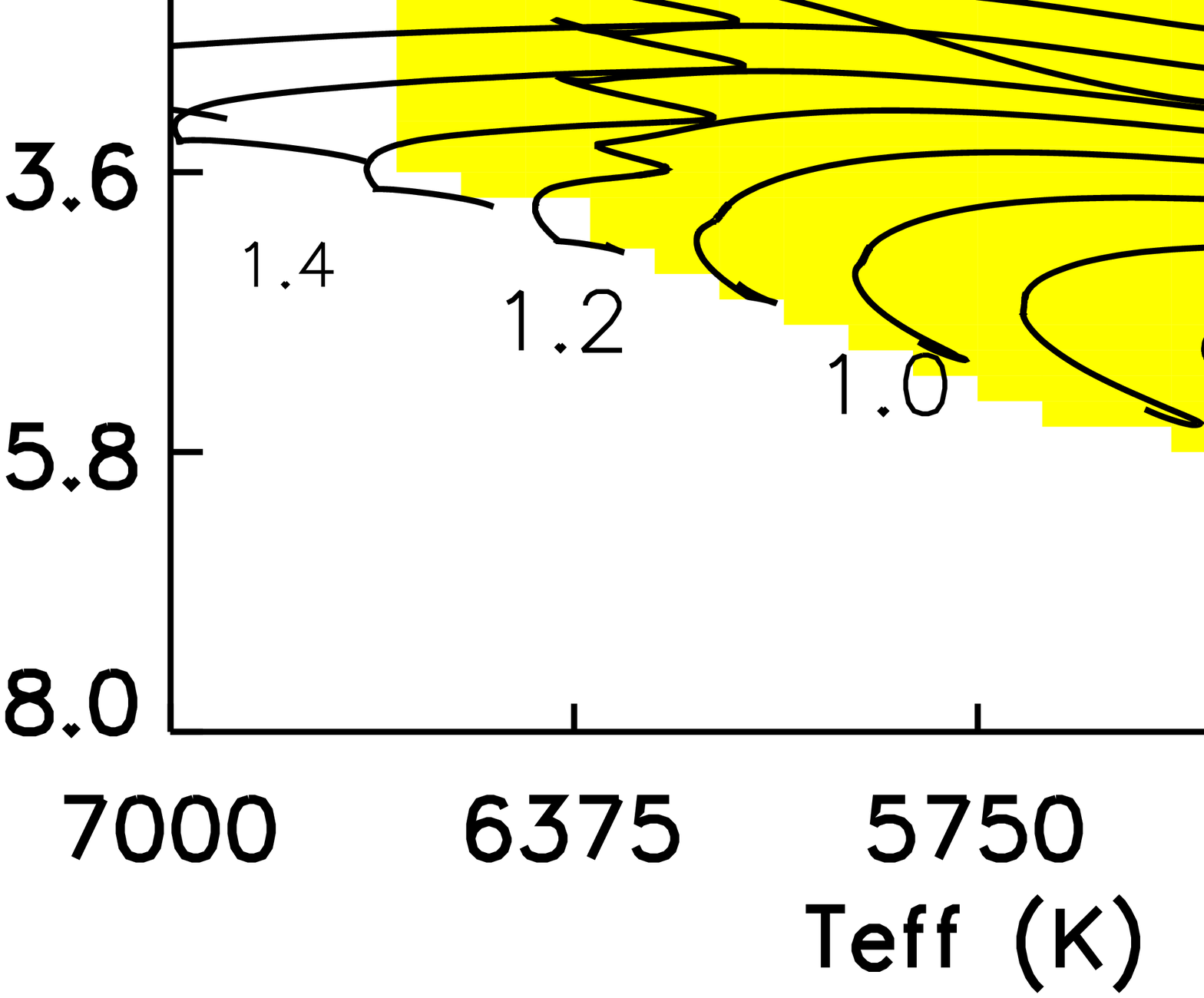}  
	\end{minipage}
}
\subfigure
{
	\begin{minipage}{0.45\linewidth}
	\centering          
	\includegraphics[width=0.96\columnwidth]{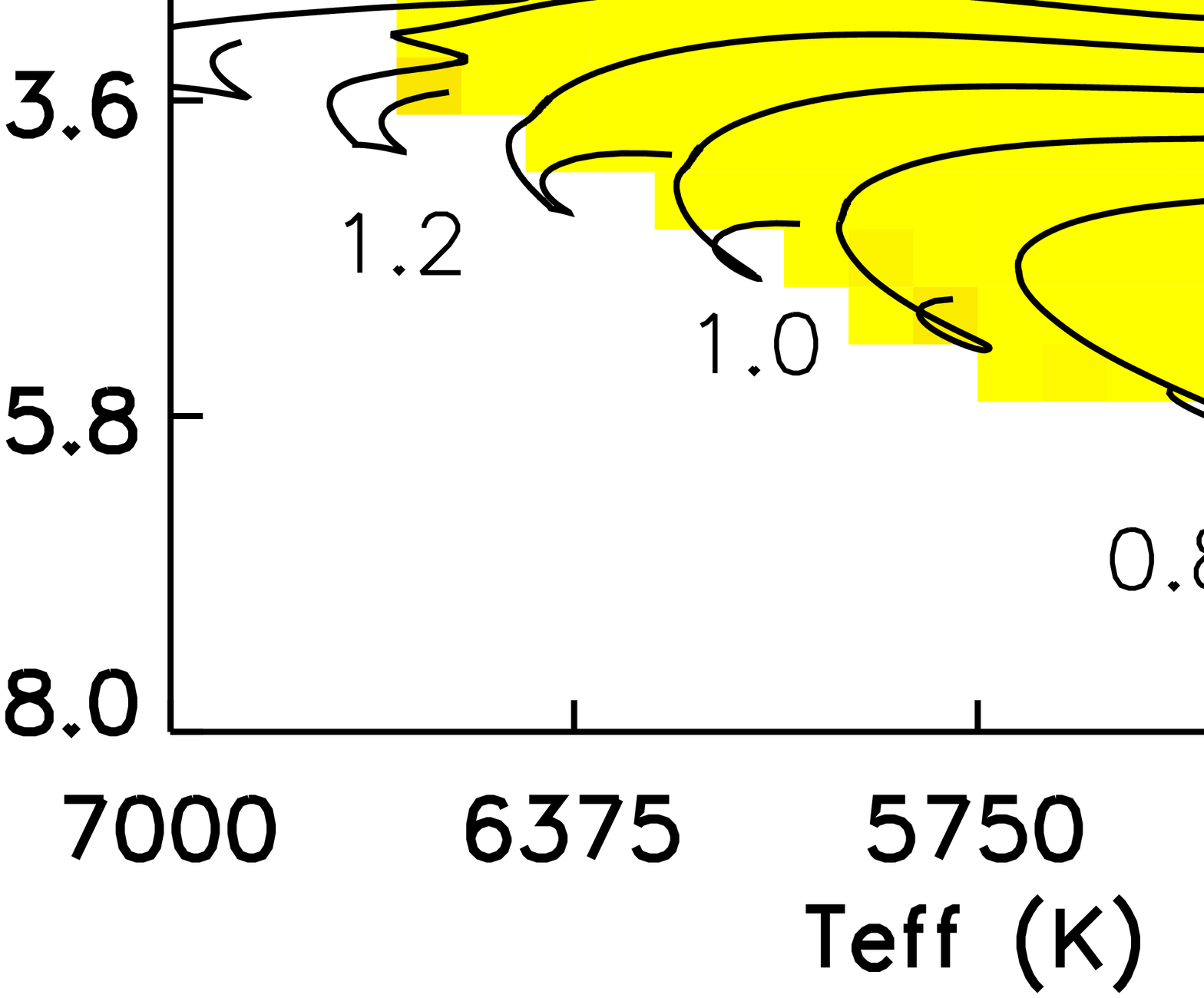}  
	\end{minipage}
}	
\subfigure
{
	\begin{minipage}{0.45\linewidth}
	\centering     
	\includegraphics[width=0.96\columnwidth]{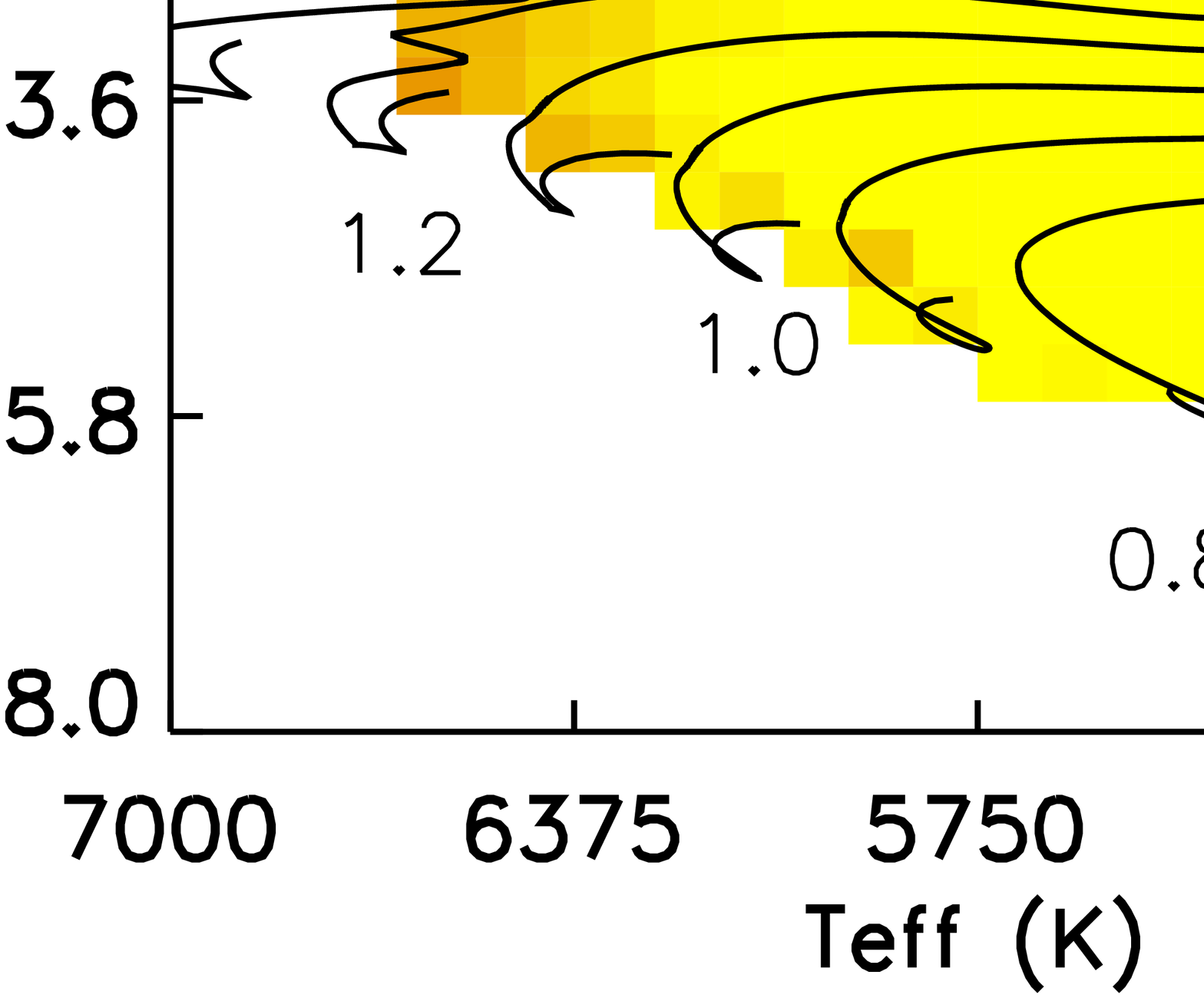}  
	\end{minipage}
}
\subfigure
{
	\begin{minipage}{0.45\linewidth}
	\centering          
	\includegraphics[width=0.96\columnwidth]{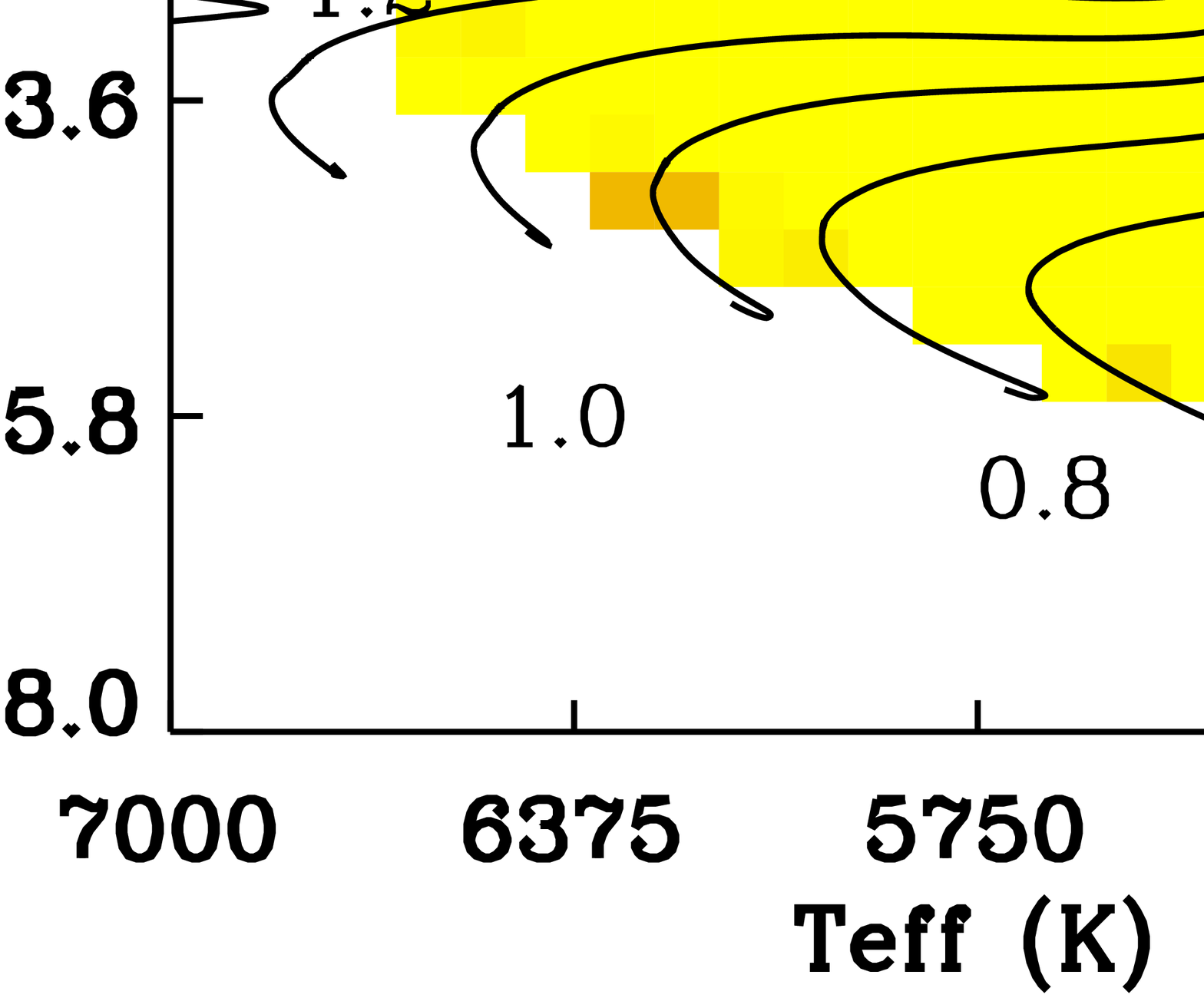}  
	\end{minipage}
}	
\vspace{1.em}
\subfigure
{
	\begin{minipage}{0.45\linewidth}
	\centering     
	\includegraphics[width=0.96\columnwidth]{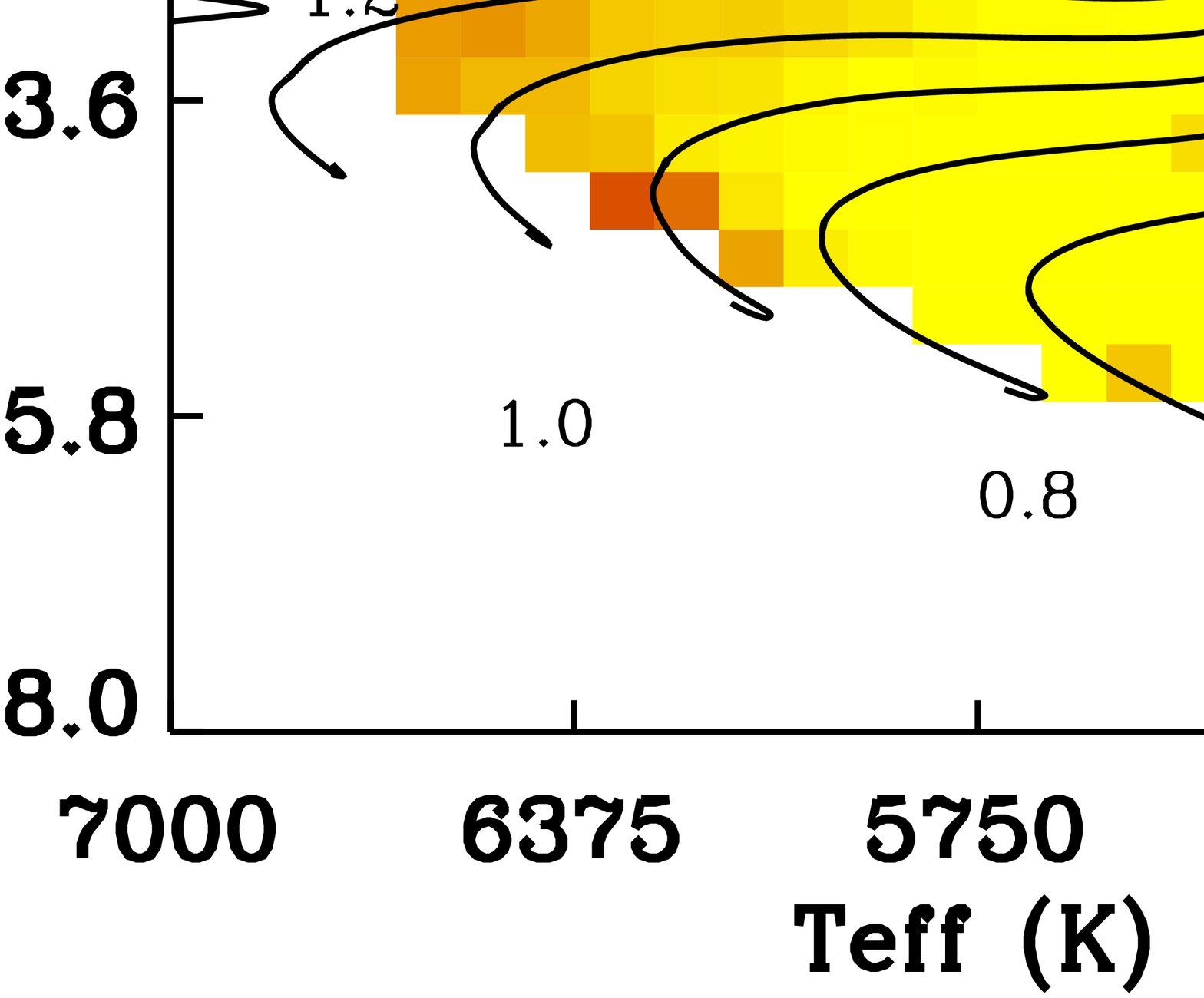}  
	\end{minipage}
}

\vspace{1.5em}
\caption{Fraction of Ba-enhanced stars across the HR diagram for different metallicity regimes: [Fe/H]$>-0.2$ ({\em top}), -0.4$<{\rm [Fe/H]}<-0.2$ ({\em middle}), and -0.6$<{\rm [Fe/H]} < -0.4$ ({\em bottom}). In the left column, colors represent the fraction of stars with ${\rm [Ba/Fe]}>1$ to all the stars, and in the right column, colors represent the fraction of stars with ${\rm [Ba/Fe]}>0.7$ to all the stars. The over-plotted black curves are the PAdova and TRieste Stellar Evolution Code (PARSEC) stellar evolution tracks of mass uniformly distributed from 0.7 to 1.5\,$M_\odot$, as well as 2.0\,$M_\odot$. For the metal-rich case ([Fe/H]$ > -0.2$\,dex), subgiant stars with $\gtrsim2M_\odot$ exhibit a high fraction of Ba-enhancement, while for the metal-poor case ([Fe/H]$ < -0.4$\,dex), subgiant stars with $\gtrsim1.4M_\odot$ can exhibit a large fraction of Ba-enhancement. 
\label{fig222}}
\end{figure}

Fig.\,\ref{fig222} shows the fraction of the Ba-enhanced stars with represent to the whole population across the Hertzsprung-Russell (HR) diagram. The PARSEC stellar evolutionary tracks \citep{Bre2012} with stellar masses uniformly distributed from 0.7\,$M_\odot$ to 1.5\,$M_\odot$, as well as 2.0\,$M_\odot$ are also shown in the figure.
For the PARSEC stellar evolutionary tracks, we have adopted the bolometric correction (BC) of \citet{Chen2019}.
The left column of the figure illustrates that the fraction of the Ba-enhanced stars increases with stellar masses.
For the metal-rich (${\rm [Fe/H]} > -0.2$) subgiant stars with $M\gtrsim2 M_\odot$, $\gtrsim20\%$ of them are Ba-enhanced stars. This fraction decreases with effective temperature (implicitly age), from $\gtrsim$50\% at the warm side ($\gtrsim6500$\,K) to $\lesssim$10\% at the cool side ($\lesssim6000$\,K). 

For more metal-poor stars, the fraction of Ba-enhanced stars can reach a large value at lower masses. For stars with $-0.6 < {\rm [Fe/H]} < -0.4$, the Ba-enhanced stars can be larger than $10\%$ for stars with $1.5\,M_\odot$. If these Ba-enhanced stars have similar [Ba/H] value and mass, the more metal-poor stars should have higher [Ba/Fe] values. There should be more stars with lower masses could have [Ba/Fe]$>1.0$.


The right column of the figure shows similar results to the left column but for stars of ${\rm [Ba/Fe]} > 0.7$. It illustrates the trend more clearly, as there are more stars of such intermediate degree of Ba-enhancement. For stars with $-0.6 < {\rm [Fe/H]} < -0.4$, about $\gtrsim20\%$ of them have ${\rm [Ba/Fe]} > 0.7$ at a mass of 1.3\,$M_\odot$.

\section{Discussion} 
\label{section4}
The above results suggest that the Ba-enhanced dwarf and subgiant sample stars can be parsed into two categories that are plausibly linked to different origin mechanisms. The Ba-enhanced stars with C and N enhancement, mostly with lower temperature and mass, are originated from binary evolution, while the other Ba-enhanced stars, mostly with warmer temperature and intermediate mass, are originated from stellar internal evolution.
\subsection{Ba-enhanced stars originated from binary interaction}
\label{subsection1}
\begin{figure}[htb!]
\centering
\vspace{1.em} 
\centering
\includegraphics[width=0.95\columnwidth]{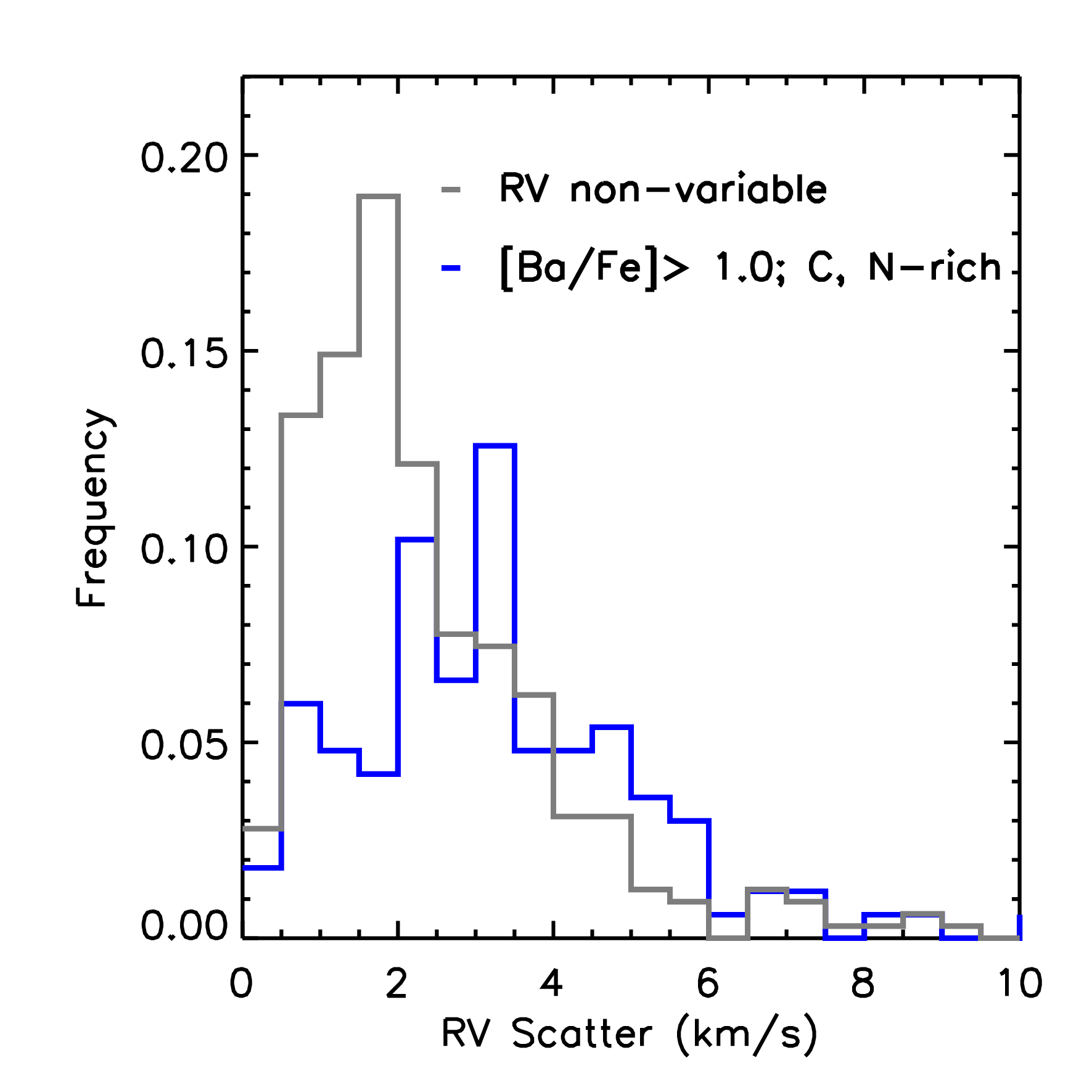}
\vspace{0.8em}
\caption{The normalized distribution of the radial velocity scatters for RV non-variable stars (grey) and Ba-enhanced stars with enhanced C and N abundances (blue) derived from LAMOST Medium-resolution surveys. The RV non-variable stars are the common stars from APOGEE DR17 with small RV dispersion ($< 1\,$km/s). Most of them are considered to be single stars. A large proportion of the Ba-enhanced stars have larger RV scatters than those of the RV no-variable stars.
\label{fig44}}
\end{figure}

The majority of the Ba-enhanced sample stars with C and N enhancement are cooler than 6000\,K. Their abundance patterns for Mg, Si, Ca, Ti, Al, Mn, Cr, Ni, and Cu exhibit no significant difference to those of the Ba-normal stars. These elemental patterns can be explained if these stars are consequences of binary interactions: they have gained Ba-rich as well as C and N-rich materials from AGB companions, so that they exhibit higher C and N abundances while abundances for other elements stay invariant. This is further supported by their bluer $(NUV-G)_0$ colors, which implies that these stars now have a WD companion, as the remnant of an AGB progenitor. 

In previous studies, the binary evolution mechanism has been widely referred to explain the existence of Ba-rich giants \citep[see e.g.,][]{Gri1980,Gri1981,McC1983,Jor1988,McC1990,Jor1998}.
The Ba-rich and carbon-rich dwarfs and subgiants with relatively cool temperatures corresponding to late F and G stars have also been suggested to be consequences of binary evolution \citep{Bond1974, Nor1994, Esc2019}.

To further validate their binary nature, we look into their radial velocity (RV) variations using data from LAMOST Medium-Resolution Spectroscopic survey \citep[MRS;][]{Liu2020}. We found 167 Ba-enhanced stars in our sample having repeat RV measurements from LAMOST MRS, each with 100 individual epochs in typical. To characterize their RV scatters among different epochs, we first define a set of RV non-variables, which are presumed to be single stars, by using the RV repeat measurements in APOGEE DR17 (\citealt{Abd2022}). For RV non-variables, we require their RV scatters among individual APOGEE measurements to be smaller than 1.0~km/s. We cross-match LAMOST MRS catalog with APOGEE DR17, and obtain 322 common stars of RV non-variables. Fig.\,\ref{fig44} shows the C and/or N-rich Ba-enhanced stars have typical RV scatters of $\sim3$km/s, which is larger than that of the RV non-variables ($\sim1.5$\,km/s). This indicates that the majority of Ba-enhanced stars with C and/or N enhancement are likely in binary systems, as expected.

\subsection{Origin of the `` high-[$\alpha$/Fe] desert"}
Our results revealed a lack of high-$\alpha$ Ba-enhanced stars, which we deemed as a ``high-$\alpha$ {\em desert}" (Sect. 3.1.3). It is well-known that a large portion ($\simeq50$\%) of stars, including the high-$\alpha$ population, are in binary systems \citep[e.g.,][]{Duchene2013, Gao2014, Yuan2015}. As the high-$\alpha$ stars are intrinsically old, a proportion of them should have gained Ba-rich materials through binary interactions. However, this is not observed by our sample stars. We thus need a mechanism to explain the presence of such a ``high-$\alpha$ {\em desert"}.

Here we propose plausible mechanisms.
\subsubsection{Mass and metallicity dependent AGB yields}
For a typical high-$\alpha$ star, its accreted materials from the AGB companion cannot result in a sufficiently high level of Ba enhancement (i.e., [Ba/Fe]$>1.0$). To illustrate this, we make a simple quantitative examination on the expected surface [Ba/Fe] from binary mass exchange.
Following \citet{Hus2009}, a star's surface element abundance ratio [X/Fe] after accreting materials from an AGB companion can be expressed as 
\begin{equation}
\left[\frac{\rm X}{\rm Fe}\right]={\rm log}\left(10^{\left[\frac{\rm X}{\rm Fe}\right]^{\rm ini}}\cdot (1-10^{-d})+ 10^{\left[\frac{\rm X}{\rm Fe}\right]^{\rm AGB}} \cdot 10^{-d}\right),  
\end{equation}
where the ${\left[\frac{\rm X}{\rm Fe}\right]^{\rm ini}}$ and ${\left[\frac{\rm X}{\rm Fe}\right]^{\rm AGB}}$ are the [X/Fe] values of stars before and after mass transfer.
The $d$ is the so-called dilution factor, which is defined as 10$^{d}$ = $M^{\rm env}_{\star}$/$M^{\rm transf}_{\rm AGB}$. Here 
$M^{\rm env}_{\star}$ is the envelope mass of the star after the mass transfer, and $M^{\rm transf}_{\rm AGB}$ is the mass transferred from the AGB.

Since the convective zone mass of a G-type main-sequence star is smaller than $10^{-1.5}$ $M_\odot$ \citep{Pin2001}, it is plausible to assume that the mass transferred from the AGB companion ($M^{\rm transf}_{\rm AGB}$) is approximately equivalent to the surface convective envelope mass of the Ba-enhanced dwarf. In this case, its dilution factor is zero, i.e., the surface abundance of the Ba-enhanced star is identical to that of the AGB companion. 

\begin{figure}[htb!]
\vspace{1.em}
\centering 
\includegraphics[width=0.95\columnwidth]{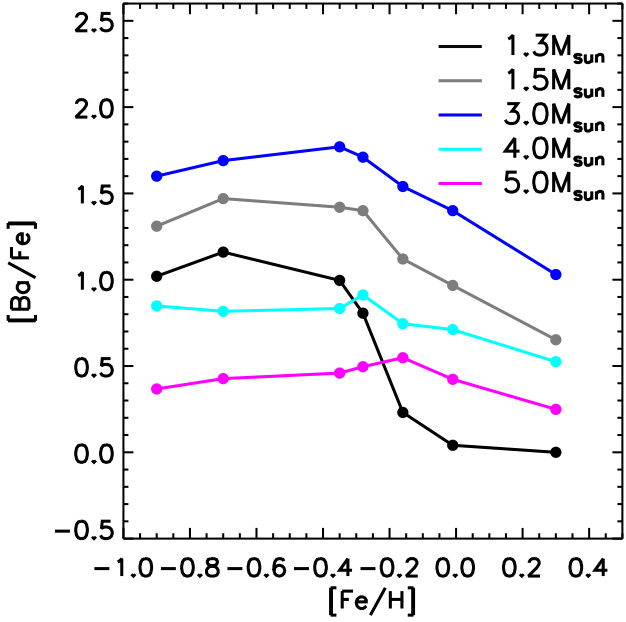} 
\vspace{2.em}
\caption{The expected stellar surface [Ba/Fe] ratios after accreting Ba-rich materials from AGB companions of different masses and metallicity, assuming a dilution factor of zero.
\label{fig555}}
\end{figure}

Fig.\,\ref{fig555} shows the expected stellar surface [Ba/Fe] values after accreting materials from AGB companions of different masses, assuming a  dilution factor of zero.
Here we have adopted the AGB yields model of the FRANEC Repository of Updated Isotopic Tables \& Yields (FRUITY)  (\citealt{Cri2009, Cri2011,Cri2015}). Our computations cover a wide range of metallicity from $Z=0.001$ to $Z=0.020$ for the mass range 1.3--6.0\,$M_\odot$.

The figure shows that, for stars with ${\rm -1.0\lesssim [Fe/H]}\lesssim0.0$, only AGB companions with mass between 1.5\,$M_\odot$ and 3\,$M_\odot$ can result in a Ba-enhancement level of ${\rm [Ba/Fe]}>1.0$. While AGB companions with mass higher than 4.0\,$M_\odot$ cannot result in ${\rm [Ba/Fe]}>1.0$ for all metallicity considered. The AGB companions of $1.3\,M_\odot$ can result in ${\rm [Ba/Fe]}>1.0$ only for stellar systems of ${\rm [Fe/H]}\lesssim-0.4$, but the resultant stellar surface [Ba/Fe] is sharply decreased to below ${\rm [Ba/Fe]}=1.0$ for systems of ${\rm [Fe/H]}\gtrsim-0.4$.

For high-$\alpha$ stellar populations, we expect the low-mass AGB stars ($M < 1.5\,M_\odot$) are the main donors of mass transfer, both because there are more low-mass stars than high-mass ones in the Galaxy due to the initial stellar mass function, and because the high-$\alpha$ populations are generally old ($\gtrsim10$\,Gyr) stars. So that the strong decrease of [Ba/Fe] above the [Fe/H] knee point (${\rm [Fe/H]}\simeq-0.4$) in Fig.\,\ref{fig555} might have played a major role to the lack of high-$\alpha$ Ba-enhanced stars in our sample.

Note that in the above discussion we have assumed a dilution faction of zero. For cases with a larger dilution factor, the resulting [Ba/Fe] will be even lower, so that the conclusion above still holds.  

\subsubsection{Low-[$\alpha$/Fe] stars with thick disk kinematics}
\begin{figure}[htb!]
\centering 
\subfigure
{
	\begin{minipage}{0.85\linewidth}
	\centering     
	\includegraphics[width=0.93\columnwidth]{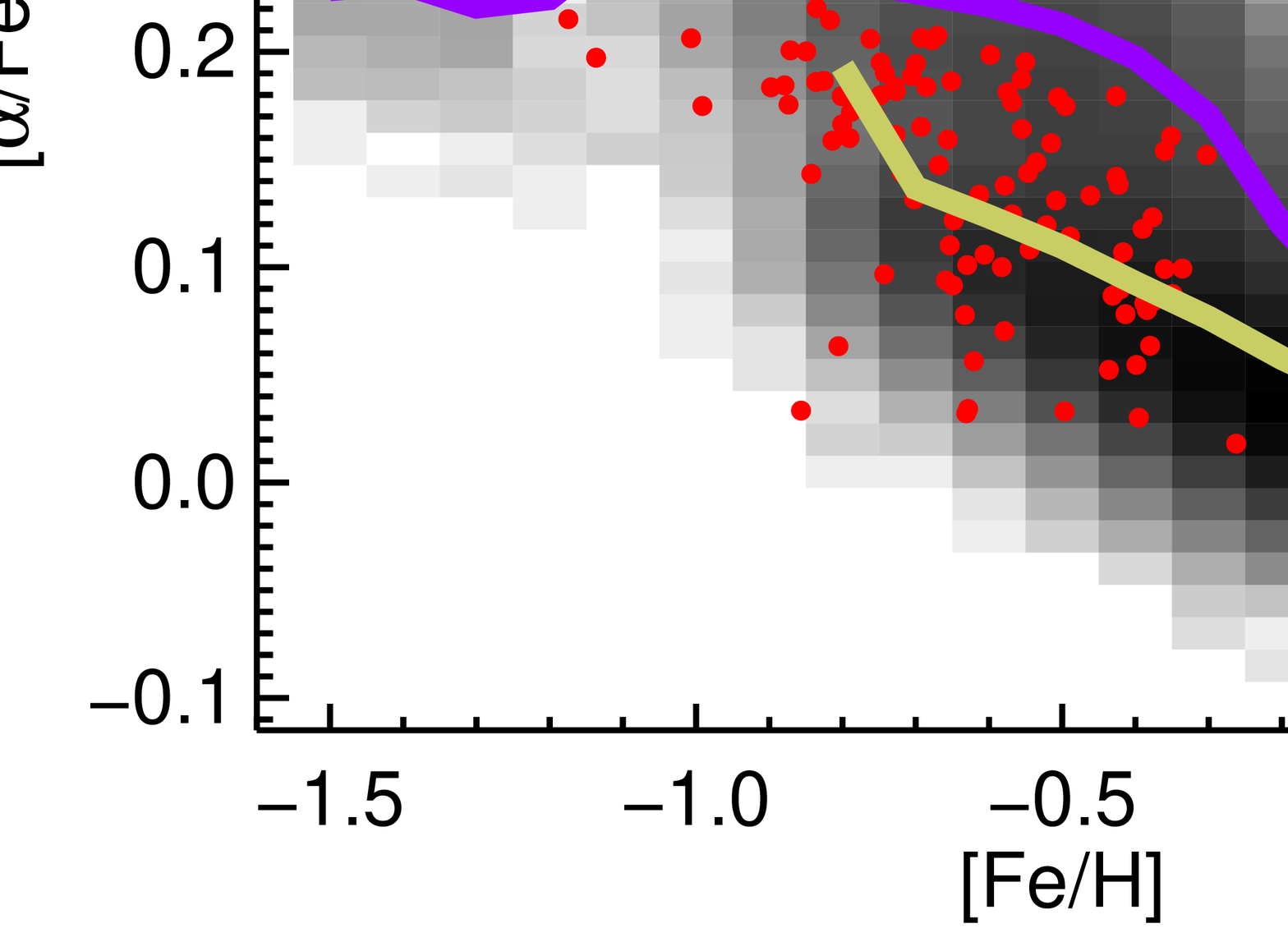}  
	\end{minipage}
}
\subfigure
{
	\begin{minipage}{0.85\linewidth}
	\centering     
	\includegraphics[width=0.93\columnwidth]{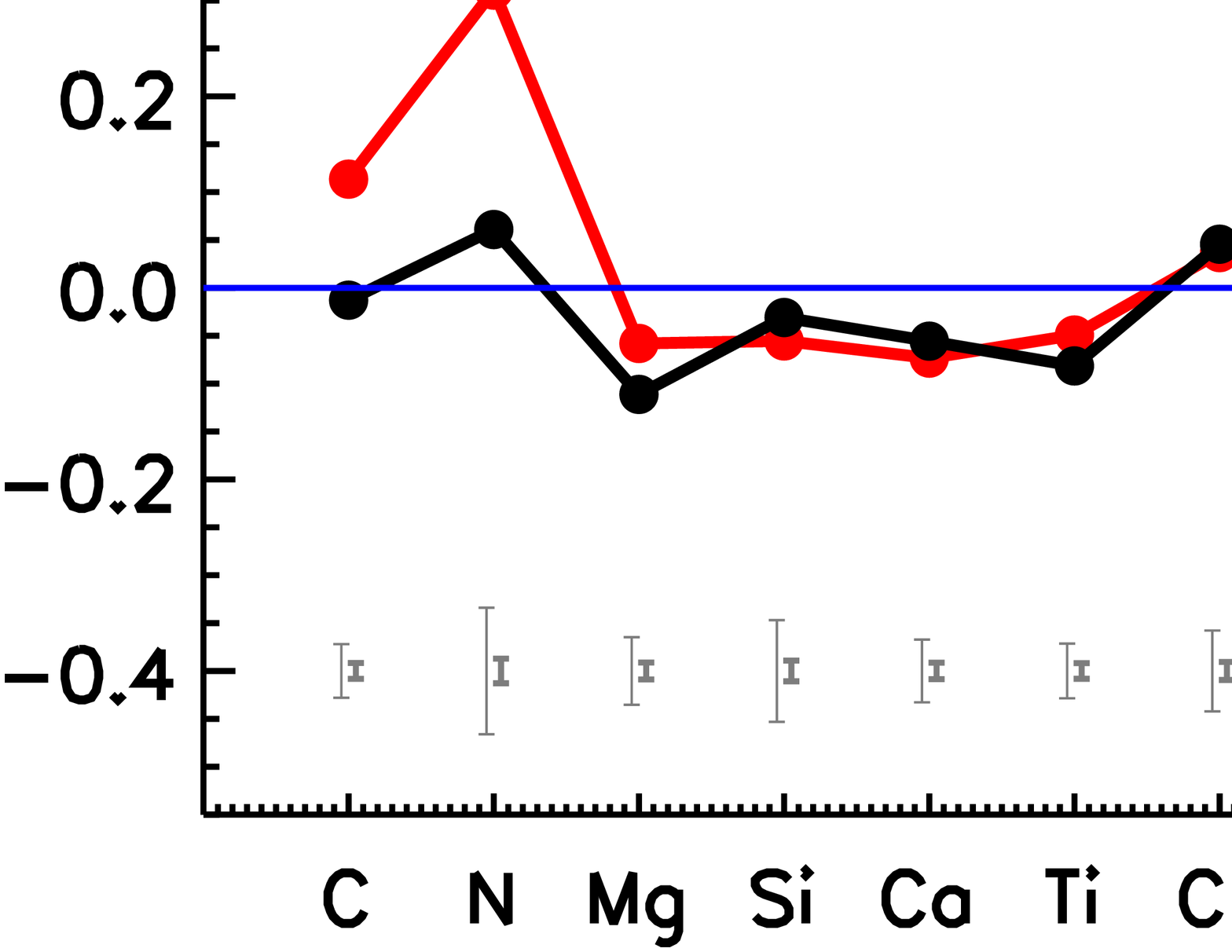}  
	\end{minipage}
}
\vspace{2.em}
\caption{{\em Top}: Distribution of stars in the [Fe/H]--[$\alpha$/Fe] plane. The red dots represents the cool ($T_{\rm eff} < 6000$\,K) Ba-enhanced stars with ``thick disk" kinematics ($500 < L_{\rm Z} < 1200$\, km/s$\cdot$ kpc). The grey-black background show density distribution of all Ba-normal stars. The mean [$\alpha$/Fe] as a function of [Fe/H] for Ba-normal stars with thick disk kinematics is shown by the purple solid line, while that for stars with thin disk kinematics is shown by the yellow solid line. 
{\em Bottom:} Abundance differences between the low-[$\alpha$/Fe], Ba-enhanced stars with thick disk kinematics and the Ba-normal, kinematically thick disk stars (red), as well as those between the low-$\alpha$, Ba-normal stars and the Ba-normal, kinematically thick disk stars (black). The median abundances in this panel are calculated by the subsamples of stars (see text). The low-[$\alpha$/Fe] Ba-enhanced stars with thick disk kinematics exhibit higher C and N abundances than those of both Ba-normal thick disk stars and the low-[$\alpha$/Fe] Ba-normal stars, while the other elemental abundances are similar to those of the low-$\alpha$ Ba-normal stars. The error bars at the bottom show the median measurement uncertainties for individual stars (slam lines) and errors of the mean abundances of the Ba-enhanced sample (thick lines), after dividing the median individual uncertainty by square root of the star number.} 
\label{fig666}
\end{figure}

The discussions above have demonstrated the existence of a substantial lack of Ba-enhanced stars in the high-[$\alpha$/Fe], chemical thick-disk population. However, a portion of the low-[$\alpha$/Fe] Ba-enhanced stars are found to exhibit kinematics similar to the thick disk population. Fig.\,\ref{fig666} shows the [$\alpha$/Fe]--[Fe/H] distribution for Ba-enhanced stars with $T_{\rm eff} < 6000$~K, $\log\,g > 3.8$, and $500 < L_{Z}< 1200$\,km/s$\cdot$\,kpc, an orbital angular momentum range for thick disk stars  (Fig.\,\ref{fig11}). These kinematically thick disk Ba-enhanced stars are also found to have small guiding center radii and large orbital eccentricity. Most of these stars have [$\alpha$/Fe] values similar to those of the low-[$\alpha$/Fe], kinematical thin-disk population (yellow line in Fig.\,\ref{fig666}), but lower than the mean value of the Ba-normal, kinematical thick-disk population (purple line in Fig.\,\ref{fig666}).

The origin of these chemically thin but kinematically thick disk Ba-enhanced stars is unclear. To understand this, we compare the elemental abundance patterns between the low-[$\alpha$/Fe], Ba-enhanced stars with thick disk kinematics and the kinematical thick-disk Ba-normal stars, as well as low-[$\alpha$/Fe] Ba-normal stars. 
The lines in bottom panel of Fig.\,\ref{fig666} show the elemental patterns of the median abundance differences. We calculate the median abundances by using the subsamples of stars with $5500<T_{\rm eff} < 5800$~K, $3.6<\log\,g <4.2$, and $-0.6<$[Fe/H]$<-0.4$. There are 40 low-[$\alpha$/Fe], Ba-enhanced stars with thick disk kinematics, 1453 kinematical thick-disk Ba-normal stars, and 1637 low-[$\alpha$/Fe] Ba-normal stars.
It shows that the former have significantly higher C and N abundances than the Ba-normal stars, suggesting they may have experienced binary evolution, through which they obtained C and/or N-rich and Ba-rich materials from an AGB companion. On the other hand, their $\alpha$-elements (Mg, Si, Ca Ti) are systematically lower than the Ba-normal thick disk stars. It seems unlikely that these low-[$\alpha$/Fe], Ba-enhanced stars with thick disk kinematics share the same origin as the Ba-normal thick disk population, unless they have experienced some poorly known surface elemental transport processes that altered their surface $\alpha$ abundances. Elemental transport process has been suggested to be able to change the surface elemental patterns (Section\,4.3).
However, it was found to happen in stars with intermediate mass ($M\gtrsim1.4 M_\odot$), but has never been found in stars with $\sim $1\,$M_\odot$, the case of those low-[$\alpha$/Fe] stars with thick disk kinematics in Fig.\,\ref{fig666}.

Chemically thin but kinematically thick disk stars have been discussed in literature, for instance, \citet{Red2006} suggested that low-[$\alpha$/Fe] stars with thick disk kinematics were subjected to greater heating due to stochastic heating processes, implying there were formed in the early thin disk. This seems to be a plausible explanation of our low-[$\alpha$/Fe] Ba-enhanced stars with thick disk kinematics. 

We have further examined the occurrence rate of the chemical thin but kinematic thick disk stars. For occurrence rate, $f$, we defined as 
\begin{equation}
f = \frac{N_{\rm TKLA}}{N_{\rm LA}}, 
\end{equation}
where $N_{\rm TKLA}$ is the number of low-[$\alpha$/Fe] stars with thick disk kinematics, $N_{\rm LA}$ the total number of low-[$\alpha$/Fe] stars. 
We find tentative evidence that the Ba-enhanced stars have higher $f$ than Ba-normal stars (see Appendix A for details). This implies binary interaction may have played a visible role for the disk heating.

\subsection{Ba-enhanced stars due to internal element transports} 
\label{subsection2}
The warmer, more massive Ba-enhanced stars do not exhibit significant C and N enhancement, but exhibit clear depletion in Mg abundance. This is similar to previous findings for Am/Fm stars, whose surface abundances are altered by element transport due to competition between gravitational settling and radiative acceleration \citep{Mic1970,Mic1982,Vic2010,Mic2011,Dea2020}. These Am/Fm stars have been found to exhibit Ba enhancement by up to 1000 times \citep[e.g.][]{Gha2018, Xiang2020}. Recently, a large sample of Ba-enhanced Am/Fm stars from the LAMOST DD-Payne catalog have been studied by \citet{Xiang2020}, and their sample stars exhibit enhanced abundances of iron-peak elements (Cr, Mn, Fe, Ni) but depleted abundances of Mg and Ca. The Am/Fm sample stars of \citet{Xiang2020} have typical effective temperature hotter than 6700\,K, which is higher than ours. Thus our sample stars are likely an extension of the Am/Fm stars to lower effective temperature. 

The radiative acceleration process of hot metal-rich stars has been studied for a long time, but detailed studies of the low-temperature ($T_{\rm eff} < 6800$\,K) or metal-poor ([Fe/H]$ < -0.2$) stars are limited so far.
We suggest other than about one-third Ba enhanced stars having enhanced [(N+C)/Fe] abundance ratios, the remaining Ba-enhanced stars could have changed their surface abundances by internal element transporting process.

\begin{figure*}[htb!]
\centering   
\subfigure
{
	\begin{minipage}{0.45\linewidth}
	\centering  
	\includegraphics[width=0.95\columnwidth]{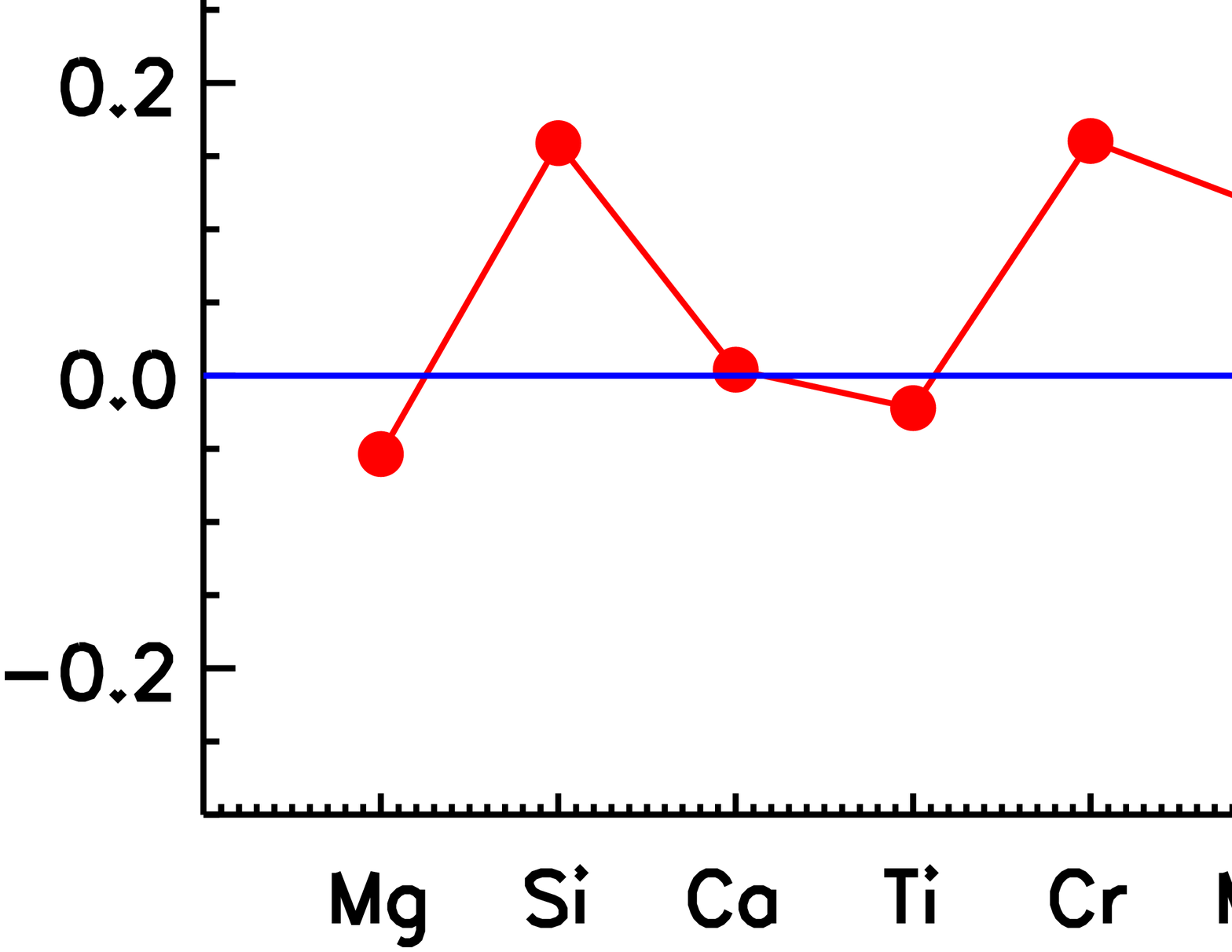} 
	\end{minipage}
}      
\subfigure
{
	\begin{minipage}{0.45\linewidth}
	\centering     
	\includegraphics[width=0.93\columnwidth]{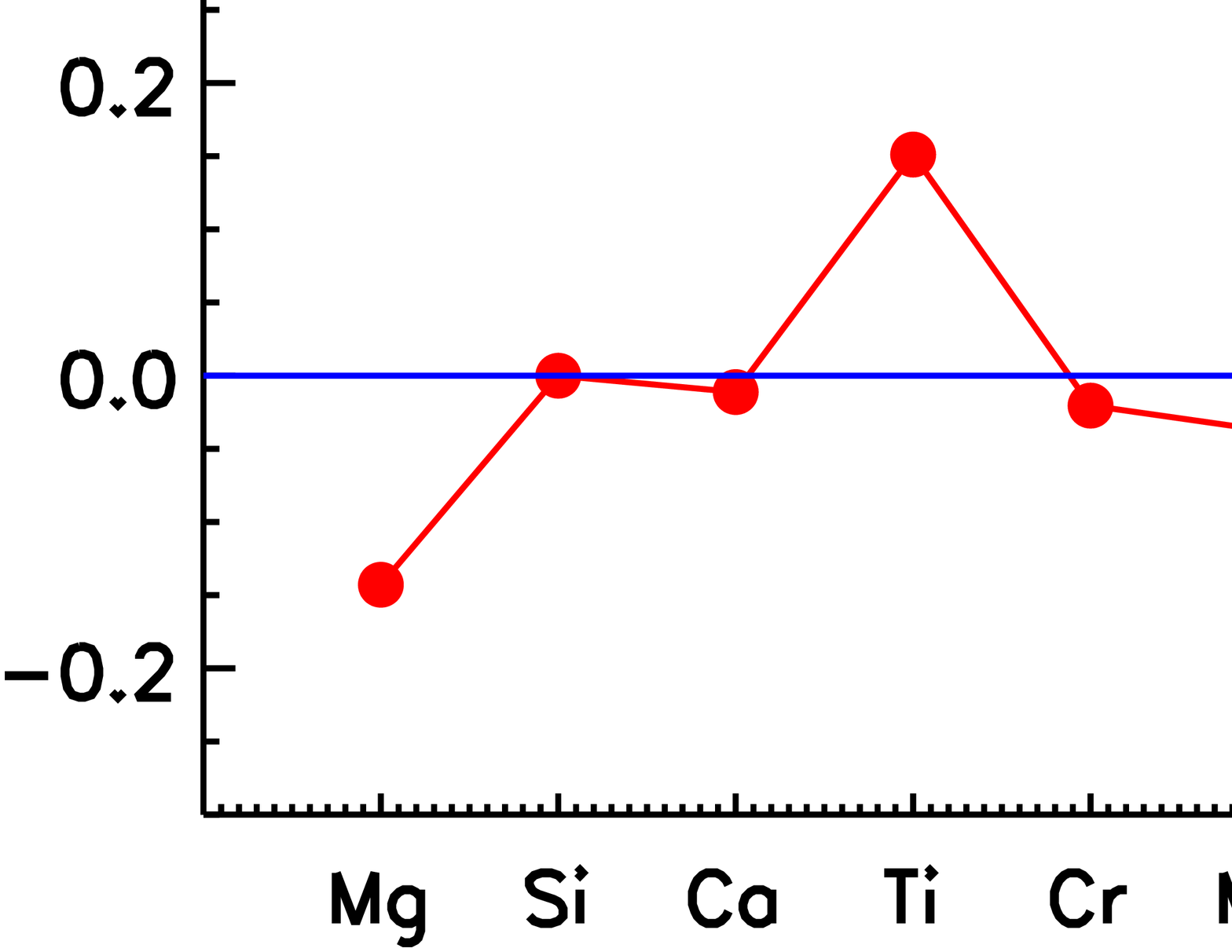}  
	\end{minipage}
}
\vspace{2.em}
\caption{The differences of [X/H] median between the Ba-enhanced and Ba-normal stars with ${\rm [Fe/H]}>-0.2$. The left and right panel shows the results for hot stars with $6700<T_{\rm eff}<7500$\,K and warm stars with $6000 < T_{\rm eff} < 6700$\,K, respectively. These two metal-rich Ba-enhanced star samples both show depleted [Mg/Fe] values, but they have different behaviors of Si, Ti, Fe, Mn, and Ni abundances.
\label{fig111a}}
\end{figure*}

In Fig.\,\ref{fig111a}, we show the peculiar elemental abundance patterns of our warm Ba-enhanced stars as well as those of the hot ($6700 < T_{\rm eff} < 7500$\,K) Ba-enhanced Am/Fm stars of \citet{Xiang2020}.
Both hot and warm Ba-enhanced sample stars have depleted Mg, which is in line with  the stellar radiative acceleration models \citep[e.g.][]{Tal2006, Vic2010}. 
However, these hot and warm Ba-enhanced metal-rich sample stars exhibit very different patterns in Si, Ca, Ti, Cr, Mn, Fe and Ni. For these warm Ba-enhanced stars, the Si, Ca, Cr, Mn, and Fe do not exhibit significant enhancement or depletion with respect to the Ba-normal stars of similar temperature and metallicity. This implies that the element transport processes in these warm stars might cause a very different elemental abundance patterns to those of the hotter Am/Fm stars. 

\begin{figure*}[htb!]
\centering   
\vspace{0.5cm}
\subfigbottomskip=15pt
\subfigure
{
	\begin{minipage}{0.45\linewidth}
	\centering          
	\includegraphics[width=0.95\columnwidth]{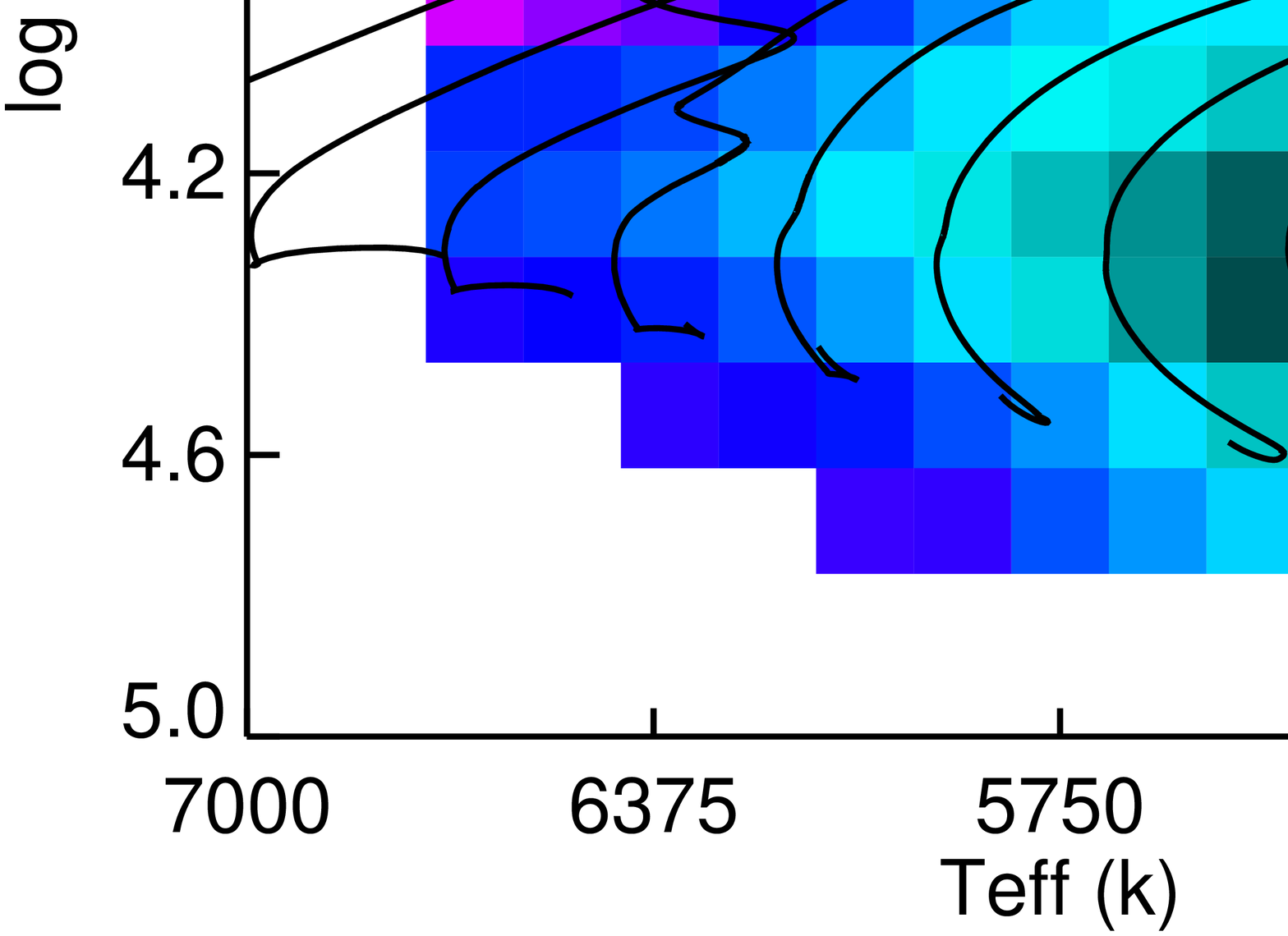}  
	\end{minipage}
}	
\subfigure
{
	\begin{minipage}{0.45\linewidth}
	\centering     
	\includegraphics[width=0.95\columnwidth]{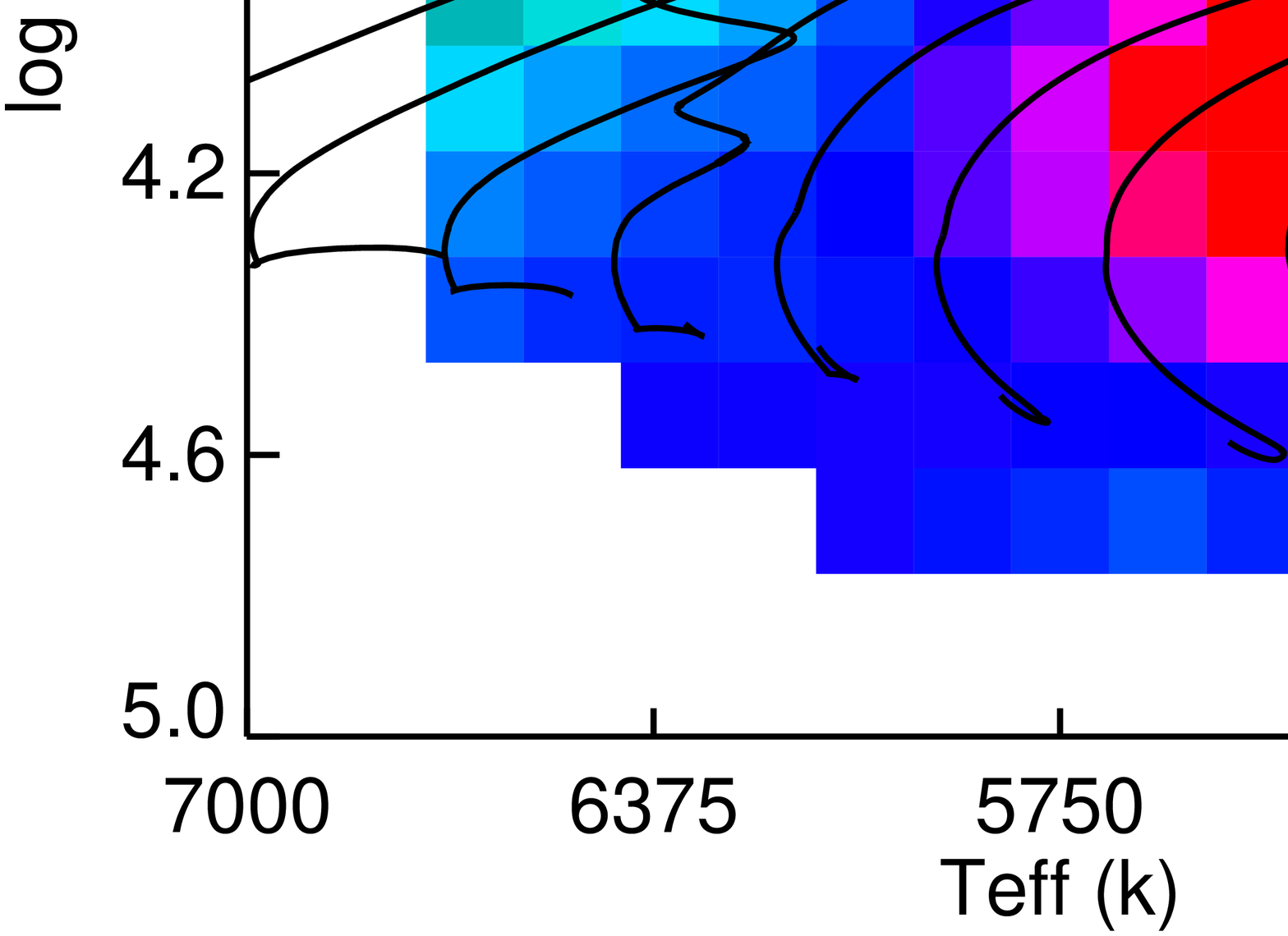}  
	\end{minipage}
}
\vspace{1.em}
\subfigure
{
	\begin{minipage}{0.45\linewidth}
	\centering    
	\includegraphics[width=0.95\columnwidth]{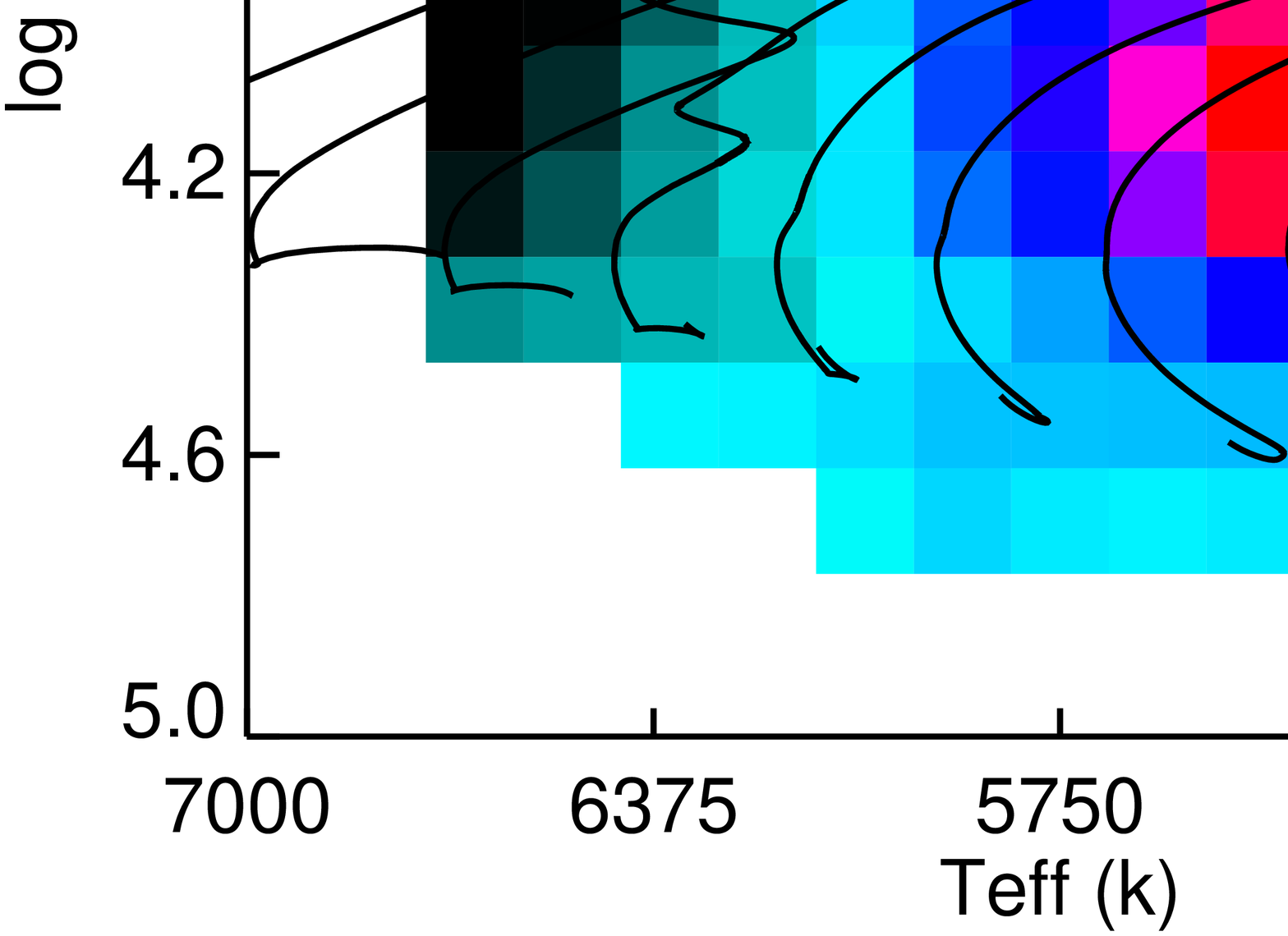}  
	\end{minipage}
}	
\subfigure
{
	\begin{minipage}{0.45\linewidth}
	\centering    
	\includegraphics[width=0.95\columnwidth]{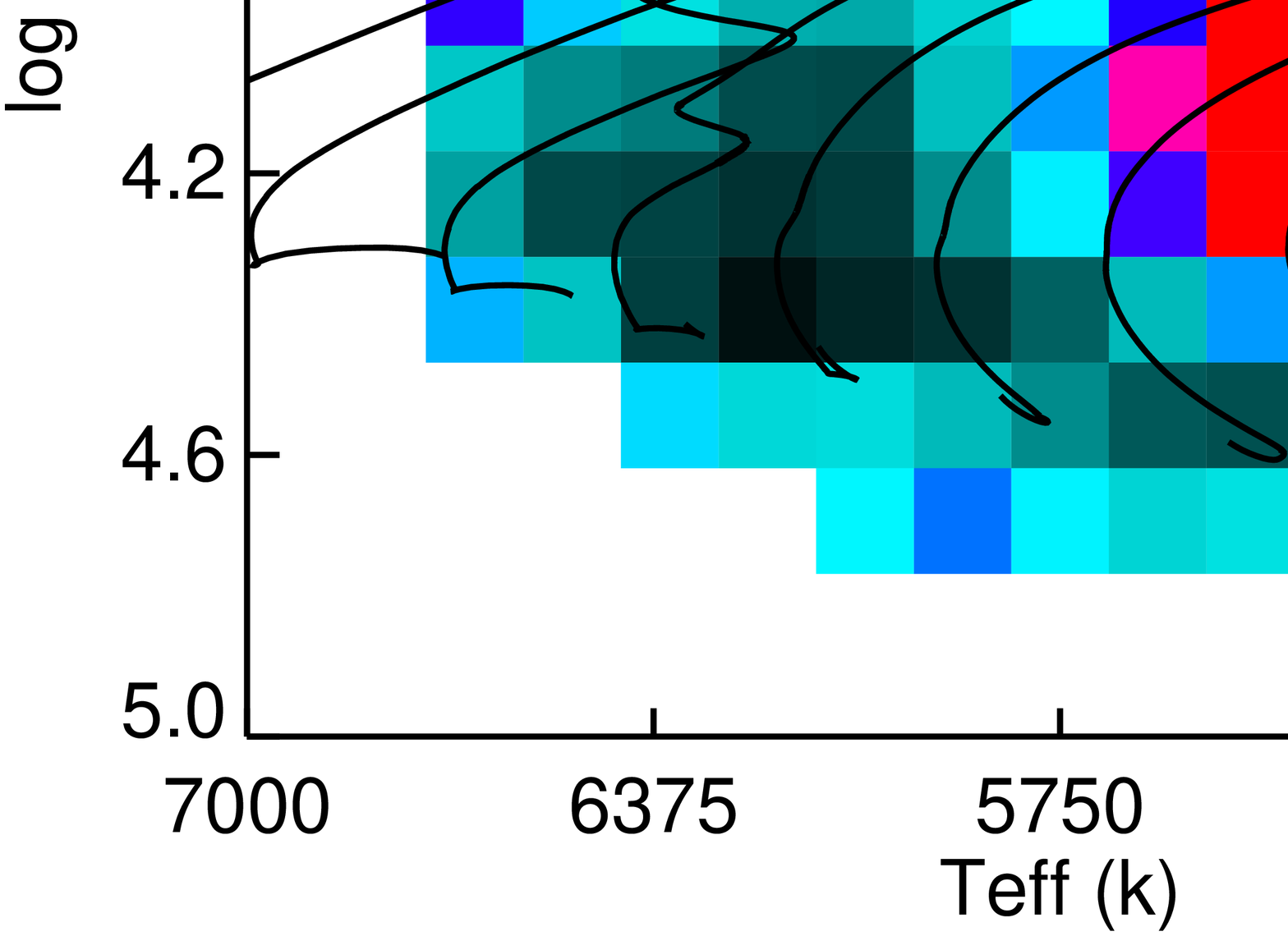} 
	\end{minipage}
}	
\vspace{1.em}
\caption{ Distributions of medium elemental abundances across the Kiel diagram for C and N normal stars ([(N+C)/Fe]$<0.1$) with  $\rm{[Fe/H]}>-0.2$. The PARSEC stellar evolution tracks with mass uniformly distributed from 0.7 to 1.4\,$M_\odot$ and Z=0.014 are overplotted.
\label{fig333}}
\end{figure*}

We also found that the median [Ba/Fe], [Mg/Fe], [Ti/Fe], and [Ni/Fe] ratios may change with their stellar masses, as shown in Fig.\,\ref{fig333}.
The more massive stars in the top-left regime have higher [Ba/Fe] and [Ti/Fe] ratios, as well as lower [Mg/Fe] and [Ni/Fe] ratios than those of low-mass stars. There could be some effect due to the Galactic chemical evolution, which will cause higher [Mg/Fe] values for older stars with smaller mass. However, for stars hotter than 6000\,K, we expect the chemical peculiarities are driven by stellar internal elemental transport processes.

In Appendix B, we also show the distributions of elemental abundances of Mg, Ni and Ti for stars from APOGEE DR17 and GALAH DR3. From these catalogues, we find similar trends to those of Fig.\,\ref{fig333}.

\begin{figure*}[htb!]
\centering   
\subfigure
{
	\begin{minipage}{0.3\linewidth}
	\centering          
	\includegraphics[width=0.95\columnwidth]{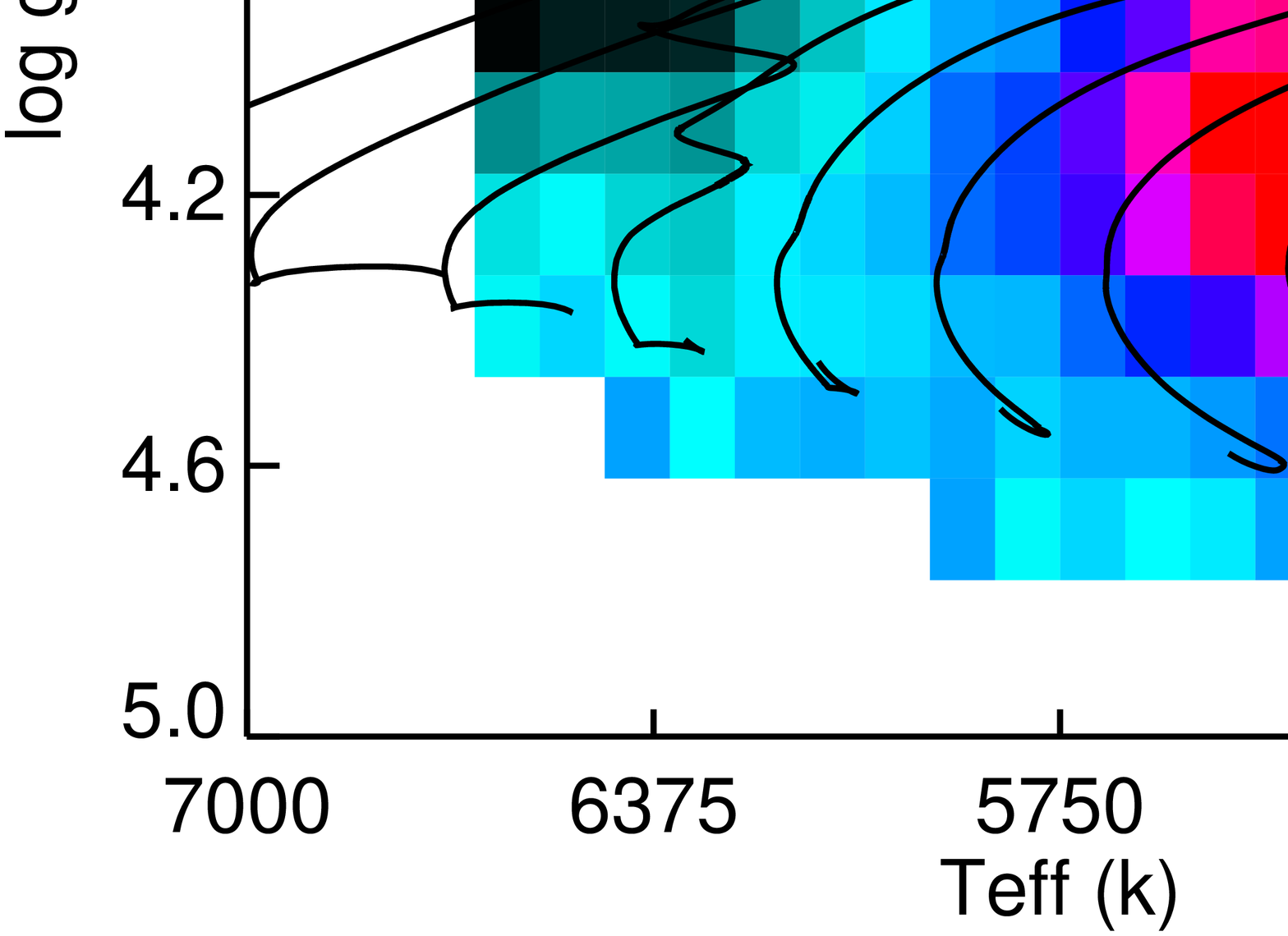}  
	\end{minipage}
}	
\subfigure
{
	\begin{minipage}{0.3\linewidth}
	\centering     
	\includegraphics[width=0.95\columnwidth]{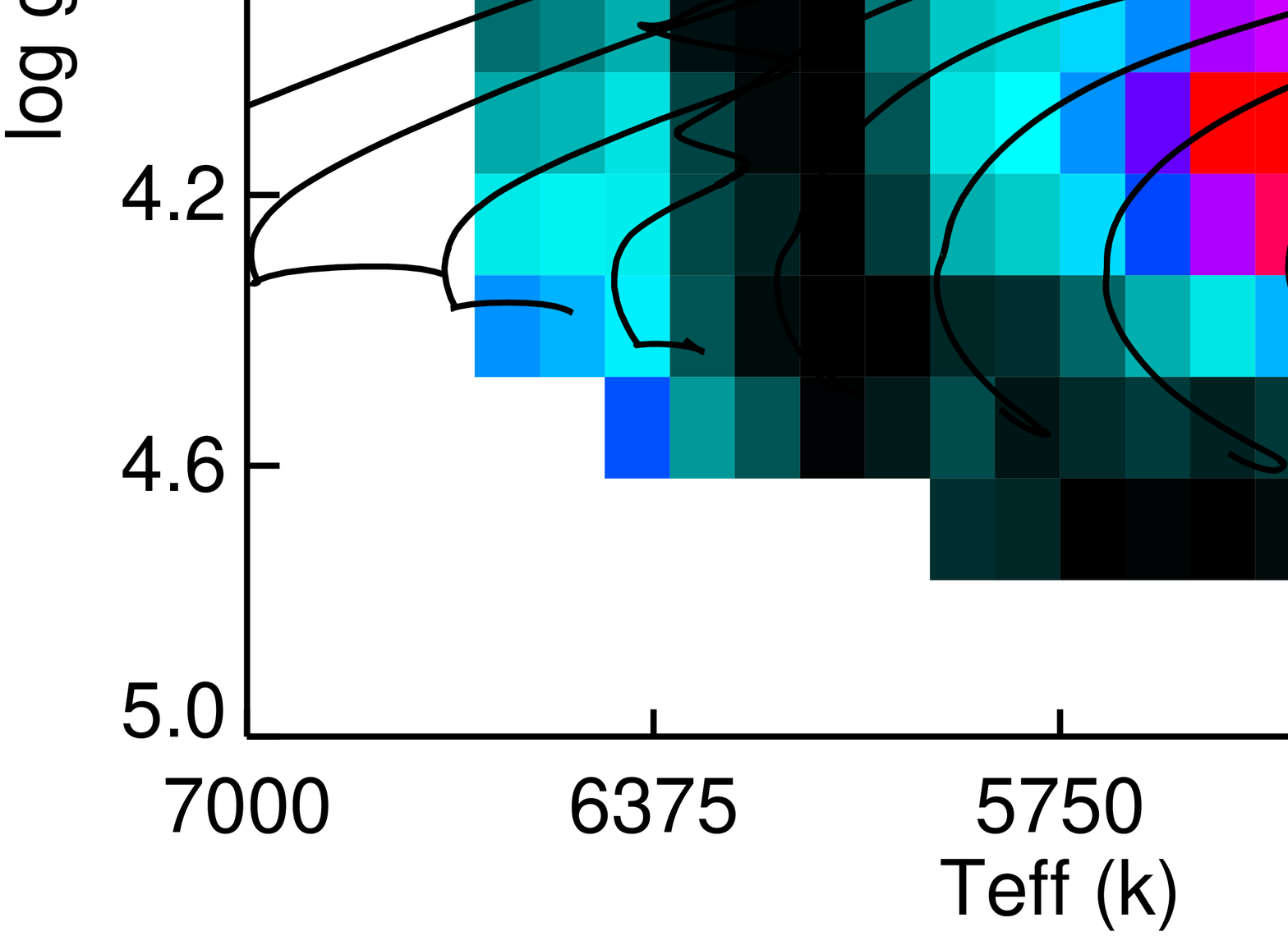}  
	\end{minipage}
}
\vspace{1.em}
\subfigure
{
	\begin{minipage}{0.3\linewidth}
	\centering    
	\includegraphics[width=0.95\columnwidth]{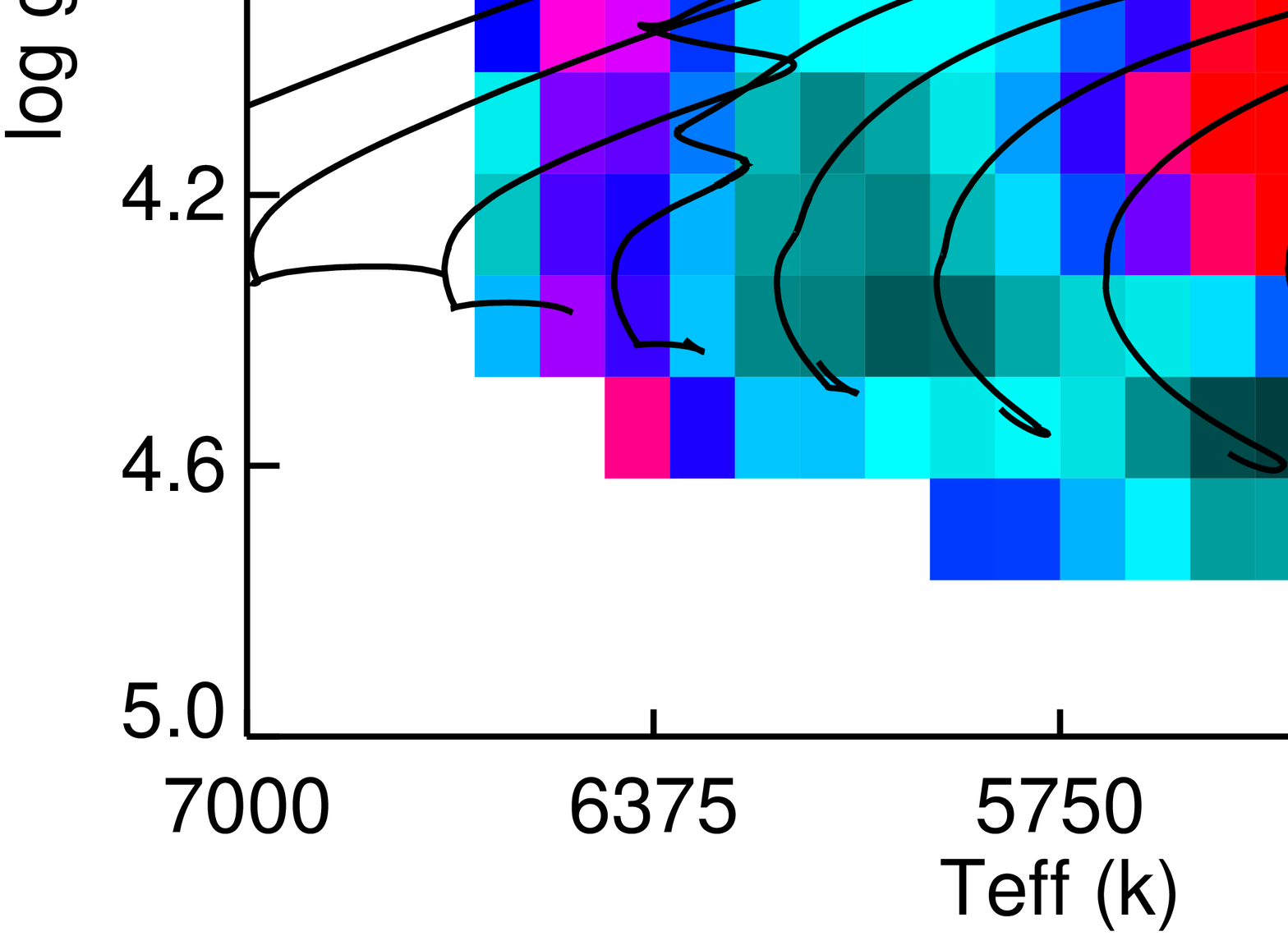}  
	\end{minipage}
}	
\vspace{2.em}
\caption{Distributions of medium elemental abundances across the Kiel diagram of stars from APOGEE DR17 with $\rm{[Fe/H]} > -0.2$. The PARSEC stellar evolution tracks with mass uniformly distributed from 0.7 to 1.4\,$M_\odot$ and Z=0.014 are overplotted.
\label{figc}}
\end{figure*}

\begin{figure*}[htb!]
\centering   
\subfigure
{
	\begin{minipage}{0.3\linewidth}
	\centering          
	\includegraphics[width=0.95\columnwidth]{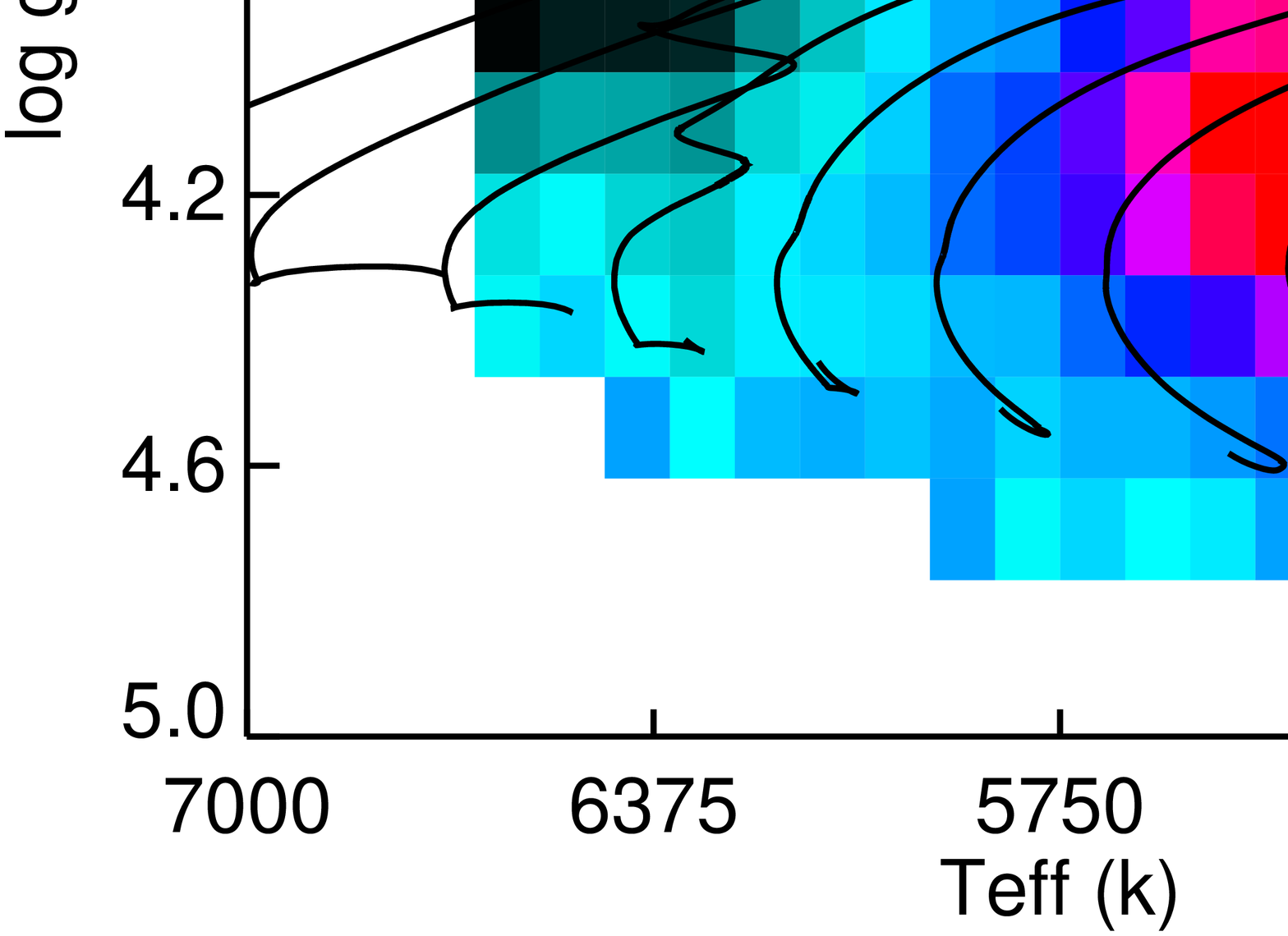}  
	\end{minipage}
}	
\subfigure
{
	\begin{minipage}{0.3\linewidth}
	\centering     
	\includegraphics[width=0.95\columnwidth]{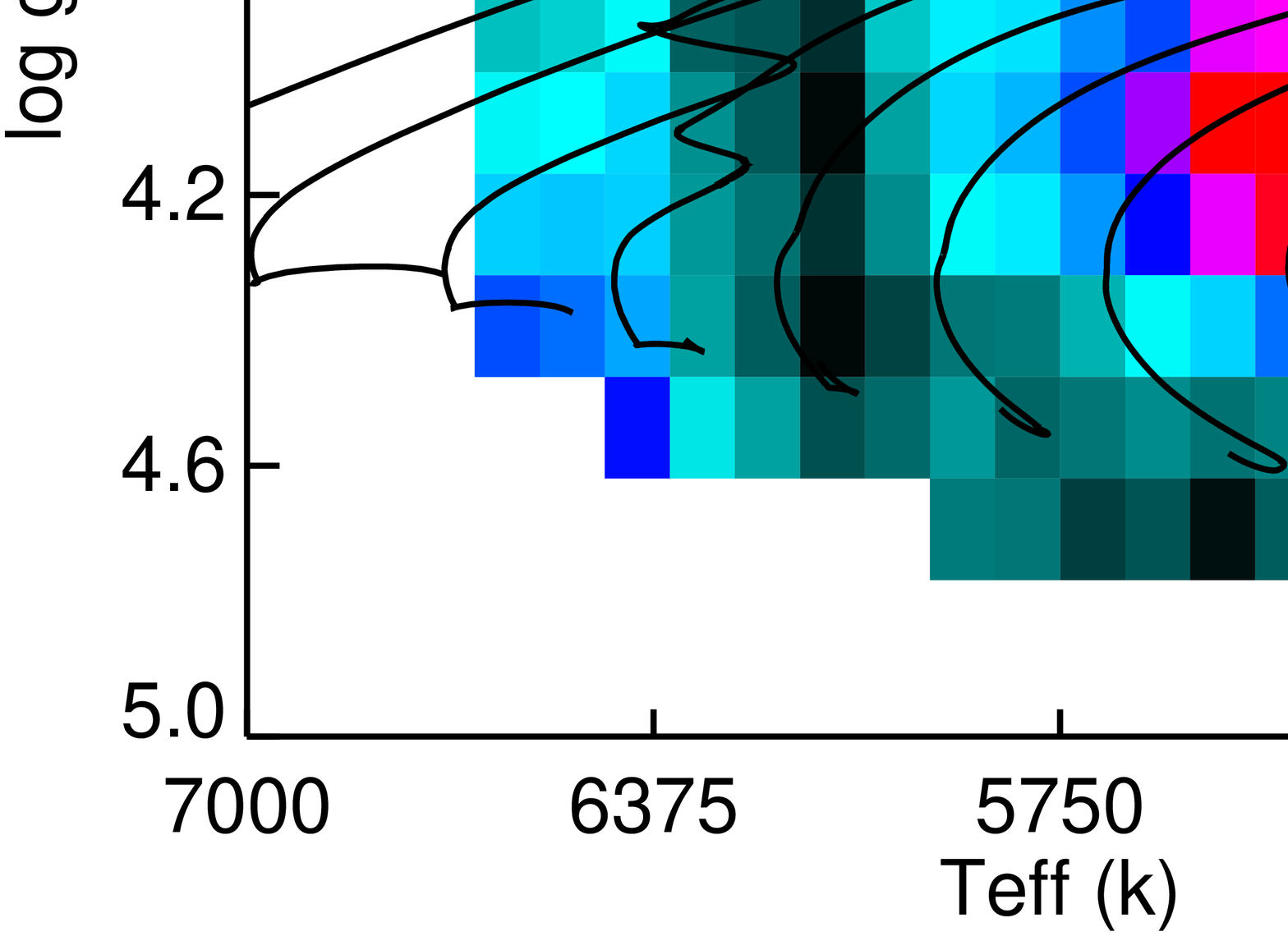}  
	\end{minipage}
}
\vspace{1.em}
\subfigure
{
	\begin{minipage}{0.3\linewidth}
	\centering    
	\includegraphics[width=0.95\columnwidth]{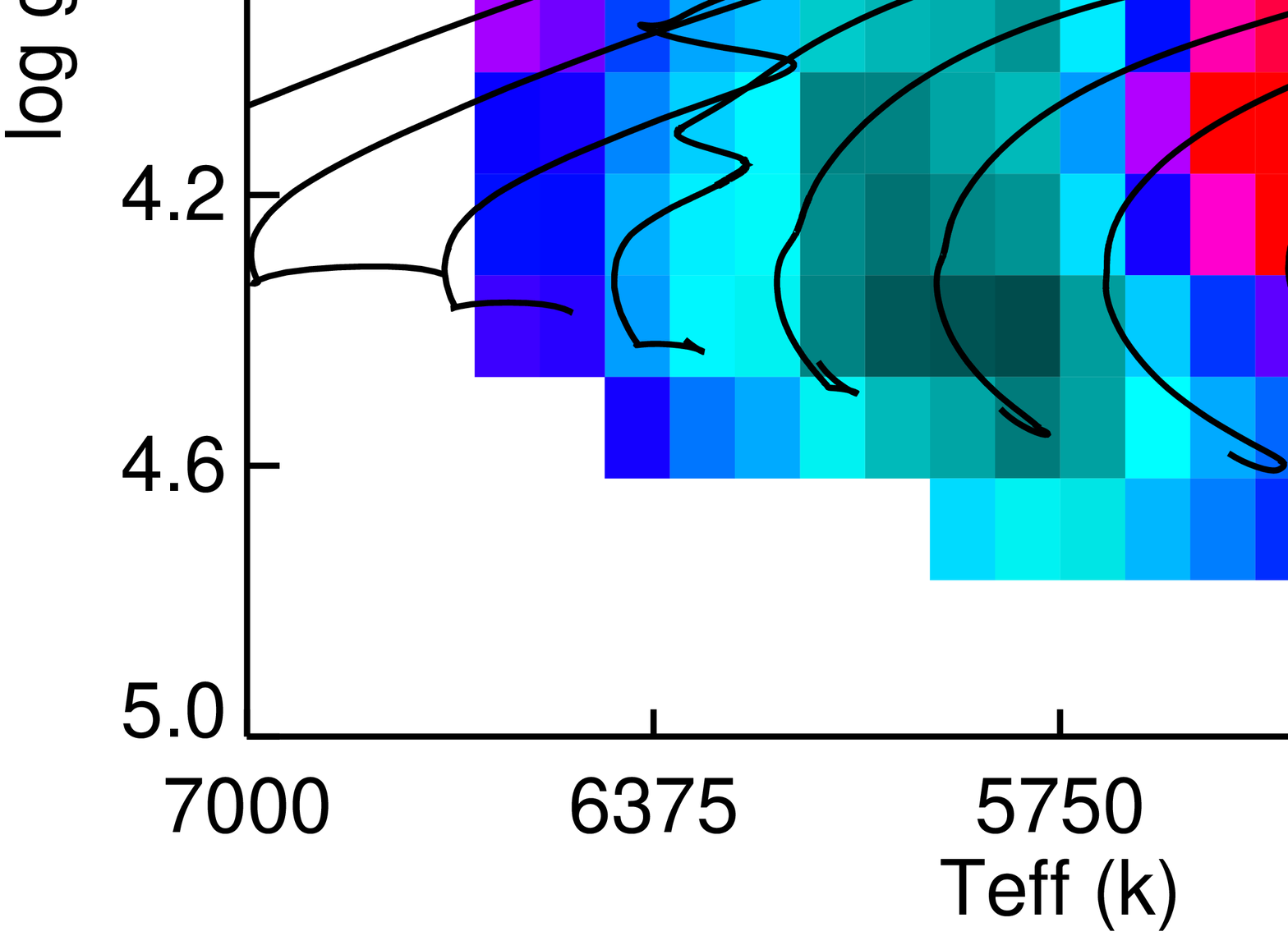}  
	\end{minipage}
}	
\vspace{1.em}
\caption{ Distributions of medium elemental abundances across the Kiel diagram of stars from GALAH DR3 with $\rm{[Fe/H]} > -0.2$. The PARSEC stellar evolution tracks with mass uniformly distributed from 0.7 to 1.4\,$M_\odot$ and Z=0.014 are overplotted.
\label{figd}}
\end{figure*}

\section{Conclusion} 
\label{section5}
We have built a sample of 8468 Ba-enhanced (${\rm [Ba/Fe]}>1.0$) dwarf and subgiant stars from LAMOST DR5, and investigated their chemical and kinematical properties to understand their origins. We found in general there are two mechanisms to account for the presence of these Ba-enhanced dwarf and subgiants, namely external accretion due to binary evolution, and internal elemental transport for singe star evolution. 

About one third of our Ba-enhanced stars exhibit enhanced C and N abundances and UV brightness excess, suggesting they have experienced external accretion event to obtain Ba-rich and C-rich materials from an AGB companion that has evolved to be a WD now. Their abundances for the other elements are, however, similar to those of the Ba-normal stars. These stars could have a broad range of effective temperatures, and many of them are cooler than 6000\,K.   

The remaining Ba-enhanced stars, mostly are metal-rich (${\rm [Fe/H]}>-0.2$), warm ($6000<T_{\rm eff}<6700$\,K) stars, exhibit C and N abundances similar to those of the Ba-normal stars. Their Ba enhancement are likely caused by internal elemental transports. Other than enhanced Ba, they also exhibit enhanced Ti, depleted Mg and Ni abundances. However, their abundance patterns are different to the hotter Am/Fm stars ($T_{\rm eff}>6700$\,K), which are also consequences of stellar internal elemental transports. 

The fraction of such Ba-enhanced stars are especially high for subgiants with intermediate mass, as more than 50\% of the subgiants of $\gtrsim1.5\,M_\odot$ and $T_{\rm eff}\gtrsim6500$\,K are Ba-enhanced stars. This confirms the previous suggestion that having peculiar surface abundances due to stellar internal elemental transport process is a ubiquitous phenomenon for stars with intermediate mass. Our data also revealed some intermediate-mass, warm Ba-enhanced stars with enhanced C and N abundances, which may have experienced both internal and external evolutionary processes for making them as chemical peculiar stars.

Furthermore, our results revealed a lack of Ba-enhanced stars with high [$\alpha$/Fe] above ${\rm [Fe/H]}\gtrsim-0.6$\,dex, which we deemed as a ``high-$\alpha$ desert" for the Ba-enhanced stars. We propose that it is caused by the low efficiency for producing Ba enhancement materials by low-mass AGB companions in this metallicity regime. We also found there are more  low-[$\alpha$/Fe] Ba-enhanced stars with thick disk kinematics in our sample.

Our results call for detailed modelling of stellar elemental transport processes, in both context of internal and external evolution. 

\vspace{2.em}
\noindent {\bf Acknowledgments}
We thank the referee for the suggestions that have improved the clarity of the manuscript.

This work has made use of data products from the Guo Shou Jing Telescope (the Large Sky Area Multi-Object Fibre Spectroscopic Telescope, LAMOST).
LAMOST is a National Major Scientific Project built by the Chinese Academy of Sciences.
Funding for the project has been provided by the National Development and Reform Commission. LAMOST is operated and managed by the National Astronomical Observatories, Chinese Academy of Sciences.

This work has made use of data from the European Space Agency (ESA) mission Gaia \footnote{https://www.cosmos.esa.int/gaia}, processed by the Gaia Data Processing and Analysis Consortium (DPAC\footnote{https://www.cosmos.esa.int/ web/gaia/dpac/consortium}). Funding for the DPAC has been provided by national institutions, in particular the institutions participating in the Gaia Multilateral Agreement.

This work was funded by the National Key R\&D Program of China (No. 2019YFA0405500) and the National Natural Science Foundation of China (NSFC Grant No.11973001, 11833006, 12090040, 12090044). M.X. acknowledges financial support from NSFC Grant No.2022000083 and the National Key R\&D Program of China No. 2022YFF0504200. Y.S.T. acknowledges financial support from the Australian Research Council through DECRA Fellowship DE220101520. Y.Q.W. acknowledges financial support from the NSFC Grant No.11903044. 
\appendix
\section{kinematically ``thick disk" Ba-enhanced stars}
\label{Appendix1}
In the text, we have found some Ba-enhanced kinematically defined ``thick disk" stars, which exhibit artificially too low [$\alpha$/Fe] values than expected. Here we verify the proportion of the kinematically ``thick disk" stars low [$\alpha$/Fe] stars for the Ba-enhanced stars is higher than that of the Ba-normal stars.

We consider the [Fe/H] window of $-0.6<$[Fe/H]$<-0.4$. In this window, there are 489 Ba-normal stars among 40,827 low-[$\alpha$/Fe] sample stars with $T_{\rm eff}\lesssim 6000$\,K exhibiting $500<$L$_{Z}<1200$\,km/s$\cdot$kpc.
This proportion is 1.20\%. This is the fraction of a low-velocity tail for the low-$\alpha$, Ba-normal stars. If we assume all the low-$\alpha$, Ba-normal stars should be (kinematic) thin disk stars, this low-velocity tail fraction can serve as an upper limit of the contamination rate for selecting kinematical thick disk stars. 

However, for the Ba-enhanced stars, we found 31 of them have $500 < L_{Z} < 1200$\,kpc$\cdot$km/s among 649 low-[$\alpha$/Fe] sample stars in the $-0.6 < $[Fe/H]$ <-0.4\,$dex window with $T_{\rm eff}\lesssim 6000$\,K. The proportion is 4.78\%, which is higher than that of the Ba-normal stars. This means larger fraction of the low [$\alpha$/Fe] Ba-enhanced stars have thick disk kinematics.
\section{Elemental abundance trends cross Keil diagram using APOGEE and GALAH}
\label{Appendix2}
We have found the median [Mg/Fe], [Ti/Fe], and [Ni/Fe] ratios may change with their stellar masses (Fig\,\ref{fig333}).
For these elemental abundance, the systematic errors in the measurements, especially the trends with effective temperature \citep[see Fig.\,14 of][]{Xiang2019} might have an impact on our results. To examine such effects, we verify that the trend of [Mg/Fe], [Ti/Fe], and [Ni/Fe] ratios with stellar mass using independent abundance measurements from both GALAH DR3 (\citealt{Bud2021}) and APOGEE DR17 (\citealt{Abd2022}).

Fig.\,\ref{figc} and Fig.\,\ref{figd} illustrate that the Mg, Ni and Ti abundance distributions from APOGEE DR17 and GALAH DR3, respectively, exhibit similar trends to those of stars from LAMOST. The more massive stars in the top-left regime have higher [Ti/Fe] ratios but lower [Mg/Fe] and [Ni/Fe] ratios than those of low-mass stars. Our conclusion in the context is robust.
\newpage
\bibliographystyle{aasjournal}
\bibliography{la_ba_paper.bib}

\begin{thebibliography}{}
\expandafter\ifx\csname natexlab\endcsname\relax\def\natexlab#1{#1}\fi
\providecommand{\url}[1]{\href{#1}{#1}}
\providecommand{\dodoi}[1]{doi:~\href{http://doi.org/#1}{\nolinkurl{#1}}}
\providecommand{\doeprint}[1]{\href{http://ascl.net/#1}{\nolinkurl{http://ascl.net/#1}}}
\providecommand{\doarXiv}[1]{\href{https://arxiv.org/abs/#1}{\nolinkurl{https://arxiv.org/abs/#1}}}

\bibitem[{{Abdurro'uf} {et~al.}(2022){Abdurro'uf}, {Accetta}, {Aerts}, {Silva
  Aguirre}, {Ahumada}, {Ajgaonkar}, {Filiz Ak}, {Alam}, {Allende Prieto},
  {Almeida}, {Anders}, {Anderson}, {Andrews}, {Anguiano}, {Aquino-Ort{\'\i}z},
  {Arag{\'o}n-Salamanca}, {Argudo-Fern{\'a}ndez}, {Ata}, {Aubert},
  {Avila-Reese}, {Badenes}, {Barb{\'a}}, {Barger}, {Barrera-Ballesteros},
  {Beaton}, {Beers}, {Belfiore}, {Bender}, {Bernardi}, {Bershady}, {Beutler},
  {Bidin}, {Bird}, {Bizyaev}, {Blanc}, {Blanton}, {Boardman}, {Bolton},
  {Boquien}, {Borissova}, {Bovy}, {Brandt}, {Brown}, {Brownstein}, {Brusa},
  {Buchner}, {Bundy}, {Burchett}, {Bureau}, {Burgasser}, {Cabang}, {Campbell},
  {Cappellari}, {Carlberg}, {Wanderley}, {Carrera}, {Cash}, {Chen}, {Chen},
  {Cherinka}, {Chiappini}, {Choi}, {Chojnowski}, {Chung}, {Clerc}, {Cohen},
  {Comerford}, {Comparat}, {da Costa}, {Covey}, {Crane}, {Cruz-Gonzalez},
  {Culhane}, {Cunha}, {Dai}, {Damke}, {Darling}, {Davidson}, {Davies},
  {Dawson}, {De Lee}, {Diamond-Stanic}, {Cano-D{\'\i}az}, {S{\'a}nchez},
  {Donor}, {Duckworth}, {Dwelly}, {Eisenstein}, {Elsworth}, {Emsellem},
  {Eracleous}, {Escoffier}, {Fan}, {Farr}, {Feng}, {Fern{\'a}ndez-Trincado},
  {Feuillet}, {Filipp}, {Fillingham}, {Frinchaboy}, {Fromenteau}, {Galbany},
  {Garc{\'\i}a}, {Garc{\'\i}a-Hern{\'a}ndez}, {Ge}, {Geisler}, {Gelfand},
  {G{\'e}ron}, {Gibson}, {Goddy}, {Godoy-Rivera}, {Grabowski}, {Green},
  {Greener}, {Grier}, {Griffith}, {Guo}, {Guy}, {Hadjara}, {Harding},
  {Hasselquist}, {Hayes}, {Hearty}, {Hern{\'a}ndez}, {Hill}, {Hogg},
  {Holtzman}, {Horta}, {Hsieh}, {Hsu}, {Hsu}, {Huber}, {Huertas-Company},
  {Hutchinson}, {Hwang}, {Ibarra-Medel}, {Chitham}, {Ilha}, {Imig}, {Jaekle},
  {Jayasinghe}, {Ji}, {Johnson}, {Jones}, {J{\"o}nsson}, {Katkov}, {Khalatyan},
  {Kinemuchi}, {Kisku}, {Knapen}, {Kneib}, {Kollmeier}, {Kong}, {Kounkel},
  {Kreckel}, {Krishnarao}, {Lacerna}, {Lane}, {Langgin}, {Lavender}, {Law},
  {Lazarz}, {Leung}, {Leung}, {Lewis}, {Li}, {Li}, {Lian}, {Liang}, {Lin},
  {Lin}, {Lin}, {Lintott}, {Long}, {Longa-Pe{\~n}a}, {L{\'o}pez-Cob{\'a}},
  {Lu}, {Lundgren}, {Luo}, {Mackereth}, {de la Macorra}, {Mahadevan},
  {Majewski}, {Manchado}, {Mandeville}, {Maraston}, {Margalef-Bentabol},
  {Masseron}, {Masters}, {Mathur}, {McDermid}, {Mckay}, {Merloni},
  {Merrifield}, {Meszaros}, {Miglio}, {Di Mille}, {Minniti}, {Minsley},
  {Monachesi}, {Moon}, {Mosser}, {Mulchaey}, {Muna}, {Mu{\~n}oz}, {Myers},
  {Myers}, {Nadathur}, {Nair}, {Nandra}, {Neumann}, {Newman}, {Nidever},
  {Nikakhtar}, {Nitschelm}, {O'Connell}, {Garma-Oehmichen}, {Luan Souza de
  Oliveira}, {Olney}, {Oravetz}, {Ortigoza-Urdaneta}, {Osorio}, {Otter},
  {Pace}, {Padilla}, {Pan}, {Pan}, {Parikh}, {Parker}, {Peirani}, {Pe{\~n}a
  Ram{\'\i}rez}, {Penny}, {Percival}, {Perez-Fournon}, {Pinsonneault},
  {Poidevin}, {Poovelil}, {Price-Whelan}, {B{\'a}rbara de Andrade Queiroz},
  {Raddick}, {Ray}, {Rembold}, {Riddle}, {Riffel}, {Riffel}, {Rix}, {Robin},
  {Rodr{\'\i}guez-Puebla}, {Roman-Lopes}, {Rom{\'a}n-Z{\'u}{\~n}iga}, {Rose},
  {Ross}, {Rossi}, {Rubin}, {Salvato}, {S{\'a}nchez}, {S{\'a}nchez-Gallego},
  {Sanderson}, {Santana Rojas}, {Sarceno}, {Sarmiento}, {Sayres}, {Sazonova},
  {Schaefer}, {Schiavon}, {Schlegel}, {Schneider}, {Schultheis}, {Schwope},
  {Serenelli}, {Serna}, {Shao}, {Shapiro}, {Sharma}, {Shen}, {Shetrone}, {Shu},
  {Simon}, {Skrutskie}, {Smethurst}, {Smith}, {Sobeck}, {Spoo}, {Sprague},
  {Stark}, {Stassun}, {Steinmetz}, {Stello}, {Stone-Martinez},
  {Storchi-Bergmann}, {Stringfellow}, {Stutz}, {Su}, {Taghizadeh-Popp},
  {Talbot}, {Tayar}, {Telles}, {Teske}, {Thakar}, {Theissen}, {Tkachenko},
  {Thomas}, {Tojeiro}, {Hernandez Toledo}, {Troup}, {Trump}, {Trussler},
  {Turner}, {Tuttle}, {Unda-Sanzana}, {V{\'a}zquez-Mata}, {Valentini},
  {Valenzuela}, {Vargas-Gonz{\'a}lez}, {Vargas-Maga{\~n}a}, {Alfaro},
  {Villanova}, {Vincenzo}, {Wake}, {Warfield}, {Washington}, {Weaver},
  {Weijmans}, {Weinberg}, {Weiss}, {Westfall}, {Wild}, {Wilde}, {Wilson},
  {Wilson}, {Wilson}, {Wolf}, {Wood-Vasey}, {Yan}, {Zamora}, {Zasowski},
  {Zhang}, {Zhao}, {Zheng}, {Zheng}, \& {Zhu}}]{Abd2022}
{Abdurro'uf}, {Accetta}, K., {Aerts}, C., {et~al.} 2022, \apjs, 259, 35,
  \dodoi{10.3847/1538-4365/ac4414}

\bibitem[{{Allen} \& {Barbuy}(2006)}]{All2006}
{Allen}, D.~M., \& {Barbuy}, B. 2006, \aap, 454, 895,
  \dodoi{10.1051/0004-6361:20064912}

\bibitem[{{Bailer-Jones} {et~al.}(2021){Bailer-Jones}, {Rybizki}, {Fouesneau},
  {Demleitner}, \& {Andrae}}]{Ba2021}
{Bailer-Jones}, C.~A.~L., {Rybizki}, J., {Fouesneau}, M., {Demleitner}, M., \&
  {Andrae}, R. 2021, \aj, 161, 147, \dodoi{10.3847/1538-3881/abd806}

\bibitem[{{Bensby} {et~al.}(2003){Bensby}, {Feltzing}, \&
  {Lundstr{\"o}m}}]{Ben2003}
{Bensby}, T., {Feltzing}, S., \& {Lundstr{\"o}m}, I. 2003, \aap, 410, 527,
  \dodoi{10.1051/0004-6361:20031213}

\bibitem[{{Bianchi} {et~al.}(2011){Bianchi}, {Herald}, {Efremova}, {Girardi},
  {Zabot}, {Marigo}, {Conti}, \& {Shiao}}]{Bia2011}
{Bianchi}, L., {Herald}, J., {Efremova}, B., {et~al.} 2011, \apss, 335, 161,
  \dodoi{10.1007/s10509-010-0581-x}

\bibitem[{{Bidelman} \& {Keenan}(1951)}]{Bid1951}
{Bidelman}, W.~P., \& {Keenan}, P.~C. 1951, \apj, 114, 473,
  \dodoi{10.1086/145488}

\bibitem[{{Boffin} \& {Jorissen}(1988)}]{Bof1988}
{Boffin}, H.~M.~J., \& {Jorissen}, A. 1988, \aap, 205, 155

\bibitem[{{B{\"o}hm-Vitense} {et~al.}(2000){B{\"o}hm-Vitense}, {Carpenter},
  {Robinson}, {Ake}, \& {Brown}}]{Boh2000}
{B{\"o}hm-Vitense}, E., {Carpenter}, K., {Robinson}, R., {Ake}, T., \& {Brown},
  J. 2000, \apj, 533, 969, \dodoi{10.1086/308678}

\bibitem[{{Bohm-Vitense} {et~al.}(1984){Bohm-Vitense}, {Proffitt}, \&
  {Johnson}}]{Boh1984}
{Bohm-Vitense}, E., {Proffitt}, C., \& {Johnson}, H. 1984, in NASA Conference
  Publication, Vol. 2349, NASA Conference Publication, ed. J.~M. {Mead}, R.~D.
  {Chapman}, \& Y.~{Kondo}, 293--296

\bibitem[{{Bond}(1974)}]{Bond1974}
{Bond}, H.~E. 1974, \apj, 194, 95, \dodoi{10.1086/153227}

\bibitem[{{Boothroyd} {et~al.}(1993){Boothroyd}, {Sackmann}, \&
  {Ahern}}]{Boo1993}
{Boothroyd}, A.~I., {Sackmann}, I.~J., \& {Ahern}, S.~C. 1993, \apj, 416, 762,
  \dodoi{10.1086/173275}

\bibitem[{{Bovy} {et~al.}(2015){Bovy}, {Bird}, {Garc{\'\i}a P{\'e}rez},
  {Majewski}, {Nidever}, \& {Zasowski}}]{Bo2015}
{Bovy}, J., {Bird}, J.~C., {Garc{\'\i}a P{\'e}rez}, A.~E., {et~al.} 2015, \apj,
  800, 83, \dodoi{10.1088/0004-637X/800/2/83}

\bibitem[{{Bressan} {et~al.}(2012){Bressan}, {Marigo}, {Girardi}, {Salasnich},
  {Dal Cero}, {Rubele}, \& {Nanni}}]{Bre2012}
{Bressan}, A., {Marigo}, P., {Girardi}, L., {et~al.} 2012, \mnras, 427, 127,
  \dodoi{10.1111/j.1365-2966.2012.21948.x}

\bibitem[{{Buder} {et~al.}(2021){Buder}, {Sharma}, {Kos}, {Amarsi},
  {Nordlander}, {Lind}, {Martell}, {Asplund}, {Bland-Hawthorn}, {Casey}, {de
  Silva}, {D'Orazi}, {Freeman}, {Hayden}, {Lewis}, {Lin}, {Schlesinger},
  {Simpson}, {Stello}, {Zucker}, {Zwitter}, {Beeson}, {Buck}, {Casagrande},
  {Clark}, {{\v{C}}otar}, {da Costa}, {de Grijs}, {Feuillet}, {Horner},
  {Kafle}, {Khanna}, {Kobayashi}, {Liu}, {Montet}, {Nandakumar}, {Nataf},
  {Ness}, {Spina}, {Tepper-Garc{\'\i}a}, {Ting}, {Traven},
  {Vogrin{\v{c}}i{\v{c}}}, {Wittenmyer}, {Wyse}, {{\v{Z}}erjal},
  {{\v{Z}}erjal}, \& {Galah Collaboration}}]{Bud2021}
{Buder}, S., {Sharma}, S., {Kos}, J., {et~al.} 2021, \mnras, 506, 150,
  \dodoi{10.1093/mnras/stab1242}

\bibitem[{{Busso} {et~al.}(1999){Busso}, {Gallino}, \& {Wasserburg}}]{Bus1999}
{Busso}, M., {Gallino}, R., \& {Wasserburg}, G.~J. 1999, \araa, 37, 239,
  \dodoi{10.1146/annurev.astro.37.1.239}

\bibitem[{{Charbonnel} {et~al.}(2021){Charbonnel}, {Borisov}, {de Laverny}, \&
  {Prantzos}}]{Cha2021}
{Charbonnel}, C., {Borisov}, S., {de Laverny}, P., \& {Prantzos}, N. 2021,
  \aap, 649, L10, \dodoi{10.1051/0004-6361/202140873}

\bibitem[{{Chen} {et~al.}(2019){Chen}, {Girardi}, {Fu}, {Bressan}, {Aringer},
  {Dal Tio}, {Pastorelli}, {Marigo}, {Costa}, \& {Zhang}}]{Chen2019}
{Chen}, Y., {Girardi}, L., {Fu}, X., {et~al.} 2019, \aap, 632, A105,
  \dodoi{10.1051/0004-6361/201936612}

\bibitem[{{Cristallo} {et~al.}(2009){Cristallo}, {Straniero}, {Gallino},
  {Piersanti}, {Dom{\'\i}nguez}, \& {Lederer}}]{Cri2009}
{Cristallo}, S., {Straniero}, O., {Gallino}, R., {et~al.} 2009, \apj, 696, 797,
  \dodoi{10.1088/0004-637X/696/1/797}

\bibitem[{{Cristallo} {et~al.}(2015){Cristallo}, {Straniero}, {Piersanti}, \&
  {Gobrecht}}]{Cri2015}
{Cristallo}, S., {Straniero}, O., {Piersanti}, L., \& {Gobrecht}, D. 2015,
  \apjs, 219, 40, \dodoi{10.1088/0067-0049/219/2/40}

\bibitem[{{Cristallo} {et~al.}(2011){Cristallo}, {Piersanti}, {Straniero},
  {Gallino}, {Dom{\'\i}nguez}, {Abia}, {Di Rico}, {Quintini}, \&
  {Bisterzo}}]{Cri2011}
{Cristallo}, S., {Piersanti}, L., {Straniero}, O., {et~al.} 2011, \apjs, 197,
  17, \dodoi{10.1088/0067-0049/197/2/17}

\bibitem[{{de Castro} {et~al.}(2016){de Castro}, {Pereira}, {Roig}, {Jilinski},
  {Drake}, {Chavero}, \& {Sales Silva}}]{Dec2016}
{de Castro}, D.~B., {Pereira}, C.~B., {Roig}, F., {et~al.} 2016, \mnras, 459,
  4299, \dodoi{10.1093/mnras/stw815}

\bibitem[{{Deal} {et~al.}(2020){Deal}, {Goupil}, {Marques}, {Reese}, \&
  {Lebreton}}]{Dea2020}
{Deal}, M., {Goupil}, M.~J., {Marques}, J.~P., {Reese}, D.~R., \& {Lebreton},
  Y. 2020, \aap, 633, A23, \dodoi{10.1051/0004-6361/201936666}

\bibitem[{{Deng} {et~al.}(2012){Deng}, {Newberg}, {Liu}, {Carlin}, {Beers},
  {Chen}, {Chen}, {Christlieb}, {Grillmair}, {Guhathakurta}, {Han}, {Hou},
  {Lee}, {L{\'e}pine}, {Li}, {Liu}, {Pan}, {Sellwood}, {Wang}, {Wang}, {Yang},
  {Yanny}, {Zhang}, {Zhang}, {Zheng}, \& {Zhu}}]{Deng2012}
{Deng}, L.-C., {Newberg}, H.~J., {Liu}, C., {et~al.} 2012, Research in
  Astronomy and Astrophysics, 12, 735, \dodoi{10.1088/1674-4527/12/7/003}

\bibitem[{{Duch{\^e}ne} \& {Kraus}(2013)}]{Duchene2013}
{Duch{\^e}ne}, G., \& {Kraus}, A. 2013, \araa, 51, 269,
  \dodoi{10.1146/annurev-astro-081710-102602}

\bibitem[{{Escorza} {et~al.}(2019){Escorza}, {Karinkuzhi}, {Jorissen}, {Siess},
  {Van Winckel}, {Pourbaix}, {Johnston}, {Miszalski}, {Oomen}, {Abdul-Masih},
  {Boffin}, {North}, {Manick}, {Shetye}, \& {Miko{\l}ajewska}}]{Esc2019}
{Escorza}, A., {Karinkuzhi}, D., {Jorissen}, A., {et~al.} 2019, \aap, 626,
  A128, \dodoi{10.1051/0004-6361/201935390}

\bibitem[{{Fitzpatrick}(1999)}]{Fit1999}
{Fitzpatrick}, E.~L. 1999, \pasp, 111, 63, \dodoi{10.1086/316293}

\bibitem[{{Gaia Collaboration} {et~al.}(2021){Gaia Collaboration}, {Brown},
  {Vallenari}, {Prusti}, {de Bruijne}, {Babusiaux}, {Biermann}, {Creevey},
  {Evans}, {Eyer}, {Hutton}, {Jansen}, {Jordi}, {Klioner}, {Lammers},
  {Lindegren}, {Luri}, {Mignard}, {Panem}, {Pourbaix}, {Randich}, {Sartoretti},
  {Soubiran}, {Walton}, {Arenou}, {Bailer-Jones}, {Bastian}, {Cropper},
  {Drimmel}, {Katz}, {Lattanzi}, {van Leeuwen}, {Bakker}, {Cacciari},
  {Casta{\~n}eda}, {De Angeli}, {Ducourant}, {Fabricius}, {Fouesneau},
  {Fr{\'e}mat}, {Guerra}, {Guerrier}, {Guiraud}, {Jean-Antoine Piccolo},
  {Masana}, {Messineo}, {Mowlavi}, {Nicolas}, {Nienartowicz}, {Pailler},
  {Panuzzo}, {Riclet}, {Roux}, {Seabroke}, {Sordo}, {Tanga}, {Th{\'e}venin},
  {Gracia-Abril}, {Portell}, {Teyssier}, {Altmann}, {Andrae}, {Bellas-Velidis},
  {Benson}, {Berthier}, {Blomme}, {Brugaletta}, {Burgess}, {Busso}, {Carry},
  {Cellino}, {Cheek}, {Clementini}, {Damerdji}, {Davidson}, {Delchambre},
  {Dell'Oro}, {Fern{\'a}ndez-Hern{\'a}ndez}, {Galluccio}, {Garc{\'\i}a-Lario},
  {Garcia-Reinaldos}, {Gonz{\'a}lez-N{\'u}{\~n}ez}, {Gosset}, {Haigron},
  {Halbwachs}, {Hambly}, {Harrison}, {Hatzidimitriou}, {Heiter},
  {Hern{\'a}ndez}, {Hestroffer}, {Hodgkin}, {Holl}, {Jan{\ss}en}, {Jevardat de
  Fombelle}, {Jordan}, {Krone-Martins}, {Lanzafame}, {L{\"o}ffler}, {Lorca},
  {Manteiga}, {Marchal}, {Marrese}, {Moitinho}, {Mora}, {Muinonen}, {Osborne},
  {Pancino}, {Pauwels}, {Petit}, {Recio-Blanco}, {Richards}, {Riello},
  {Rimoldini}, {Robin}, {Roegiers}, {Rybizki}, {Sarro}, {Siopis}, {Smith},
  {Sozzetti}, {Ulla}, {Utrilla}, {van Leeuwen}, {van Reeven}, {Abbas}, {Abreu
  Aramburu}, {Accart}, {Aerts}, {Aguado}, {Ajaj}, {Altavilla}, {{\'A}lvarez},
  {{\'A}lvarez Cid-Fuentes}, {Alves}, {Anderson}, {Anglada Varela}, {Antoja},
  {Audard}, {Baines}, {Baker}, {Balaguer-N{\'u}{\~n}ez}, {Balbinot}, {Balog},
  {Barache}, {Barbato}, {Barros}, {Barstow}, {Bartolom{\'e}}, {Bassilana},
  {Bauchet}, {Baudesson-Stella}, {Becciani}, {Bellazzini}, {Bernet}, {Bertone},
  {Bianchi}, {Blanco-Cuaresma}, {Boch}, {Bombrun}, {Bossini}, {Bouquillon},
  {Bragaglia}, {Bramante}, {Breedt}, {Bressan}, {Brouillet}, {Bucciarelli},
  {Burlacu}, {Busonero}, {Butkevich}, {Buzzi}, {Caffau}, {Cancelliere},
  {C{\'a}novas}, {Cantat-Gaudin}, {Carballo}, {Carlucci}, {Carnerero},
  {Carrasco}, {Casamiquela}, {Castellani}, {Castro-Ginard}, {Castro Sampol},
  {Chaoul}, {Charlot}, {Chemin}, {Chiavassa}, {Cioni}, {Comoretto}, {Cooper},
  {Cornez}, {Cowell}, {Crifo}, {Crosta}, {Crowley}, {Dafonte}, {Dapergolas},
  {David}, {David}, {de Laverny}, {De Luise}, {De March}, {De Ridder}, {de
  Souza}, {de Teodoro}, {de Torres}, {del Peloso}, {del Pozo}, {Delbo},
  {Delgado}, {Delgado}, {Delisle}, {Di Matteo}, {Diakite}, {Diener},
  {Distefano}, {Dolding}, {Eappachen}, {Edvardsson}, {Enke}, {Esquej}, {Fabre},
  {Fabrizio}, {Faigler}, {Fedorets}, {Fernique}, {Fienga}, {Figueras},
  {Fouron}, {Fragkoudi}, {Fraile}, {Franke}, {Gai}, {Garabato},
  {Garcia-Gutierrez}, {Garc{\'\i}a-Torres}, {Garofalo}, {Gavras}, {Gerlach},
  {Geyer}, {Giacobbe}, {Gilmore}, {Girona}, {Giuffrida}, {Gomel}, {Gomez},
  {Gonzalez-Santamaria}, {Gonz{\'a}lez-Vidal}, {Granvik},
  {Guti{\'e}rrez-S{\'a}nchez}, {Guy}, {Hauser}, {Haywood}, {Helmi}, {Hidalgo},
  {Hilger}, {H{\l}adczuk}, {Hobbs}, {Holland}, {Huckle}, {Jasniewicz},
  {Jonker}, {Juaristi Campillo}, {Julbe}, {Karbevska}, {Kervella}, {Khanna},
  {Kochoska}, {Kontizas}, {Kordopatis}, {Korn}, {Kostrzewa-Rutkowska},
  {Kruszy{\'n}ska}, {Lambert}, {Lanza}, {Lasne}, {Le Campion}, {Le Fustec},
  {Lebreton}, {Lebzelter}, {Leccia}, {Leclerc}, {Lecoeur-Taibi}, {Liao},
  {Licata}, {Lindstr{\o}m}, {Lister}, {Livanou}, {Lobel}, {Madrero Pardo},
  {Managau}, {Mann}, {Marchant}, {Marconi}, {Marcos Santos}, {Marinoni},
  {Marocco}, {Marshall}, {Martin Polo}, {Mart{\'\i}n-Fleitas}, {Masip},
  {Massari}, {Mastrobuono-Battisti}, {Mazeh}, {McMillan}, {Messina},
  {Michalik}, {Millar}, {Mints}, {Molina}, {Molinaro}, {Moln{\'a}r},
  {Montegriffo}, {Mor}, {Morbidelli}, {Morel}, {Morris}, {Mulone}, {Munoz},
  {Muraveva}, {Murphy}, {Musella}, {Noval}, {Ord{\'e}novic}, {Orr{\`u}},
  {Osinde}, {Pagani}, {Pagano}, {Palaversa}, {Palicio}, {Panahi}, {Pawlak},
  {Pe{\~n}alosa Esteller}, {Penttil{\"a}}, {Piersimoni}, {Pineau}, {Plachy},
  {Plum}, {Poggio}, {Poretti}, {Poujoulet}, {Pr{\v{s}}a}, {Pulone}, {Racero},
  {Ragaini}, {Rainer}, {Raiteri}, {Rambaux}, {Ramos}, {Ramos-Lerate}, {Re
  Fiorentin}, {Regibo}, {Reyl{\'e}}, {Ripepi}, {Riva}, {Rixon}, {Robichon},
  {Robin}, {Roelens}, {Rohrbasser}, {Romero-G{\'o}mez}, {Rowell}, {Royer},
  {Rybicki}, {Sadowski}, {Sagrist{\`a} Sell{\'e}s}, {Sahlmann}, {Salgado},
  {Salguero}, {Samaras}, {Sanchez Gimenez}, {Sanna}, {Santove{\~n}a},
  {Sarasso}, {Schultheis}, {Sciacca}, {Segol}, {Segovia}, {S{\'e}gransan},
  {Semeux}, {Shahaf}, {Siddiqui}, {Siebert}, {Siltala}, {Slezak}, {Smart},
  {Solano}, {Solitro}, {Souami}, {Souchay}, {Spagna}, {Spoto}, {Steele},
  {Steidelm{\"u}ller}, {Stephenson}, {S{\"u}veges}, {Szabados}, {Szegedi-Elek},
  {Taris}, {Tauran}, {Taylor}, {Teixeira}, {Thuillot}, {Tonello}, {Torra},
  {Torra}, {Turon}, {Unger}, {Vaillant}, {van Dillen}, {Vanel}, {Vecchiato},
  {Viala}, {Vicente}, {Voutsinas}, {Weiler}, {Wevers}, {Wyrzykowski}, {Yoldas},
  {Yvard}, {Zhao}, {Zorec}, {Zucker}, {Zurbach}, \& {Zwitter}}]{Gaia2021}
{Gaia Collaboration}, {Brown}, A.~G.~A., {Vallenari}, A., {et~al.} 2021, \aap,
  649, A1, \dodoi{10.1051/0004-6361/202039657}

\bibitem[{{Gao} {et~al.}(2014){Gao}, {Liu}, {Zhang}, {Justham}, {Deng}, \&
  {Yang}}]{Gao2014}
{Gao}, S., {Liu}, C., {Zhang}, X., {et~al.} 2014, \apjl, 788, L37,
  \dodoi{10.1088/2041-8205/788/2/L37}

\bibitem[{{Gentile Fusillo} {et~al.}(2021){Gentile Fusillo}, {Tremblay},
  {Cukanovaite}, {Vorontseva}, {Lallement}, {Hollands}, {G{\"a}nsicke},
  {Burdge}, {McCleery}, \& {Jordan}}]{Gen2021}
{Gentile Fusillo}, N.~P., {Tremblay}, P.~E., {Cukanovaite}, E., {et~al.} 2021,
  \mnras, 508, 3877, \dodoi{10.1093/mnras/stab2672}

\bibitem[{{Ghazaryan} {et~al.}(2018){Ghazaryan}, {Alecian}, \&
  {Hakobyan}}]{Gha2018}
{Ghazaryan}, S., {Alecian}, G., \& {Hakobyan}, A.~A. 2018, \mnras, 480, 2953,
  \dodoi{10.1093/mnras/sty1912}

\bibitem[{{Gray} {et~al.}(2011){Gray}, {McGahee}, {Griffin}, \&
  {Corbally}}]{Gra2011}
{Gray}, R.~O., {McGahee}, C.~E., {Griffin}, R.~E.~M., \& {Corbally}, C.~J.
  2011, \aj, 141, 160, \dodoi{10.1088/0004-6256/141/5/160}

\bibitem[{{Griffin} \& {Griffin}(1980)}]{Gri1980}
{Griffin}, R., \& {Griffin}, R. 1980, \mnras, 193, 957,
  \dodoi{10.1093/mnras/193.4.957}

\bibitem[{{Griffin} \& {Herbig}(1981)}]{Gri1981}
{Griffin}, R.~F., \& {Herbig}, G.~H. 1981, \mnras, 196, 33,
  \dodoi{10.1093/mnras/196.1.33}

\bibitem[{{Guo} \& {Li}(2021)}]{Guo2021}
{Guo}, F., \& {Li}, Y. 2021, \apj, 910, 34, \dodoi{10.3847/1538-4357/abe1c5}

\bibitem[{{Han} {et~al.}(1995){Han}, {Eggleton}, {Podsiadlowski}, \&
  {Tout}}]{Han1995}
{Han}, Z., {Eggleton}, P.~P., {Podsiadlowski}, P., \& {Tout}, C.~A. 1995,
  \mnras, 277, 1443, \dodoi{10.1093/mnras/277.4.1443}

\bibitem[{{Hayden} {et~al.}(2015){Hayden}, {Bovy}, {Holtzman}, {Nidever},
  {Bird}, {Weinberg}, {Andrews}, {Majewski}, {Allende Prieto}, {Anders},
  {Beers}, {Bizyaev}, {Chiappini}, {Cunha}, {Frinchaboy},
  {Garc{\'\i}a-Her{\'n}and ez}, {Garc{\'\i}a P{\'e}rez}, {Girardi}, {Harding},
  {Hearty}, {Johnson}, {M{\'e}sz{\'a}ros}, {Minchev}, {O'Connell}, {Pan},
  {Robin}, {Schiavon}, {Schneider}, {Schultheis}, {Shetrone}, {Skrutskie},
  {Steinmetz}, {Smith}, {Wilson}, {Zamora}, \& {Zasowski}}]{Hay2015b}
{Hayden}, M.~R., {Bovy}, J., {Holtzman}, J.~A., {et~al.} 2015, \apj, 808, 132,
  \dodoi{10.1088/0004-637X/808/2/132}

\bibitem[{{Haywood} {et~al.}(2013){Haywood}, {Di Matteo}, {Lehnert}, {Katz}, \&
  {G{\'o}mez}}]{Hay2013}
{Haywood}, M., {Di Matteo}, P., {Lehnert}, M.~D., {Katz}, D., \& {G{\'o}mez},
  A. 2013, \aap, 560, A109, \dodoi{10.1051/0004-6361/201321397}

\bibitem[{{Herwig}(2005)}]{Her2005}
{Herwig}, F. 2005, \araa, 43, 435,
  \dodoi{10.1146/annurev.astro.43.072103.150600}

\bibitem[{{Huang}(2015)}]{Huang2015b}
{Huang}, Y. 2015, in IAU General Assembly, Vol.~29, 2251839

\bibitem[{{Huang} {et~al.}(2020){Huang}, {Sch{\"o}nrich}, {Zhang}, {Wu},
  {Chen}, {Wang}, {Xiang}, {Wang}, {Yuan}, {Li}, {Sun}, {Li}, \&
  {Liu}}]{Huang2020}
{Huang}, Y., {Sch{\"o}nrich}, R., {Zhang}, H., {et~al.} 2020, \apjs, 249, 29,
  \dodoi{10.3847/1538-4365/ab994f}

\bibitem[{{Husti} {et~al.}(2009){Husti}, {Gallino}, {Bisterzo}, {Straniero}, \&
  {Cristallo}}]{Hus2009}
{Husti}, L., {Gallino}, R., {Bisterzo}, S., {Straniero}, O., \& {Cristallo}, S.
  2009, \pasa, 26, 176, \dodoi{10.1071/AS08065}

\bibitem[{{Jorissen} {et~al.}(2019){Jorissen}, {Boffin}, {Karinkuzhi}, {Van
  Eck}, {Escorza}, {Shetye}, \& {Van Winckel}}]{Jor2019}
{Jorissen}, A., {Boffin}, H.~M.~J., {Karinkuzhi}, D., {et~al.} 2019, \aap, 626,
  A127, \dodoi{10.1051/0004-6361/201834630}

\bibitem[{{Jorissen} {et~al.}(1992){Jorissen}, {Manfroid}, \&
  {Sterken}}]{Jor1992}
{Jorissen}, A., {Manfroid}, J., \& {Sterken}, C. 1992, \aap, 253, 407

\bibitem[{{Jorissen} \& {Mayor}(1988)}]{Jor1988}
{Jorissen}, A., \& {Mayor}, M. 1988, \aap, 198, 187

\bibitem[{{Jorissen} {et~al.}(1998){Jorissen}, {Van Eck}, {Mayor}, \&
  {Udry}}]{Jor1998}
{Jorissen}, A., {Van Eck}, S., {Mayor}, M., \& {Udry}, S. 1998, \aap, 332, 877.
\newblock \doarXiv{astro-ph/9801272}

\bibitem[{{K{\"a}ppeler} {et~al.}(2011){K{\"a}ppeler}, {Gallino}, {Bisterzo},
  \& {Aoki}}]{Kap2011}
{K{\"a}ppeler}, F., {Gallino}, R., {Bisterzo}, S., \& {Aoki}, W. 2011, Reviews
  of Modern Physics, 83, 157, \dodoi{10.1103/RevModPhys.83.157}

\bibitem[{{Lee} {et~al.}(2011){Lee}, {Beers}, {An}, {Ivezi{\'c}}, {Just},
  {Rockosi}, {Morrison}, {Johnson}, {Sch{\"o}nrich}, {Bird}, {Yanny},
  {Harding}, \& {Rocha-Pinto}}]{Lee2011}
{Lee}, Y.~S., {Beers}, T.~C., {An}, D., {et~al.} 2011, \apj, 738, 187,
  \dodoi{10.1088/0004-637X/738/2/187}

\bibitem[{{Liu} {et~al.}(2020){Liu}, {Fu}, {Shi}, {Wu}, {Han}, {Chen}, {Dong},
  {Zhao}, {Chen}, {Zhang}, {Bai}, {Chen}, {Cui}, {Du}, {Hsia}, {Jiang}, {Hou},
  {Hou}, {Li}, {Li}, {Li}, {Liu}, {Liu}, {Luo}, {Ren}, {Tian}, {Tian}, {Wang},
  {Wu}, {Xie}, {Yan}, {Yang}, {Yu}, {Zhang}, {Zhang}, {Zhang}, {Zhang}, {Zhao},
  {Zhong}, {Zong}, \& {Zuo}}]{Liu2020}
{Liu}, C., {Fu}, J., {Shi}, J., {et~al.} 2020, arXiv e-prints,
  arXiv:2005.07210.
\newblock \doarXiv{2005.07210}

\bibitem[{{Liu} {et~al.}(2014){Liu}, {Yuan}, {Huo}, {Deng}, {Hou}, {Zhao},
  {Zhao}, {Shi}, {Luo}, {Xiang}, {Zhang}, {Huang}, \& {Zhang}}]{Liu2014}
{Liu}, X.~W., {Yuan}, H.~B., {Huo}, Z.~Y., {et~al.} 2014, in IAU Symposium,
  Vol. 298, Setting the scene for Gaia and LAMOST, ed. S.~{Feltzing},
  G.~{Zhao}, N.~A. {Walton}, \& P.~{Whitelock}, 310--321,
  \dodoi{10.1017/S1743921313006510}

\bibitem[{{MacConnell} {et~al.}(1972){MacConnell}, {Frye}, \&
  {Upgren}}]{Mac1972}
{MacConnell}, D.~J., {Frye}, R.~L., \& {Upgren}, A.~R. 1972, \aj, 77, 384,
  \dodoi{10.1086/111298}

\bibitem[{{McClure}(1983)}]{McC1983}
{McClure}, R.~D. 1983, \apj, 268, 264, \dodoi{10.1086/160951}

\bibitem[{{McClure} \& {Woodsworth}(1990)}]{McC1990}
{McClure}, R.~D., \& {Woodsworth}, A.~W. 1990, \apj, 352, 709,
  \dodoi{10.1086/168573}

\bibitem[{{Michaud}(1970)}]{Mic1970}
{Michaud}, G. 1970, \apj, 160, 641, \dodoi{10.1086/150459}

\bibitem[{{Michaud}(1982)}]{Mic1982}
---. 1982, \apj, 258, 349, \dodoi{10.1086/160083}

\bibitem[{{Michaud} {et~al.}(2015){Michaud}, {Alecian}, \& {Richer}}]{Mic2015}
{Michaud}, G., {Alecian}, G., \& {Richer}, J. 2015, {Atomic Diffusion in
  Stars}, \dodoi{10.1007/978-3-319-19854-5}

\bibitem[{{Michaud} {et~al.}(2011){Michaud}, {Richer}, \& {Vick}}]{Mic2011}
{Michaud}, G., {Richer}, J., \& {Vick}, M. 2011, \aap, 534, A18,
  \dodoi{10.1051/0004-6361/201116999}

\bibitem[{{Norfolk} {et~al.}(2019){Norfolk}, {Casey}, {Karakas}, {Miles},
  {Kemp}, {Schlaufman}, {Ness}, {Ho}, {Lattanzio}, \& {Ji}}]{Nor2019}
{Norfolk}, B.~J., {Casey}, A.~R., {Karakas}, A.~I., {et~al.} 2019, \mnras, 490,
  2219, \dodoi{10.1093/mnras/stz2630}

\bibitem[{{North} {et~al.}(1994){North}, {Berthet}, \& {Lanz}}]{Nor1994}
{North}, P., {Berthet}, S., \& {Lanz}, T. 1994, \aap, 281, 775

\bibitem[{{North} {et~al.}(2000){North}, {Jorissen}, \& {Mayor}}]{Nor2000}
{North}, P., {Jorissen}, A., \& {Mayor}, M. 2000, in IAU Symposium, Vol. 177,
  The Carbon Star Phenomenon, ed. R.~F. {Wing}, 269

\bibitem[{{North} \& {Kobi}(1991)}]{Nor1991}
{North}, P., \& {Kobi}, D. 1991, in Precision Photometry: Astrophysics of the
  Galaxy, ed. A.~G.~D. {Philip}, A.~R. {Upgren}, \& K.~A. {Janes}, 327

\bibitem[{{North} \& {Lanz}(1991)}]{Nor1991A}
{North}, P., \& {Lanz}, T. 1991, \aap, 251, 489

\bibitem[{{Pinsonneault} {et~al.}(2001){Pinsonneault}, {DePoy}, \&
  {Coffee}}]{Pin2001}
{Pinsonneault}, M.~H., {DePoy}, D.~L., \& {Coffee}, M. 2001, \apjl, 556, L59,
  \dodoi{10.1086/323531}

\bibitem[{{Reddy} {et~al.}(2006){Reddy}, {Lambert}, \& {Allende
  Prieto}}]{Red2006}
{Reddy}, B.~E., {Lambert}, D.~L., \& {Allende Prieto}, C. 2006, \mnras, 367,
  1329, \dodoi{10.1111/j.1365-2966.2006.10148.x}

\bibitem[{{Talon} {et~al.}(2006){Talon}, {Richard}, \& {Michaud}}]{Tal2006}
{Talon}, S., {Richard}, O., \& {Michaud}, G. 2006, \apj, 645, 634,
  \dodoi{10.1086/504066}

\bibitem[{{Ting} {et~al.}(2017){Ting}, {Conroy}, {Rix}, \&
  {Cargile}}]{Ting2017a}
{Ting}, Y.-S., {Conroy}, C., {Rix}, H.-W., \& {Cargile}, P. 2017, \apj, 843,
  32, \dodoi{10.3847/1538-4357/aa7688}

\bibitem[{{Udry} {et~al.}(1998{\natexlab{a}}){Udry}, {Jorissen}, {Mayor}, \&
  {Van Eck}}]{Ud1998b}
{Udry}, S., {Jorissen}, A., {Mayor}, M., \& {Van Eck}, S. 1998{\natexlab{a}},
  \aaps, 131, 25, \dodoi{10.1051/aas:1998249}

\bibitem[{{Udry} {et~al.}(1998{\natexlab{b}}){Udry}, {Mayor}, {Van Eck},
  {Jorissen}, {Prevot}, {Grenier}, \& {Lindgren}}]{Ud1998a}
{Udry}, S., {Mayor}, M., {Van Eck}, S., {et~al.} 1998{\natexlab{b}}, \aaps,
  131, 43, \dodoi{10.1051/aas:1998250}

\bibitem[{{Vick} {et~al.}(2010){Vick}, {Michaud}, {Richer}, \&
  {Richard}}]{Vic2010}
{Vick}, M., {Michaud}, G., {Richer}, J., \& {Richard}, O. 2010, \aap, 521, A62,
  \dodoi{10.1051/0004-6361/201014307}

\bibitem[{{Wang} {et~al.}(2019){Wang}, {Liu}, {Xiang}, {Huang}, {Chen}, {Yuan},
  {Ren}, {Zhang}, \& {Tian}}]{Wang2019}
{Wang}, C., {Liu}, X.~W., {Xiang}, M.~S., {et~al.} 2019, \mnras, 482, 2189,
  \dodoi{10.1093/mnras/sty2797}

\bibitem[{{Webbink}(1986)}]{Web1986}
{Webbink}, R.~F. 1986, Highlights of Astronomy, 7, 185

\bibitem[{{Wu} {et~al.}(2019){Wu}, {Xiang}, {Zhao}, {Bi}, {Liu}, {Shi},
  {Huang}, {Yuan}, {Wang}, {Chen}, {Huo}, {Ren}, {Tian}, {Liu}, {Zhang}, {Li},
  \& {Zhang}}]{Wu2019}
{Wu}, Y., {Xiang}, M., {Zhao}, G., {et~al.} 2019, \mnras, 484, 5315,
  \dodoi{10.1093/mnras/stz256}

\bibitem[{{Xiang} \& {Rix}(2022)}]{Xiang2022}
{Xiang}, M., \& {Rix}, H.-W. 2022, \nat, 603, 599,
  \dodoi{10.1038/s41586-022-04496-5}

\bibitem[{{Xiang} {et~al.}(2019){Xiang}, {Ting}, {Rix}, {Sand ford}, {Buder},
  {Lind}, {Liu}, {Shi}, \& {Zhang}}]{Xiang2019}
{Xiang}, M., {Ting}, Y.-S., {Rix}, H.-W., {et~al.} 2019, \apjs, 245, 34,
  \dodoi{10.3847/1538-4365/ab5364}

\bibitem[{{Xiang} {et~al.}(2020){Xiang}, {Rix}, {Ting}, {Ludwig}, {Coronado},
  {Zhang}, {Zhang}, {Buder}, \& {Dal Tio}}]{Xiang2020}
{Xiang}, M., {Rix}, H.-W., {Ting}, Y.-S., {et~al.} 2020, arXiv e-prints,
  arXiv:2006.03329.
\newblock \doarXiv{2006.03329}

\bibitem[{{Xiang} {et~al.}(2017){Xiang}, {Liu}, {Shi}, {Yuan}, {Huang}, {Luo},
  {Zhang}, {Zhao}, {Zhang}, {Ren}, {Chen}, {Wang}, {Li}, {Huo}, {Zhang},
  {Wang}, {Zhang}, {Hou}, \& {Wang}}]{Xiang2017}
{Xiang}, M.~S., {Liu}, X.~W., {Shi}, J.~R., {et~al.} 2017, \mnras, 464, 3657,
  \dodoi{10.1093/mnras/stw2523}

\bibitem[{{Yang} {et~al.}(2016){Yang}, {Liang}, {Spite}, {Chen}, {Zhao},
  {Zhang}, {Liu}, {Liu}, {Liu}, {Deng}, {Spite}, {Hill}, \& {Zhang}}]{Yang2016}
{Yang}, G.-C., {Liang}, Y.-C., {Spite}, M., {et~al.} 2016, Research in
  Astronomy and Astrophysics, 16, 19, \dodoi{10.1088/1674-4527/16/1/019}

\bibitem[{{Yuan} {et~al.}(2013){Yuan}, {Liu}, \& {Xiang}}]{Yuan2013}
{Yuan}, H.~B., {Liu}, X.~W., \& {Xiang}, M.~S. 2013, \mnras, 430, 2188,
  \dodoi{10.1093/mnras/stt039}

\bibitem[{{Yuan} {et~al.}(2015){Yuan}, {Liu}, {Huo}, {Xiang}, {Huang}, {Chen},
  {Zhang}, {Sun}, {Wang}, {Zhang}, {Zhao}, {Luo}, {Shi}, {Li}, {Yuan}, {Dong},
  {Li}, {Hou}, \& {Zhang}}]{Yuan2015}
{Yuan}, H.~B., {Liu}, X.~W., {Huo}, Z.~Y., {et~al.} 2015, \mnras, 448, 855,
  \dodoi{10.1093/mnras/stu2723}

\bibitem[{{Zhang} {et~al.}(2021){Zhang}, {Xiang}, {Zhang}, {Ting}, {Rix}, {Wu},
  {Huang}, {Sun}, {Tian}, {Wang}, \& {Liu}}]{Zhang2021}
{Zhang}, M., {Xiang}, M., {Zhang}, H.-W., {et~al.} 2021, \apj, 922, 145,
  \dodoi{10.3847/1538-4357/ac22a5}

\bibitem[{{Zhao} {et~al.}(2012){Zhao}, {Zhao}, {Chu}, {Jing}, \&
  {Deng}}]{Zhao2012}
{Zhao}, G., {Zhao}, Y.-H., {Chu}, Y.-Q., {Jing}, Y.-P., \& {Deng}, L.-C. 2012,
  Research in Astronomy and Astrophysics, 12, 723,
  \dodoi{10.1088/1674-4527/12/7/002}

\end{thebibliography}
\end{document}